\documentclass[12pt]{iopart}
\usepackage{graphicx}

 \def\Pom{{ I\!\!P}}  
 \def\Reg{{ I\!\!R}}  
 \def\gsim{\mathrel{\rlap{\lower4pt\hbox{\hskip1pt$\sim$}}
 \raise1pt\hbox{$>$}}}

 \renewcommand{\floatsep} {-3cm}
 
 \newcommand\la{\langle}
 \newcommand\ra{\rangle}
 \newcommand\beq{\begin{equation}}
 
 \newcommand\eeq{\end{equation}}
 \newcommand\beqn{\begin{eqnarray}}
 \newcommand\eeqn{\end{eqnarray}}
\def\mb{\,\mbox{mb}}
\def\fm{\,\mbox{fm}}
\def\GeV{\,\mbox{GeV}}

\def\lsim{\mathrel{\rlap{\lower4pt\hbox{\hskip1pt$\sim$}} 
    \raise1pt\hbox{$<$}}}         %less than or approx. symbol
\def\gsim{\mathrel{\rlap{\lower4pt\hbox{\hskip1pt$\sim$}}
    \raise1pt\hbox{$>$}}}         %greater than or approx. symbol

\newsavebox{\fmbox}

\begin{document}

%\review{Applied High Energy QCD}
\title[Applied High Energy QCD ]{Applied High Energy QCD}
\author{B. Z. Kopeliovich  \dag\ddag ~ and A. H. Rezaeian\dag}
\address{\dag~ ~Departamento de F\'\i sica y Centro de Estudios Subat\'omicos, \\
~~~Universidad T\'ecnica Federico Santa Mar\'\i a, Casilla 110-V,
Valpara\'\i so, Chile \\ 
\ddag~Joint Institute for Nuclear Research,
Dubna, Russia}

\begin{abstract}
 These lectures stress the theoretical elements that underlie a wide range of phenomenological
studies of high-energy QCD, which include both soft and hard
processes. After a brief introduction to the basics of QCD, various aspects of
QCD-based phenomenology are covered: colour transparency, hadronization
of colour charges, Regge phenomenology, parton model, Bjorken scaling and
its violation, DGLAP evolution equation, BFKL formalism, GLR-MQ evolution
equation and saturation. In the last part of the lecture, we employ the light-cone
dipole formalism to describe deep inelastic lepton scattering, Drell-Yan
processes, direct photon production, diffraction, quark and gluon shadowing
in nuclei, the Cronin effect and nuclear broadening.
\end{abstract}

%\submitto{\RPP}

\maketitle
%\vspace*{.4cm} 
%\nopagebreak
\tableofcontents 

%%%%%%%%%%%%%%%%%%%%%%%%%%%%%%%%%%%%%%%%
\section{The theory of strong interaction} 
Strong interactions are described by a quantum field theory known as
quantum chromodynamics (QCD).  In many ways QCD is a unique
theory. Quantum electrodynamics (QED), and its expansion to the
electroweak Standard Model of particle physics, is also a quantum
field theory. QED is a renormalizable theory but it loses all its
credibility as we approach the energy scale, the so-called Landau pole position, where the strength of the
coupling constant (the strength of the interaction) becomes
infinite\footnote{The dependence of coupling constants on the energy
scale is one of the basic ideas behind the renormalization group which
will be discussed in Section 1.3.}. On the other hand, if the
cutoff goes to infinity, QED becomes trivial. QED is not the only
theory with a Landau pole problem; every theory which is not
asymptotically free suffers from this problem. QCD is the only known
theory which is free from such problems. QCD needs only a few
parameters to be defined completely: one universal coupling strength
and one mass for each kind of quark.

Despite more than half a century of attempts, our knowledge about many aspects of QCD is still
rudimentary. This is mainly due to the fact that QCD evolves from a few-body theory of free quarks
and gluons at short distances to an extremely complicated infinite-body theory of objects like hadrons
and nuclei, giving rise to a variety of complex physical systems and their interactions. The aim of this
manuscript is to bring together various aspects of high-energy nuclear physics as tools for studying QCD
itself.

\subsection{QCD Lagrangian and it symmetries} 
\begin{figure}[!t]
       \centerline{\includegraphics[width=9.5 cm] {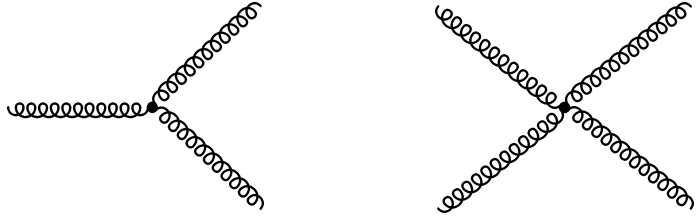}}
        \caption{Gluons carry colour charge and interact with each other via these vertices.\label{nona}}
\end{figure}

The Lagrangian of QCD is given by
%,
\begin{equation}
\mathcal{L}=\bar{q}(i\gamma^{\mu}\partial_{\mu}-m^{0})q-\frac{1}{4}(F^{a}_{\mu\nu})^{2}+g\bar{q}\gamma^{\mu}A_{\mu}q, \label{qcd1}
\end{equation}
where $q$ is the quark field which is defined in the fundamental
representation of the colour and flavor group, and the conjugate Dirac
field is defined as $\bar{q}=q^{\dag}\gamma^{0}$. The gluon field
matrix $A^{\mu}=A^{a}_{\mu}\lambda^{a}/2$ is defined in the
fundamental $SU(N_{c}=3)$ representation where $N_c$ denotes the
number of colour, $\lambda^{a}$ being the generators of the gauge
group which satisfies
$[\lambda^{a}/2,\lambda^{b}/2]=if^{abc}\lambda^{c}/2$ where $f^{abc}$
are the structure constants of $SU(3)$. We define $g$ as the strong
coupling constant. The field strength $F^{a}_{\mu\nu}$ is given by
\begin{equation}
F^{a}_{\mu\nu}=\partial_{\mu}A^{a}_{\nu}-\partial_{\nu}A^{a}_{\mu}+gf^{abc}A^{b}_{\mu}A^{c}_{\nu}.
\end{equation}
The non-Abelian nature of QCD is manifested by the quadratic term in
the gauge field strength, which gives rise to gluon-gluon interactions
shown in Fig.~\ref{nona}. The crucial difference between QCD and QED
is the presence of this quadratic term which makes the QCD field equations non-linear. These nonlinearities
give rise to a non-trivial dynamics and various rich structures which are unique properties of
the strong interaction. The colour and flavour indices of the quark field are suppressed. $m^0$ is the current
quark mass which is not directly observable if QCD confines quarks. The current quark mass is colour
independent and can be brought diagonal in flavour space. There are six flavours of quarks, each of
which has a different mass. The three light quarks are called up (u), down (d) and strange (s), while the
three heavy quarks are called charm (c), bottom (b) and top (t). The following values for the light current
quark masses are found in the Particle Data tables \cite{qcd-data},
\begin{equation}
\fl m^{0}_{u}=2~~\hbox{to}~~ 8~~\hbox{MeV}, \hspace{1cm}  m^{0}_{d}=5~~\hbox{to}~~ 15~~\hbox{MeV},\hspace{1cm}  m^{0}_{s}=100~~\hbox{to}~~300~~\hbox{MeV}.
\end{equation}

Notice that the quark masses are renormalization-scheme dependent. The
above values are obtained in a subtraction scheme at a renormalization
scale $\mathcal{O}(1\hbox{GeV})$.  In addition to flavour, quarks carry
another quantum number known as colour. Each quark comes in three
colours which, based on a convention, are called red, green and blue.

The Lagrangian  Eq.~(\ref{qcd1}) has a large classical symmetry: we have the local gauge symmetry $SU(N_{c})$ by construction,
\begin{eqnarray}
&&q\to U_{c}q, \hspace{2cm} \bar{q}\to\bar{q}U^{\dag}_{c}, \hspace{2cm} U_{c}(x)=\exp(i\theta^{a}(x)(\frac{\lambda^{a}}{2})_{c}),\nonumber\\
&& A_{\mu}\to U_{c}A_{\mu}U^{\dag}_{c}-\frac{1}{g}U_{c}i\partial_{\mu}U^{\dag}_{c}. \nonumber\
\end{eqnarray}
In QED, there is only one electric charge, and the gauge transformation
involves a single phase factor $U=\exp(i\alpha(x))$. The QCD
Lagrangian Eq.~(\ref{qcd1}) has also a global flavour symmetry which
does not affect the gluon fields,
\begin{equation}
q\to U_{V}q, \hspace{2cm} \bar{q}\to\bar{q}U^{\dag}_{V},\hspace{2cm}  U_{V}=\exp(i\theta^{a}_{V}(\frac{\lambda^{a}}{2})_{F}). \label{qcd2}
\end{equation}
where $(\frac{\lambda^{a}}{2})_{F}$ denotes the generators of the
flavour group $U(N_{f})$ and $N_{f} $ denotes the number of
flavors. The above symmetry is referred to as vector flavor symmetry
$U_{V}(N_{f})$. When the generator is the unit matrix, we have
$U_{V}(1)$ symmetry associated with conservation of baryon number. There is
another global symmetry which is exact at $m^{0}=0$, namely chiral
symmetry. This symmetry is very similar to vector flavor symmetry,
apart from an extra factor of $\gamma_{5}$ in the generator of
the transformation.
\begin{equation}
q\to U_{A}q, \hspace{2cm} \bar{q}\to\bar{q}U_{A}, \hspace{2cm} U_{A}=\exp\left(i\gamma_{5}\theta_{A}^{a}(\frac{\lambda^{a}}{2})_{F}\right). \label{refax}
\end{equation}
Notice that due to the factor $\gamma_{5}$ the quark field and its
conjugate partner are transformed by the same matrix in contrast
to vector transformation Eq.~(\ref{qcd2}). This transformation Eq.~(\ref{refax}) is
called the axial-vector transformation and can be combined with
the vector transformation to define a bigger symmetry at chiral
$m^{0}=0$ which is then called chiral symmetry $U_{V}(N_{f})\times
U_{A}(N_{f})$. One may alternatively define right- and left-handed 
quark fields by following transformation
\begin{equation}
q_{L}=\frac{1-\gamma_{5}}{2}q,
\hspace{2cm}q_{R}=\frac{1+\gamma_{5}}{2}q,
\end{equation}
The right- and left-handed massless fermions are eigenvalues of the
helicity or chirality (with eigenvalue $\pm 1$) and are not mixed
together. The chiral symmetry can be equivalently written as
$U_{L}(N_{f})\times U_{R}(N_{f})$.  

Not all the above-mentioned symmetries survive quantization. Particles with opposite helicity are
related by a parity transformation, therefore in a chirally symmetric world, the hadrons should come in
parity doublets. However, in real life we do not observe such degeneracy. Therefore one can conclude
that chiral symmetry is not realized in the ground state and chiral symmetry is spontaneously broken.
A theory where the vacuum has less symmetry than the Lagrangian is called a theory with spontaneous symmetry breaking. 
The Goldstone theorem \cite{qcd1} tell us that the spontaneous
breaking of a continuous global symmetry implies the existence of
associated massless spinless particles. This indeed was
confirmed due to the existence of the light pseudoscalar mesons in
nature (pions, kaons and etas) which may be assigned as pseudo-Goldstone
bosons \cite{qcd2}. Moreover, the existence of a quark condensate $\langle\bar{q}q\rangle$
implies that the $SU(N_{f})_{L}\times SU(N_{f})_{R}$ symmetry is
spontaneously broken down to $SU(N_{f})_{V}$. Therefore one may
conceive QCD quark condensate as an order parameter for
chiral symmetry breaking. The concept of
spontaneous broken chiral symmetry is the cornerstone in the
understanding of the low-energy hadronic spectrum.
\begin{figure}[!t]
       \centerline{\includegraphics[width=6 cm] {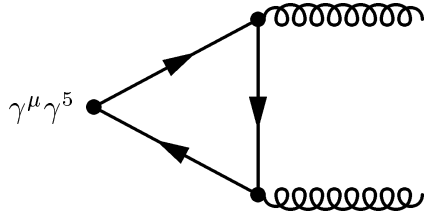}}
        \caption{The diagram corresponding to the $U_{A}(1)$-anomaly. \label{tri}}
\end{figure}

 The $U(1)_{A}$ symmetry implies that all hadrons should come with
 opposite parity partners. However, this is not the case, therefore
 this symmetry must be broken somehow. If the spontaneous symmetry
 breaking mechanism works here, then one should observe a Goldstone
 boson associated with $U(1)_{A}$, namely an $I=0$ pseudoscalar meson
 having roughly the same mass as the pion. Surprisingly there is no
 such Goldstone boson. This problem is sometime called $U(1)_{A}$
 puzzle. It turned out that the $U(1)_{A}$ symmetry is explicitly
 broken by quantum effects. This effect is known as the axial anomaly
 \cite{qcd3}. The axial charge corresponding to the axial current
 $j_{\mu}^{5}=\bar{q}\gamma_{\mu}\gamma^{5}q$ is not conserved because
 of the contribution of the triangle graph in Fig.~\ref{tri}. The four-divergence of the axial current is given by
\begin{equation}
\partial^{\mu}J_{\mu}^{5}=\sum_{q} 2im_{q}\bar{q}\gamma^{5}q +\frac{N_{f}}{8\pi^{2}}tr G^{\mu\nu}\bar{G}_{\mu\nu},
\end{equation}
where $\bar{G}_{\mu\nu}=\epsilon_{\mu\nu k\lambda}G^{k\lambda}/2$ is
the dual field strength tensor. The last term (gluonic part) is a full
divergence, and one may expect that this term not to have any
physical effect if the QCD vacuum were trivial. It was shown by 't Hoof that
 due to instanton effects, the $U(1)_{A}$ symmetry is not manifested
 in nature \cite{qcd3}.

Finally, at $m^{0}=0$, the QCD Lagrangian is
invariant under a scale transformation which is called
dilatational symmetry:
\begin{equation}
\fl q(x)\to \epsilon^{3/2}q(\epsilon^{-1}x), \hspace{2cm} A^{a}_{\mu}(x)\to\epsilon A^{a}_{\mu}(\epsilon^{-1}x), \hspace{2cm} x_{\mu}\to \epsilon^{-1}x_{\mu}.
\end{equation}
This symmetry is again broken at the quantum level due to the trace
anomaly \cite{qcd4}.

\subsection{QCD versus QED}
Let us remember the main differences between QCD and QED. QCD is an
extended version of QED which now, instead of one charge, has three
different kinds of charge called colour. Similar to the photon in QED,
here massless spin-one particles, the gluons, respond to the presence
of colour charge.  The colour charged quarks emit and absorb gluons in
the same way as electrically charged leptons do.  However, radiation
of a photon does not change the charge of the electron, while a gluon
can change the quark colour. The response of gluons to colour charge,
as measured by the QCD coupling constant, is much more drastic than
the response of photons to electric charge. Gluons, unlike photons,
interact directly with each other, although the colour charges, like
electric charge in QED, are conserved in all physical
processes. Therefore gluons must be able to carry unbalanced colour
charges in contrast to their counterpart the photon in QED.

\begin{figure}[!t]
       \centerline{\includegraphics[width=14 cm] {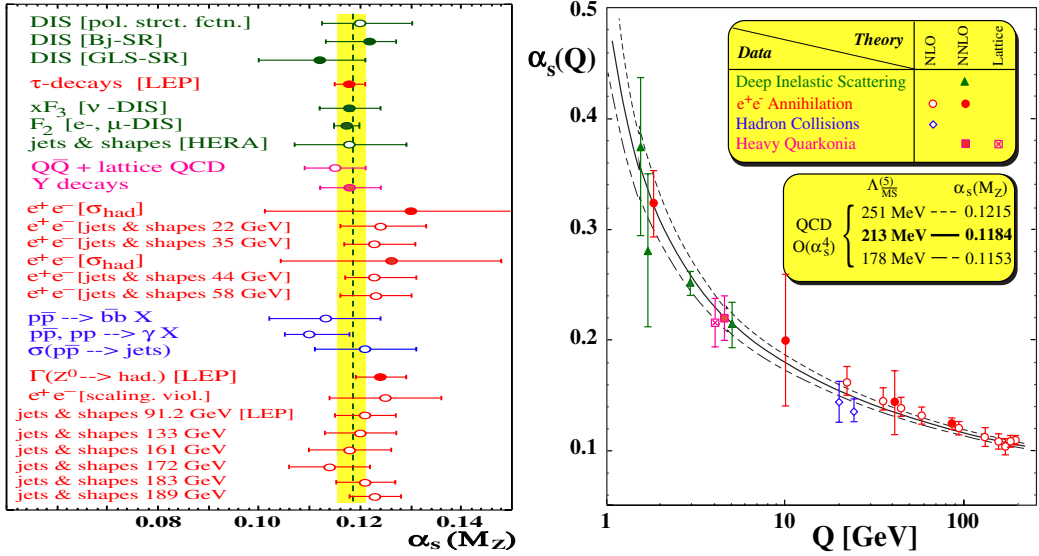}}
       \caption{Right: The running coupling constant as a function of
       momentum transfer $Q^{2}$ determined from different
       processes. Left: Summary of $\alpha_{s}$ \cite{alphb}. \label{al}}
\end{figure} 

%\section{Non-perturbative features of QCD}
In the following sections we shall also recapitulate the most important features of QCD which are not accessible perturbatively.
These non-perturbative features are unique for QCD and should be traced back to the main
differences between QCD and QED.

\subsection{Asymptotic Freedom}
Having introduced the gauge fixing term and an associated
ghost term by means of the Faddeev-Popov procedure
\cite{qcd5,qcd1}, one can carry out perturbation theory in terms
of coupling. Similar to QED, a dimensionless physical quantities
$\mathcal{R}$ can be expressed by a perturbation series in powers of
the coupling parameter $\alpha_{s}$ ($\alpha_{s}$ is the notation for
$g^{2}/4\pi$).  Owing to the renormalization process, a renormalization
scale
 $\mu$ enters the algebra
\cite{qcd6} in order to remove the
ultraviolet divergence. Therefore, one can write the dimensionless
quantities $\mathcal{R}$ in terms of other available dimensionless
parameters $Q^2 / \mu^2 $ and the renormalized coupling
$\alpha_{s}(\mu^2 )$.  However, the physical quantity $\mathcal{R}$
cannot depend on the arbitrary $\mu$. This means that $\mathcal{R}$ should
be renormalization scale invariant
\begin{equation} 
\mu^{2}\frac{d\mathcal{R}}{d\mu^{2}}=\Big[\mu^{2}\frac{\partial}{\partial \mu^{2}}
+\mu^{2}\frac{d\alpha_{s}}{d\mu^{2}} \frac{d\partial}{d\alpha_{s}}\Big]\mathcal{R}\left(\alpha_{s}(\mu^{2}),Q^{2}/\mu^{2}\right)=0.
\end{equation}
This equation explicitly shows that any dependence of $\mathcal{R}$ on
$\mu$ must be cancelled by an appropriate $\mu$-dependence of $\alpha_{s}$.
It is also natural to identify the renormalization scale with the
physical energy scale of the process, i.e.  $\mu^2 = Q^2$. 
The running coupling is described by the renormalization group equation \cite{qcd6},
\begin{equation}
Q^2 \frac{\partial \alpha_{s}}{\partial Q^2} = \beta \left( \alpha_{s}(Q^{2})\right) \ .
\end{equation}
Whenever the coupling is small, the $\beta$ function can be computed perturbatively,
\begin{equation}
\beta(\alpha_{s})= - \beta_{0} \alpha_{s}^2(Q^2) - \beta_{1} \alpha_{s}^3(Q^2)+...,
\end{equation}
with 
\begin{equation}
\beta_{0}=\frac{33 - 2 N_f}{12 \pi}, \hspace{2cm} \beta_{1}=\frac{153 - 19 N_f}{24 \pi^2}.
\end{equation}
Therefore one can readily calculate the effective running coupling at one-loop level ignoring the $\beta_{1}$ term, 
\begin{equation}
\alpha_{s}(Q^{2})= \frac{1}{\beta_{0} \ln{\frac{Q^{2}}{\Lambda^{2}}}},
\end{equation}
where $\Lambda$ is a scale parameter of QCD and depends on the subtraction scheme and the number of active flavours. 
The present world average for $\alpha_{s}$ at the $Z^{0}$ mass is $
\alpha_{s}(M_{Z})=0.118\pm 0.002$ which leads to
\begin{equation}
\Lambda_{\overline{MS}}^{(5)}=(208^{+25}_{-23}) \hbox{MeV},
\end{equation}
where the symbol $\overline{MS}$ stands for minimal subtraction scheme
\cite{qcd6} and the superscript indicates the number of active
flavours. This value is taken from an analysis of various high
energy processes \cite{qcd7,alphb}, see also Fig.~\ref{al}. The most striking feature of
the running coupling is that it decreases logarithmically with $Q^{2}$
for $N_{f}<17$ when $\beta_{0}>0$. This originates from the
self-interaction of gluons which leads to anti-screening, in contrast
to QED where the sign of $\beta_{0}$ is negative. Therefore 
perturbation theory works very well for large $Q^{2}$.  This
phenomenon is called asymptotic freedom
\cite{qcd8}.  However, if $Q^{2}$ is near $\Lambda_{\overline{MS}}$,
perturbation theory does not work anymore and non-perturbative
phenomena enter the stage. One of the biggest challenges of QCD is to connect these two domains. 
Admittedly, there is yet no unambiguous
method to connect small and large distances in QCD.

\subsection{Chiral symmetry breaking}
In the first section, we introduced the symmetries of the QCD Lagrangian. In the
limit of massless quarks, QCD possesses chiral symmetry
$U_{L}(N_{f})\times U_{R}(N_{f})$ which means that left- and
right-handed quarks are not mixed, 
\begin{equation}
q_{L}\to V_{L}q_{l}; \hspace{1cm} q_{R}\to V_{R}q_{R};\hspace{1cm} V_{L},V_{R}\in U(N_{f}).
\end{equation}
As we already discussed, owing to the presence of the quark condensate
$\langle \bar{q} q\rangle$, chiral symmetry is spontaneously
broken and left- and right-handed quarks and antiquarks can transform into each other: 
\begin{equation}
\langle \bar{q} q\rangle =\langle \bar{q}_{L} q_{R}\rangle +\langle \bar{q}_{R} q_{L}\rangle.
\end{equation}
Dynamical chiral symmetry breaking is one of the important
non-perturbative features of QCD which is responsible for the generation
of quark masses\footnote{There is another very different
way to generate mass from vacuum, the so-called Casimir effect
\cite{qcd-new81}, which originates from the response of the vacuum in the
presence of non-perturbative boundary conditions. The existence of boundary conditions in quantum
field theory is not always free of problems (see, for example,  
Ref.~\cite{qcd-new82}).}. In order to show that this phenomenon is purely non-perturbative, we
employ the QCD gap equation \cite{asy},
\begin{equation}
S(p)^{-1}=(i\gamma .p+m^{0})+\int \frac{d^{4}q}{(2\pi)^{4}} g^{2}D_{\mu\nu}(p-q)\frac{\lambda^{a}}{2}\gamma_{\mu}S(q)\Gamma^{a}_{\nu}(p,q), \label{qcd-gap}
\end{equation}
where $m^{0}$ and $g$ are the current-quark bare mass and the coupling constant,
respectively. $D_{\mu\nu}(p-q)$ is the dressed-gluon
propagator and $\Gamma^{a}_{\nu}(p,q)$ is the dressed-quark-gluon vertex.
The general solution of the gap equation is a dressed-quark propagator of the form
\begin{equation}
S(p)=\frac{1}{i\gamma .pA(p^{2})+B(p^{2})}=\frac{Z(p^{2})}{i\gamma .p+M(p^{2})}. \label{qcd-s}
\end{equation}
The functions $A(p^2)$ and $B(p^2)$ contain the effects of vector and
scalar quark-dressing induced by the quark interaction with the
gluon field. The function $M(p^{2})$ denotes the quark mass. One may
now use the gap equation to work out the fermion self-energy
perturbatively \cite{qcd9}. One obtains,
\begin{equation}
B(p^{2})=m^{0}\left(1-\frac{\alpha}{\pi}\ln (p^{2}/m^{2})+...\right).
\end{equation}
It is observed that at all orders of the loop expansion, terms are
proportional to the current-quark mass and consequently vanish as
$m^{0}\to 0$. 
The quark mass is defined as a pole of the dressed-quark propagator;
therefore no mass is generated at a current-quark mass equal to zero, i.e., the dynamical chiral symmetry
breaking is impossible in perturbation theory and there is no mixing between left- and right-handed
quarks at the perturbative level. Notice that, apart from the
trivial solution $B(p^{2})=0$ at $m=0$, a non-trivial solution
$B(p^{2})\ne 0$ can indeed be found at the chiral point, albeit
accessible non-perturbatively. The renormalization effect is not
included in Eq.~(\ref{qcd-gap}), but it does not change the above
argument \cite{qcd9}. The quark condensate\footnote{Note that, at finite density and
temperature, the formation of a quark cooper pair condensate $\langle
qq\rangle\ne 0$ is also possible, leading to colour
symmetry breaking, the so-called colour superconductivity phenomenon(BCS)
\cite{bcs} and diquark Bose-Einstein condensation(BEC) \cite{bec}.} in
QCD is given by the trace of the full quark propagator
Eq.~(\ref{qcd-s}),
\begin{equation}
\langle\bar{q}q\rangle=-i\lim_{y\to x} \Tr S(x,y).
\end{equation}
Notice that since $\bar{q}q$ is a gauge invariant object, one may take any
gauge to obtain the dressed quark propagator which has a general
form as equation (\ref{qcd-s}). It is obvious that when
$B(p^{2})= 0$, the quark condensate does not take place, simply
because of the identity $\Tr\gamma_{\mu}=0$. It has been shown in many
non-perturbative approaches that the emergence of a dynamical quark mass
leads to the non-vanishing of quark condensate and vice versa, see, for example, Refs.~\cite{mech,phdme}.

\subsection{Confinement}
Another important non-perturbative feature of QCD is colour confinement
\cite{qcd8}. Loosely speaking, confinement is defined as the absence of any free coloured objects in nature. But it is possible that there exists
a composite coloured particle which can form colourless bound states with another coloured particle like
quarks. Colour confinement is still not properly understood, and a clear and indisputable mechanism responsible
for this effect remains yet to be discovered. The basic property of confinement can be explored
by looking at heavy $q\bar{q}$ propagation at a large distance $R$ in a time interval
$T$. The behaviour of such a system can be described by the Wilson loop,
\begin{equation}
W(R,T)=Tr[P\exp\left(i\int_{C}A_{\mu}^{a}T^{a}dx^{\mu}\right)],
\end{equation}
where $T^{a}$ denotes the generator of
$SU(3)$. One can show that at large interval of time $T$,
\begin{equation}
W(R,T\to \infty)=\exp\left(-T V(R)\right),
\end{equation}
where $V(R)$ is the static potential between the heavy quarks. At large distances this potential grows  linearly:
\begin{equation}
V(R\to \infty)=\sigma R.
\end{equation}
Therefore the Wilson loop at large $R$ and $T$ behaves as
$W(R\to \infty,T\to \infty)=\exp\left(-\sigma T R\right)$,
which is the so-called area low and indicates confinement.

Confinement originates non-perturbatively, since it is associated with
a linear potential with a string tension
\begin{equation}
\sigma\propto \Lambda^{2}exp\left(-\int\frac{dg}{\beta(g)}\right),
\end{equation}
which is obviously non-perturbative in the coupling. Note that the
string picture of quark confinement is not free of flaws, since string
breaking will occur once the potential energy approaches the quark
pair creation threshold.

It is well-known that for confinement it is sufficient that no
coloured Schwinger function possesses a spectral representation. This
is equivalent to say that all coloured Schwinger functions violate
reflection positivity \cite{asy}. Another way of realization of QCD
confinement is due to Gribov theory in which colour confinement is
determined by the existence of very light (almost
massless) quarks \cite{g99}.  There are in fact many different ways
that the confinement can be realized, such as monopole condensation, infrared enhancement of the ghost propagator, etc. For a review
of this subject see Ref.~\cite{qcd12}.

One may wonder if there is a non-trivial solution for the gap equation
$B(p^{2})\ne 0$ which gives rise to a pole of the quark propagator,
this might contradict QCD confinement since the quark is
coloured. Indeed this is one of the subtle point in every QCD model and
cannot be easily resolved. In principle, there will be a long-range
force between massive quarks to confine them and also a short range
spin-spin interaction between massive dressed quarks. The former will
modify the low momentum part of the propagator to remove the quark
from being on-shell. Actually, this describes a phenomenologically
motivated picture of a constituent quark model based on the dynamical
symmetry breaking.  Having said that, it is very hard to incorporate
the dynamical symmetry breaking and the confinement into a QCD model.
In fact, many models constructed to describe the low-energy properties
of hadrons \cite{phdme,fs4} are assumed to be only dominated by the quark flavor
dynamics and dynamical symmetry breaking and are indeed reliable only
at intermediate scales, between confinement scale few hundred MeV up
to a scale about 1 GeV.

\section{Evidences for coloured quarks} 
Historically, the idea of colour degree of freedom emerged as a viable
solution to the problem of how to construct the wave function for the
doubly charged $\Delta^{++}$ baryons \cite{greenb}. The wave function of
$\Delta^{++}$ in space, spin and flavour is symmetric and violates
the Pauli exclusion principle since $\Delta^{++}$ is fermion with spin $3/2$. This problem was resolved by introducing a new degree
of freedom, the colour degree of freedom, and requiring that
the $\Delta^{++}$ wave function to be antisymmetric in the colour degree of
freedom. 
\begin{figure}[!t]
       \centerline{\includegraphics[width=10.0 cm] {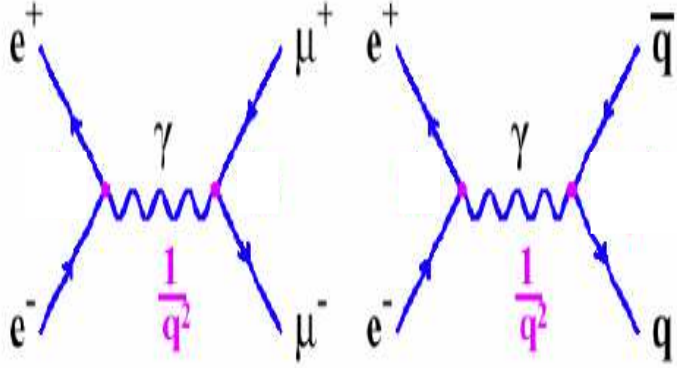}}
       \caption{$e^{+}e^{-}$ annihilation to $q\bar{q}$ or
       $\mu^{+}\mu^{-}$ pair. \label{color0}}
\end{figure}

%\begin{figure}[thb]
%       \centerline{\includegraphics[width=12.0 cm] {loop1.ps}}
%        \caption{ Hadronic vacuum polarization\label{color3}}
%\end{figure}

Although coloured states are not detected in experiments, and only
colour singlet states exist in nature, there is much
experimental evidence in favour of a colour degree of freedom.  One of the
direct experimental test for a colour degree of freedom comes from
$e^{+}e^{-}$ annihilation into hadrons.  In the $e^{+}e^{-}$
annihilation process, first a pair of quarks $e^{+}e^{-}\to q\bar{q}$
is produced which then fragment into hadrons. The cross-section for
producing a free $q\bar{q}$ pair is the same as for producing a
$\mu^{+}\mu^{-}$ pair except for the quark charge and colour number which
should be replaced with the muon charge, see Fig.~\ref{color0}. Therefore in order to
extract information about the QCD content of $e^{+}e^{-}$ annihilation, in
particular, the colour degree of freedom, it is convenient to express
the total cross-section of $e^{+}e^{-}\to q\bar{q}$ annihilation in
units of the cross-section of $\mu$ production,
\begin{equation}
R=\frac{e^{+}e^{-}\to \hbox{Hadrons}}{e^{+}e^{-}\to\mu^{+}\mu^{-}}.
\end{equation}
\begin{figure}[!t]
       \centerline{\includegraphics[width=12.0 cm] {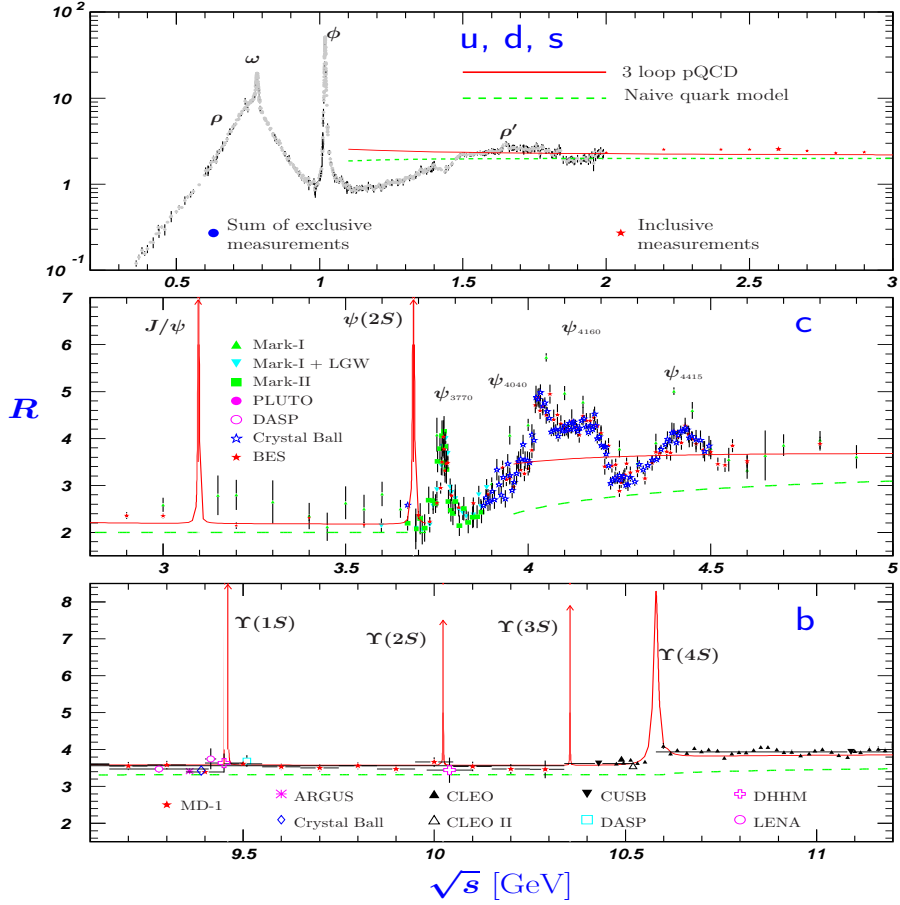}}
        \caption{ R in the light-flavor, charm, and beauty threshold regions. Data errors are total below 2 GeV and statistical above 2 GeV.
 The full list of references to the original data and the details of the R ratio extraction from them can be found in \cite{qcolor}. \label{color1}}
\end{figure}

The cross-section to produce any number of hadrons is
proportional to that to produce a $\mu^{+}\mu^{-}$ pair. 
This is because a highly virtual photons decays to
quarks in a time scale $t\sim 1/\sqrt{s}$ (where  $\sqrt{s}$ is the center of mass energy), while a hadron with mass
$M_{h}$ needs a formation time $t\sim 1/M_{h}$. Therefore, there is not
enough time for confinement to affect the annihilation cross-section and one can assume that the produced $q\bar{q}$ pair fragments
into hadrons with unit probability,

\begin{equation}
\sigma(e^{+}e^{-}\to \hbox{Hadrons})\propto \sigma(e^{+}e^{-}\to\mu^{+}\mu^{-}). 
\end{equation}
Therefore one finds
\begin{equation}
R=N_{c}\sum_{q=u,d, ..}e^{2}_{q},
\end{equation}
where the factor $N_{c}$ is the number of colour and $e_{q}$ denotes the quark
charge. The summation in the above equation is over all flavours that
are kinematically allowed. Depending on energy, various flavour degrees of
freedom contribute,  
\begin{equation}
R=
\cases{
 \frac{2}{3} N_{c} & (\hbox{u, d, s}),\\
 \frac{10}{9} N_{c} & (\hbox{u, d, s, c}),\\
\frac{11}{9} N_{c} & (\hbox{u, d, s, c,b}),\\}
\label{nc}
\end{equation}
up to about $3$ GeV only u, d and s contribute, while at higher energies charm
and b quarks start contributing as well. If one assumes that $N_{c}=3$, then
Eq.~(\ref{nc}) predicts $R=2, \frac{10}{3}$ and $\frac{11}{3}$,
respectively. If we ignore nonperturbative effects close to threshold,
such as the formation of bound states, we expect $R$ to present
a series of steps as a function of $\sqrt{s}$. In Fig. \ref{color1} we show various experimental data
which shows remarkable agreement with more detailed perturbative QCD
calculation based on the assumption that $N_{c}=3$.  

\begin{figure}[!t]
       \centerline{\includegraphics[width=4.0 cm] {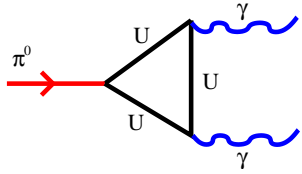}}
        \caption{ Decay of pion to photons.\label{color2}}
\end{figure}

Another strong evidence of colour degree of freedom is the measurement of
the neutral pion decay into photons $\pi^{0}\to \gamma \gamma$. The pion decay rate is
computed from the triangle diagram shown in Fig.~\ref{color2}. Because of
the quark loops, the decay rate is proportional to $N_{c}^{2}$. The experimental value of the pion decay rate can only be described by
$N_{c}=3$ \cite{pion11}.
%%%%%%%%%%%%%%%%%%%%%%%%%%%%%%%%%%%%%%%%%%%%%%%%%%%%%%%%%%%%%%%
\section{Colour transparency (CT)}

So far we have treated {\it colour} as just a new quantum number, a new degree of freedom. Is there any
evidence that this {\it colour} is responsible for the strong interactions?

If an interaction is controlled by colour, how can colourless hadrons interact? Apparently, only
due to the spatial distribution of colour (carried by quarks and gluons) inside the hadrons, i.e., due to the
existence of hadronic colour-dipole momentum.

This observation immediately leads to experimentally observable consequences. Since colourless
dipoles of vanishing size cannot interact, the interaction cross-section of such a dipole (say, quark-antiquark) with other hadrons should vanish when the transverse dipole separation goes to zero \cite{zkl},
 \beq
\sigma(r_T)\propto r_T^2\ .
\label{100}
 \eeq
 This remarkable relation deserves commenting upon:
(i) only transverse dipole separation matters, since at high energies longitudinal 
momentum transfer in exclusive
reactions vanishes. For example, if the beam particle of mass $m_1$ and
energy $E$ (in the target rest frame) is excited to mass $m_2$, while the
target remains intact, the longitudinal momentum transfer reads,
$q_L=(m_2^2-m_1^2)/2E$; (ii) the quadratic $r_T$-dependence is dictated by 
dimension counting, no other dimension parameters can be used here (the QCD scale 
$\Lambda_{QCD}$ may enter only via the coupling $\alpha_s$); (iii) an additional 
logarithmic dependence on $r_T$ may and does exist \cite{zkl}; (iv) such a 
small-$r_T$ behaviour is common for QED and QCD; however, in the former case
the total cross-section is predominantly elastic, while in the latter case is 
inelastic.

\subsection{Quasielastic scattering off nuclei}

\begin{figure}[t]
\centering\includegraphics[width=.5\linewidth]{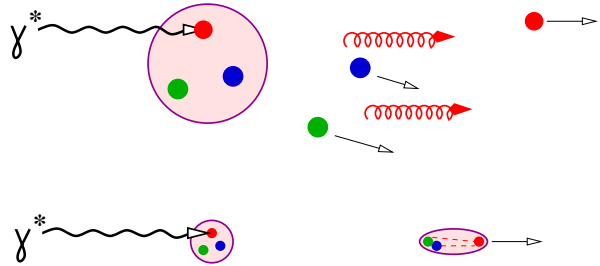}
 \caption{{\it Top}: deep-inelastic electron proton scattering, $ep\to eX$, at
large Bjorken $x$: the virtual photon knocks a valence quark out of the proton,
whose remnants form the final hadronic state $X$. {\it Bottom}: elastic $ep\to
e'p'$: the initial proton caught in a small size configuration survives a strong
kick with increased probability.}
 \label{eep}
 \end{figure}

The experimentally measured total hadronic cross-sections is a result of the interplay of different dipole
sizes whose probabilities are controlled by the hadronic wave functions. In some cases the probability
of small size configurations in a hadron can be enhanced leading to a reduced interaction cross-section
of such a hadron. An example is elastic electron-proton scattering, $ep\to e'p'$, with high
momentum transfer \cite{stan,al}, as illustrated in Fig,~\ref{eep}.

When the recoil proton has a reduced size\footnote{Strictly speaking this is 
not a proton. This state can be projected either into a proton (as in the present 
case), or proton excitations.}, it should interact more weaker than a regular proton 
with other targets. Such a possibility exists, if the elastic $ep$ scattering is 
embedded into a nucleus, i.e. in quasielastic $A(e,e'p)A^*$ reaction. The benchmark to compare with in this case is the expectations based on Glauber model 
calculations, where the recoil proton attenuates exponentially with the path length 
in the nucleus and with the normal proton-nucleon cross-section.

\begin{figure}[htb]
\includegraphics[width=.65\linewidth]{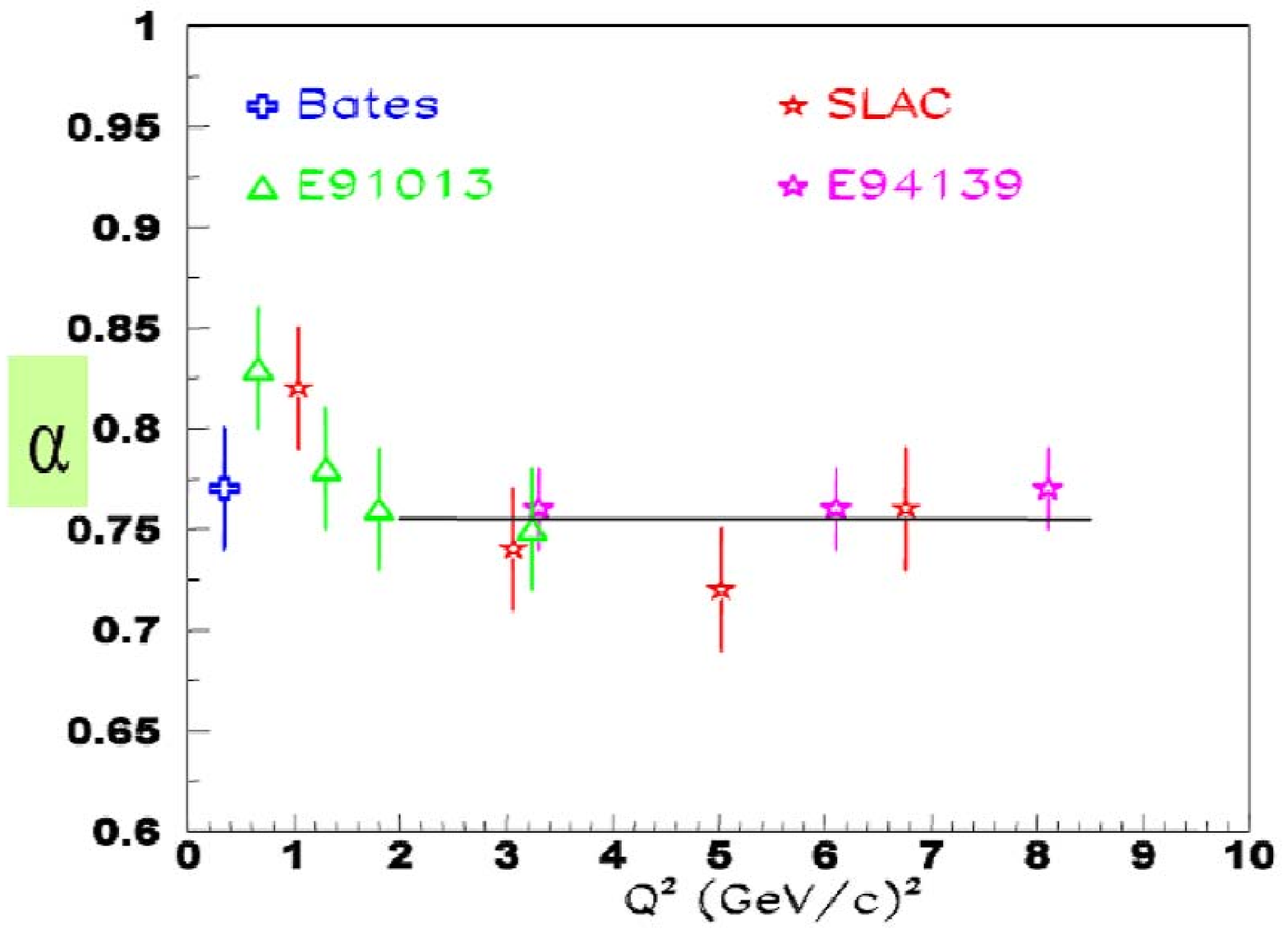}
\includegraphics[width=.35\linewidth]{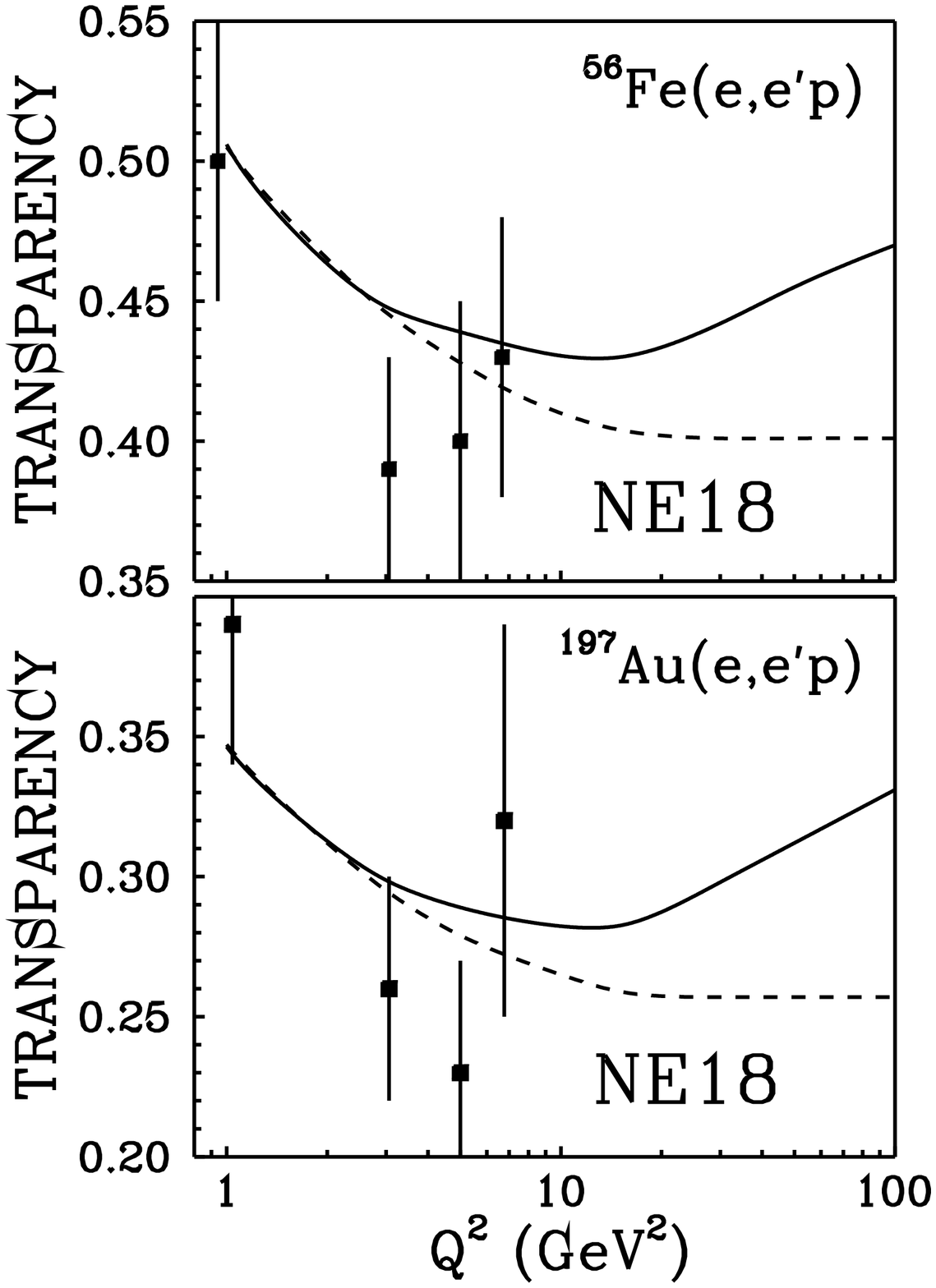}
 \caption{Right: Nuclear transparency measured in quasielastic scattering on iron (upper 
panel) and gold (lower panel) in NE18 experiment at SLAC \cite{e18}. Dashed and 
solid curves present expectations based on the Glauber model and CT
\cite{k-nem}.  Left: Data from different 
experiments on quasielastic electron scattering, $A(e,e'p)A^*$, for
A-dependence of the cross-section fitted by $A^\alpha$
\cite{kawtar}. The curve is $\alpha=0.75$. }
 \label{fig:e18}
 \end{figure}

The observable usually measured in such experiments is nuclear transparency, defined as,
 \beq
Tr=\frac{\sigma(eA\to e'pA^*)}
{Z\,\sigma(ep\to e'p')}.
\label{200}
 \eeq Basing on the above ideas of colour transparency one should
 expect a deviation rising with $Q^2$ from the Glauber model
 predictions. Unfortunately no experiment performed so far has
 provided a clear evidence for such an effect. The results of the
 dedicated experiment NE18 at SLAC \cite{e18} are depicted in
 Fig.~\ref{fig:e18}(right).  

Apparently data show no preference either for Glauber, or CT based models \cite{k-nem}.  
Other measurements of $A(e,e'p)A^*$ reactions were not successful either, when searching
for a CT signal. To fit the cross-section of this reaction by power 
$A$-dependence, $\sigma(e,e'p)\propto A^\alpha$, a rise of $\alpha$ with 
$Q^2$ would be a signal of CT. However, the collection of data \cite{kawtar} 
depicted in  Fig.~\ref{fig:e18}(left) versus $Q^2$ show no rise. 
Moreover, the value $\alpha= 0.75$ agrees with what one should expect
from the Glauber model.

Analogously, experiment in quasielastic proton-proton scattering, $A(p,2p)A^*$,
performed at BNL \cite{eva} did not provide any clear signal of CT. Although
data deviates from the Glauber model predictions, at higher momentum transfers
the agreement is restored.

% \begin{figure}[htb]
%\centering\includegraphics[width=.6\linewidth]{ct-results.eps}
% \caption{Data 
%from different 
%experiments on quasielastic electron 
%scattering, $A(e,e'p)A^*$, for A-dependence of the cross section 
%fitted by $A^\alpha$ \cite{kawtar}. The curve is $\alpha=0.75$.} 
% \label{others}
% \end{figure}

Why did these experiments fail to observe a CT effect? It turns out that it is not
enough to produce a small-sized configuration in a hard reaction. The produced
hadron has to maintain this small size during propagation through the nucleus.
It is clear that in a sufficiently long time interval the hadron will develop its 
wave function and restore the regular size. The time scale controlling this 
process is called formation time and for a recoil proton is given by,
 \beq
l_f < \frac{2E_p}{m_{p^*}^2-m_p^2}
\approx 0.4\fm\times E_p(\GeV)\ .
\label{300}
 \eeq
 Here $E_p$ is the energy of the recoil proton , and $m_{p^*}$ is the mass of the 
first proton excitation. In order to have $l_f\gg R_A$ for heavy nuclei 
$R_A\approx 5\fm$, the proton energy should be much higher than $10\GeV$.
The highest energy of recoil protons in the NE18 experiment \cite{e18} was $E_p\approx 4\GeV$
which is too low to keep the size of the produced hadron small within the nuclear 
range.

This is the principal problem of quasielastic scattering where the photon energy 
$\nu$ and virtuality are strongly correlated, $2m_p\nu=Q^2$. Thus the recoil 
proton energy is $E_p\approx Q^2/2m_p$. Therefore one must go to extremely high 
virtualities, $Q^2\geq 20\GeV^2$, just in order to increase $E_p$. However, the 
cross-section becomes vanishingly small.

\subsection{Diffractive electroproduction of vector mesons}

\begin{figure}[htbp]
\centering\includegraphics[width=.6\linewidth]{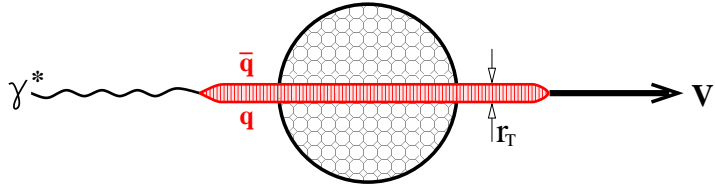}
 \caption{Virtual diffractive photoproduction of vector mesons.
A virtual photon fluctuates into a $\bar qq$ pair of transverse separation 
$r_T^2\sim Q^2$ which propagates through a nucleus, interacts diffractively, and 
being brought to the mass shell develops the wave function of the vector meson.}
 \label{gamma-v}
 \end{figure}

Diffractive virtual photoproduction of vector mesons is free of this problem.
The space-time development of this reaction is illustrated in Fig.~\ref{gamma-v}

\begin{figure}[!t]
\includegraphics[width=.54\linewidth]{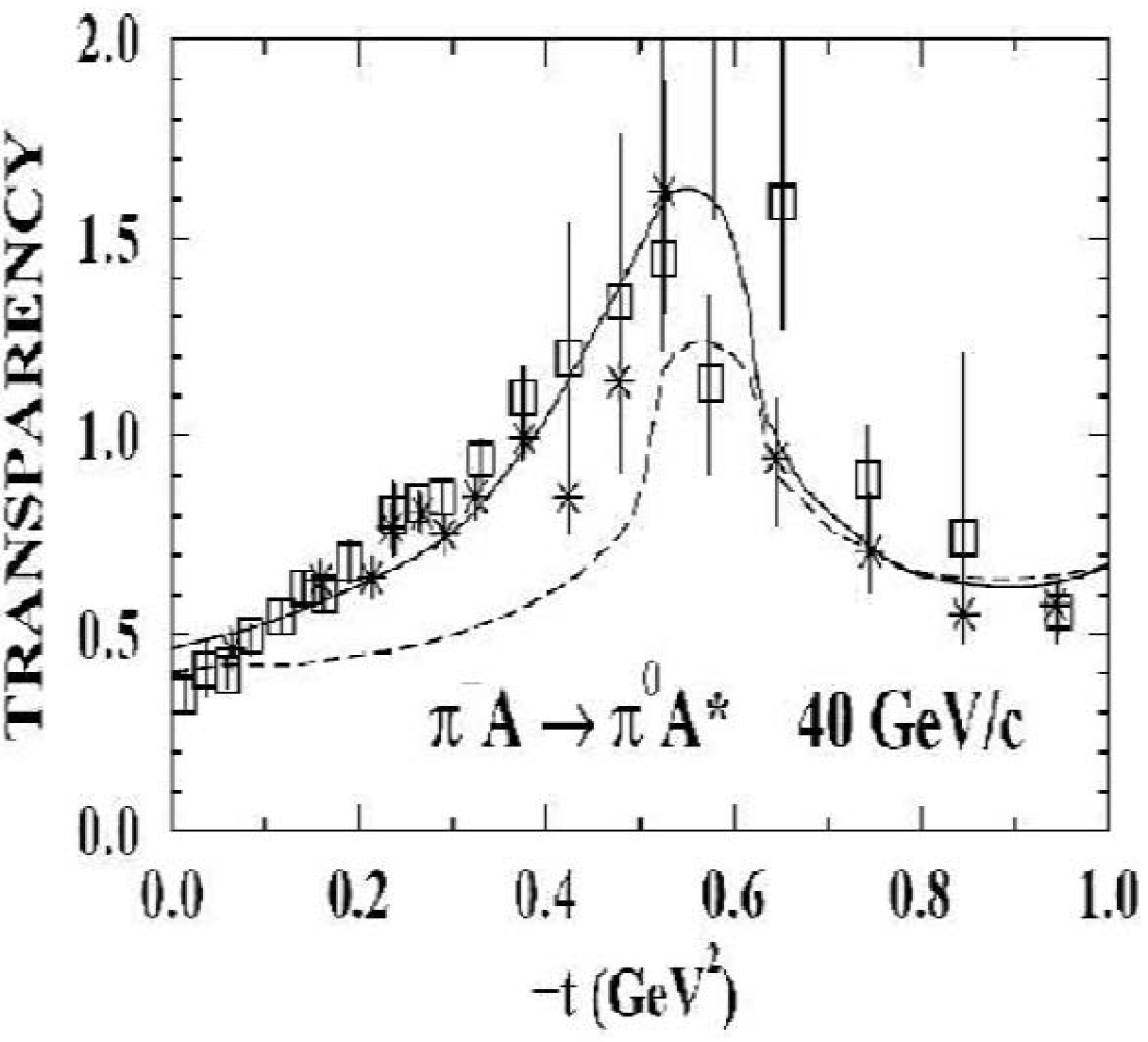}
\includegraphics[width=.54\linewidth]{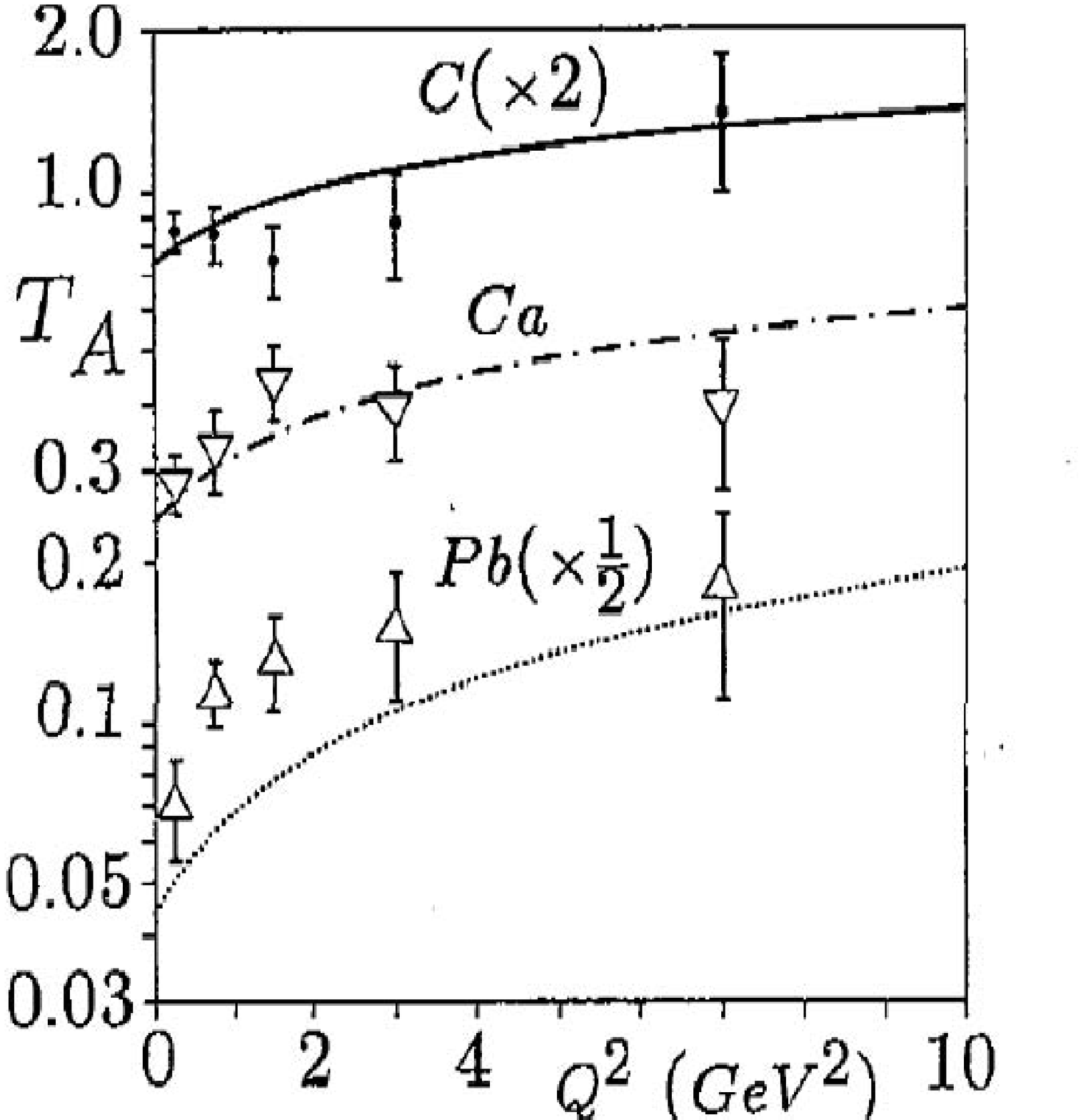}
\caption{ Right: Nuclear transparency as function of $Q^2$ for carbon, calcium and
lead. Data points from the E665 experiment at Fermilab \cite{e665} are compared
with calculations \cite{knnz2}. Left: Nuclear transparency in quasi-free charge exchange of pions on carbon.
Data are from \cite{proza}. Solid and dashed curves represent calculations
\cite{kz} including or disregarding CT effects, respectively. }
\label{fig:e665} 
\end{figure}

 At very high energies, the $\bar qq$ fluctuation lifetime,
 \beq
t_c=\frac {2E_\gamma}{Q^2+M_{\bar qq}^2},
\label{400}
 \eeq
 (which is also called coherence time), becomes very long. So one can treat the
$\bar qq$ dipole propagating through the nucleus as "frozen" by Lorentz time
dilation at the initial size $r_T^2\sim 1/Q^2$. Thus, one can keep the scale 
$Q^2$ finite, while the photon (and vector meson) energy can be increased with no 
restriction. This is the main advantage of this process for the search for CT effects
compared to quasielastic reactions. The first measurements proposed 
in 
\cite{knnz1} and performed by the E665 collaboration \cite{e665} confirmed the 
theoretical expectations \cite{knnz2} of CT effects depicted in 
Fig.~\ref{fig:e665}(right). The high photon energy 
in this experiment led to $l_c\gg R_A$ and allowed us to greatly simplify theoretical 
calculations. It turns out that at the opposite limiting case of $l_c\ll R_A$
and high $Q^2$, the photon energy may be still high enough to keep the formation 
time scale Eq.~(\ref{300}) sufficiently long to observe CT effects. In this case 
a 
signal of CT would be a rising energy dependence of nuclear transparency.
Corresponding measurements are under way at Jefferson Lab \cite{kawtar}.

Similarly to diffraction, quasi-free hadron scattering off a nucleus can be
performed at high energies, while the hadron size can be controlled by
transverse momentum $p_T^2\approx -t$ \cite{kz}. In the case of Reggeon
exchange, the pion formfactor formfactor suppresses large-sized configurations in
the hadronic wave function at rather small $t$. Measurements were performed
by the PROZA collaboration \cite{proza} with $40\GeV$ pions in quasi-free
charge exchange scattering $\pi^-A\to\pi^0A^*$. The results are depicted in 
Fig.~\ref{fig:e665}(left) in comparison with Glauber model expectations (dashed 
curve) and calculations including CT effects \cite{kz}.

%\begin{figure}%[!t]
%\includegraphics[width=.45\linewidth]{cex.eps}
%\caption{
%Nuclear transparency in quasi-free charge exchange of pions on carbon.
%Data are from \cite{proza}. Solid and dashed curves represent calculations
%\cite{kz} including or disregarding CT effects, respectively.}
% \label{fig:cex}}
%\end{figure}}
Notice that both models predict a peak at $-t\approx 0.6\GeV^2$, because the
cross-section of free scattering, $\pi^-p\to\pi^0n$, has a minimum at this
momentum transfer, and the position of the minimum in quasi-free scattering
is shifted by multiple interactions in the nucleus.

 \section{Bags, strings...}

\begin{figure}[b]
\centering\includegraphics[width=.7\linewidth]{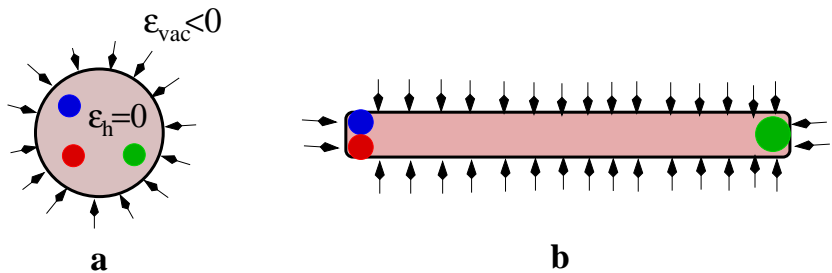}
 \caption{{\bf a:} Pictorial illustration for the MIT bag model; {\bf b:} a
stretched bag becomes a tube of a constant cross-section, which can be treated as a 
string.}
 \label{bag}
 \end{figure}

Gluonic condensate in vacuum pushes the energy density below the
perturbative level, $\epsilon_{vac}<0$. If the colour field of the
valence quarks suppresses vacuum fluctuations, then the energy density
inside the hadron is higher than outside. Therefore the vacuum tries
to squeeze the hadron. However, the chromo-electromagnetic energy
$(E^2+H^2)/2$ inside a smaller volume rises leading to an
equilibrium, as illustrated in Fig.~\ref{bag}(a).
Thus hadrons look like bubbles in the QCD vacuum, this is the key idea of the
the MIT bag model \cite{mit}. 

\begin{figure}[htbp]
\centering\includegraphics[width=.7\linewidth]{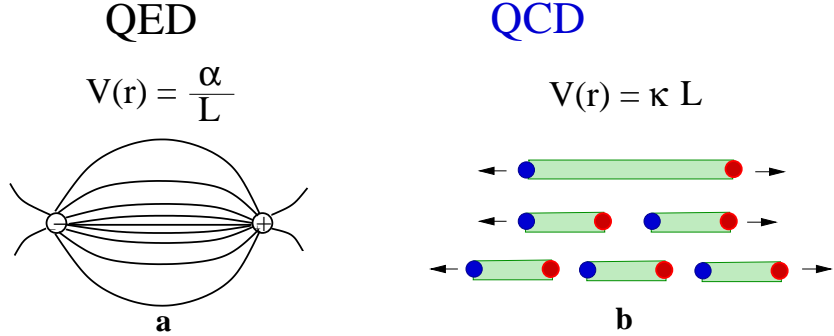}
 \caption{{\bf a:} Electric field pattern in QED: the potential falls 
with charge separation as 
$1/L$; {\bf b:} Colour field pattern in QCD: the field is squeezed into a 
tube which breaks up by production of $\bar qq$ pairs tunnelling from vacuum.}  
 \label{string}
 \end{figure}

What happens if a quark is knocked out with a high momentum? On account of the same properties
of the QCD vacuum the chromo-electric flux is squeezed into a tube of a constant cross-section,
 \beq
\pi r^2=\frac{g^2}{8\kappa}\,,
\label{500}
 \eeq
 as illustrated in Fig.~\ref{bag}(b). Here $g$ is the colour charge at the
ends of the tube; $\kappa$ is the energy density stored in the tube per unit of
length. This pattern of colour fields is quite different from that in QED, as illustrated in Fig.~\ref{string}.

The potential between two electric charges falls with distance as $1/L$,
while in QCD it rises linearly. In fact, the rising with distance of a string potential
explains the observed linearity of Regge trajectories (see below).

Usually the transverse size is not important, so the tube may be treated as a
one-dimensional string, and $\kappa$ is called string tension. It can be either 
calculated on the lattice, or related to the universal slope of Regge
trajectories $\alpha_\Reg^\prime=0.9\GeV^{-2}$ \cite{cnn},
 \beq
\kappa=\frac{1}{2\pi\alpha_\Reg^\prime}\approx 
1\,\frac{\GeV}{\fm}\,.
\label{600}
 \eeq
 This energy is sufficient for the creation of a couple of constituent quarks via
tunnelling from the vacuum. One can hardly stretch a string longer than 1fm, since it breaks
into pieces, as illustrated in Fig.~\ref{string}(b). The $\bar qq$ pairs produced
from vacuum via the Schwinger mechanism completely screen the field of the
end-point colour charges due to the linearity of the string potential \cite{cnn}.
The Schwinger phenomenon and existence of light quarks are the main reasons for not observing free quarks and gluons (colour screening).
\begin{figure}[htbp]
\centering\includegraphics[width=.8\linewidth]{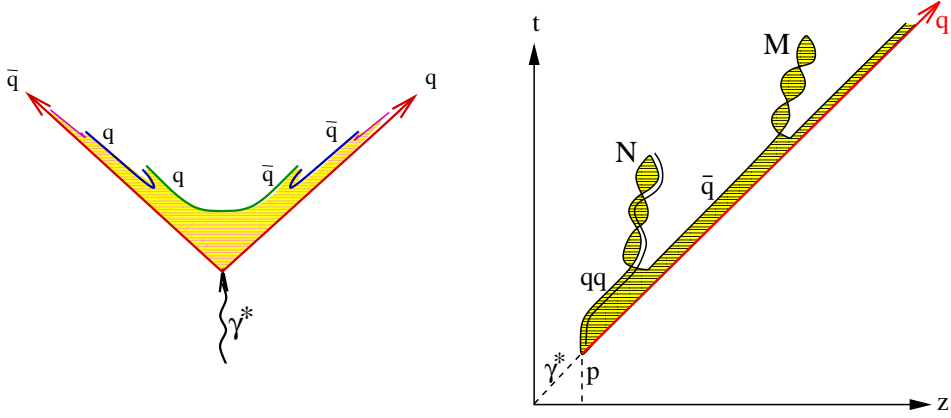}
 \caption{Time-coordinate development of string fragmentation in the center
of mass (left) and target rest (right) frames.}
 \label{z-t}
 \end{figure}

 \section{Hadronization of colour charges}
Thus a colour charge is always accompanied by an anti-charge neutralizing its 
colour. The colour field in between forms a tube/string which is a very unstable 
construction, $\bar qq$ pairs pop up via tunnelling from vacuum, as shown in 
Fig.~\ref{string}(b), and the string is 
never much longer than $1\fm$.  The probability of such a string breaking over time 
interval $T$ is given by,
 \beq
P(T)=1-\exp\left[-w\int\limits_0^T 
dt\,L(t)\right]\,,
\label{700}
 \eeq
where $L(t)$ is the time-dependent length of the string, and the probability density 
for the creation of a $\bar qq$ pair per unit time per unit  
length is given by the Schwinger formula \cite{cnn},
 \beq
w=\left(\frac{\kappa r}{\pi}\right)^2
\exp\left(-\frac{2\pi m_q^2}{\kappa}\right)
\approx 2\fm^{-2}\,.
\label{800}
 \eeq 

 The string length $L(t)$ is getting shorter after each break, thus
 delaying the next pair production.  Therefore, hadron momenta rise in
 geometric progression, i.e. the rapidity distribution of produced
 hadrons is constant. Notice that such a plateau in rapidity has been
 predicted by models with multiperipheral dynamics and for gluon
 radiation in perturbative QCD \cite{dkmt}.  This process is
 illustrated on a time-coordinate plot in the c.m. frame of the
 initial $\bar qq$ pair (e.g. $e^+e^-$ annihilation), and in the
 target rest frame (e.g. in DIS) in Fig.~\ref{z-t} on the left and
 right correspondingly. Since both ends of the string are moving in
 the same direction and with the same velocity (the speed of light),
 the length of the string is independent of time. Its maximal possible
 value is $L_{max}=m_q/\kappa$. However, after each break of the
 string it becomes about twice as short, as illustrated in
 Fig.~\ref{z-t}.

Notice that the leading quark loses energy at a constant rate, 
$dE_q/dz=-\kappa$, though the whole hadronization process, until the creation of 
a hadron that includes this quark. It is interesting to notice that in perturbative QCD the 
leading quark loses energy for gluon radiation also at a constant rate, 
$dE_q/dz=-(2\alpha_s/3\pi)Q^2$ \cite{feri}.
%%%%%%%%%%%%%%%%%%%%%%%%%%%%%%%%%%%%%%%%%%%%%%%%%%%%%%%%%%%%%%%%%%%%%%%%%%%%%%%
%%%%%%%%%%%%%%%%%%%%%%%%%%%%%%%%%%%%%%%%%%%%%%%%%%%%%%%%%%%%%%%%%%%%%%%%%%%%%%

%%%%%%%%%%%%%%%%%%%%%%%%%%%%%%%%%%%%%%%%%%%%%%%%%%%%%%%%%%%%%%%%%%%%%%%%%%%%%%%%%%%%%%%%%%%%%
%%%%%%%%%%%% ct-2
\section{Regge phenomenology}
{\it The theory of Regge poles is a quite dormant topic. It does not
seem to be taught very much anymore. In addition there is often found an attitude that the subject is obsolete, because it is
identified so strongly with the pre-quark, pre-parton era of the
S-matrix, dispersion-relations approach to strong interactions.
This point of view is just plain wrong}. The Chew, Frautschi, Regge, et. al., {\it description of high energy
behaviour in terms of singularities in the complex angular momentum plane is completely general. 
And the basic technique of Watson-Sommerfeld transform should
be a standard part of the training in theoretical particle physics.}  \\
\hspace*{12cm} ---James Bjorken\cite{quo}

 \begin{figure}[!t]
\centering\includegraphics[width=.5\linewidth]{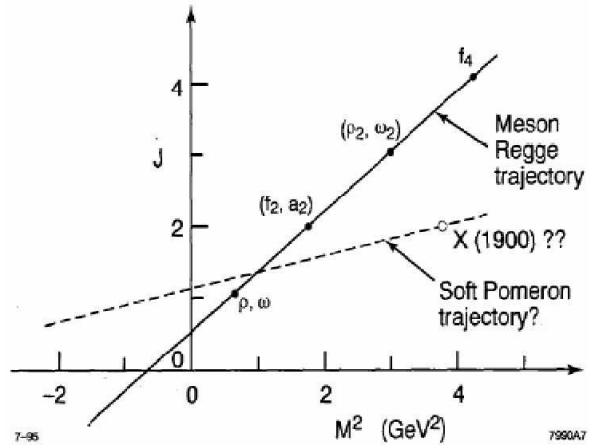}
 \caption{Regge trajectories for mesons and soft Pomeron.}
 \label{trajectories}
 \end{figure} 

\subsection{Poles in angular momentum plane}
The energy dependence of the amplitude is governed 
by poles (or cuts) in the complex angular momentum
plane \cite{collins},
 \beq
A(s,t)=\sum\limits_r h_r(t)\,\xi_r(t)\,
\left(\frac{s}{s_0}\right)^{\alpha_r(t)}\,,
\label{850}
 \eeq
where we sum over different Regge poles $r$, and $h_r(t)$ is a
phenomenological residue function which is not given by the theory, but is
fitted to data. It depends on $t$, but not energy, and correlates with the
choice of the parameter $s_0$.

The phase factor $\xi_r(t)$ depends on the Regge pole signature
$\sigma=(-1)^{J}$, where $J$ are spins (even or odd) of mesons lying on the
trajectory.
 \beq
\xi_r(t)=\left\{ 
\begin{array}{cc}
i+{\rm ctg}\left[\frac{\pi}{2}\alpha_r(t)\right]& {\rm if}\ \ \sigma=-1\\
-i+{\rm tg}\left[\frac{\pi}{2}\alpha_r(t)\right]& {\rm if}\ \ \sigma=+1\\
\end{array}
\right. 
\label{900-c}
 \eeq

The energy dependent factor $(s/s_0)^{\alpha(t)}$ is controlled by the Regge
trajectory $\alpha(t)$ which is nearly straight,
$\alpha(t)=\alpha(0)+\alpha't$, as is demonstrated in on the Chew-Frautschi
plot in Fig.~\ref{trajectories}. This is the miracle of Regge theory:  the linear Regge
trajectories bridge the low-energy physics of resonances ($t=M^2>0$) with
high-energy scattering ($t<0$).

High energies are dominated by Reggeons with highest trajectories $\alpha_r(t)$,
the Pomeranchuk pole (Pomeron),
 \beqn
\alpha_{\Pom}(0) &\approx& 1.1; \nonumber\\
\alpha_{\Pom}^\prime &\approx& 0.25\GeV^{-2}\,,
\label{1000}
 \eeqn
 and leading Reggeons,
 \beqn
\alpha_f(0)&\approx&\alpha_\omega(0)\approx\alpha_\rho(0)
\approx\alpha_{a_2}(0)\approx0.5\,;
\nonumber\\
\alpha_{\Reg}^\prime &=& 0.9\GeV^{-2}\,.
\label{1100}
 \eeqn

The first important prediction of the Regge pole theory was shrinkage
of the elastic slope with energy. The slope parameter controls the
$t$-dependence of the elastic cross-section, $d\sigma_{el}/dt\propto
e^{Bt}$. According to (\ref{850}) the slope parameter $B$ rises with
energy as, \beq B(s)=B_0+2\alpha_\Pom^\prime\ln(s/s_0)\,,
\label{1200-c}
 \eeq
 where $B_0$ is a phenomenological parameter.

\begin{figure}[!t]
\centering\includegraphics[width=.4\linewidth]{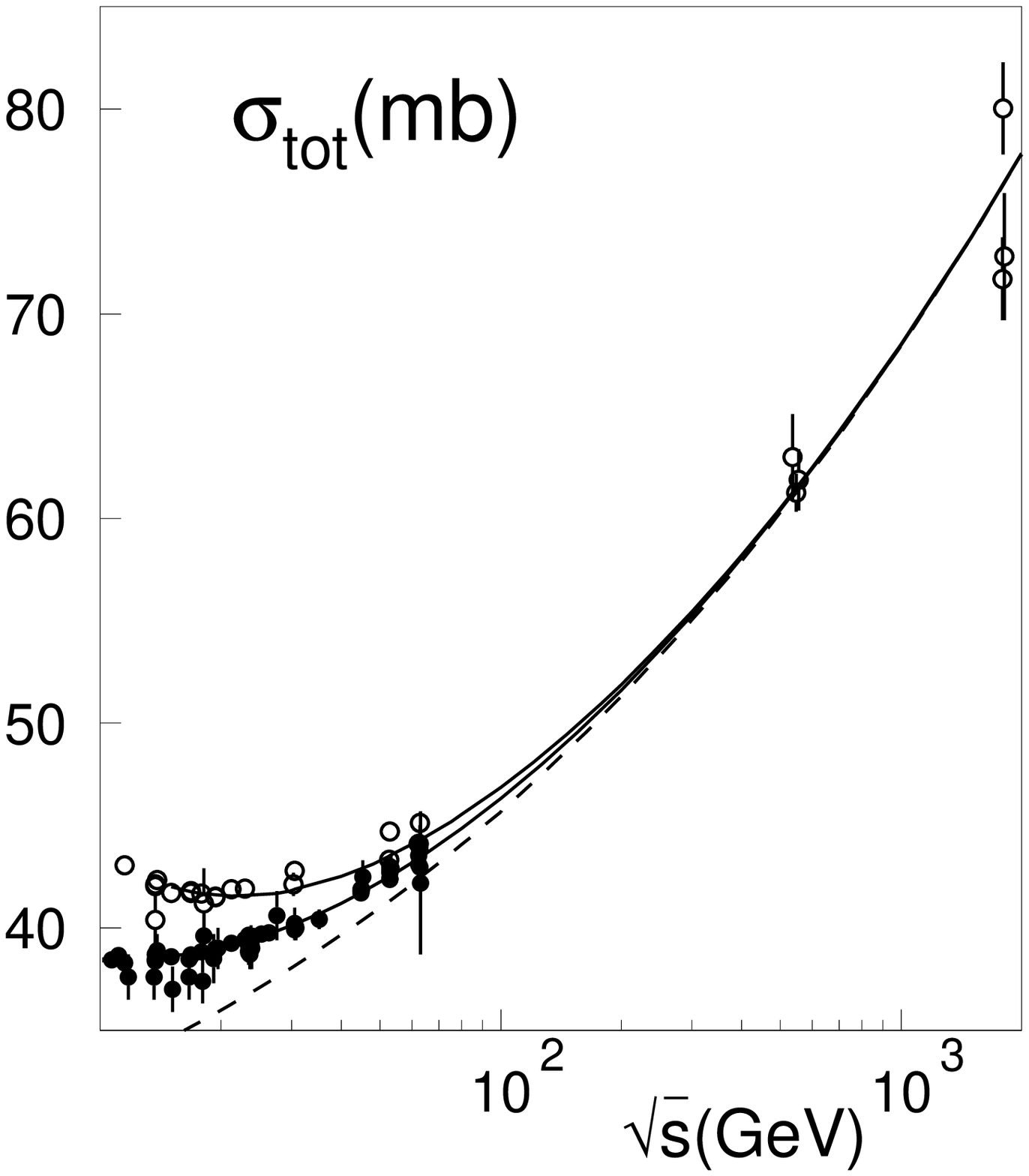}
\hspace*{1cm}
\centering\includegraphics[width=.4\linewidth]{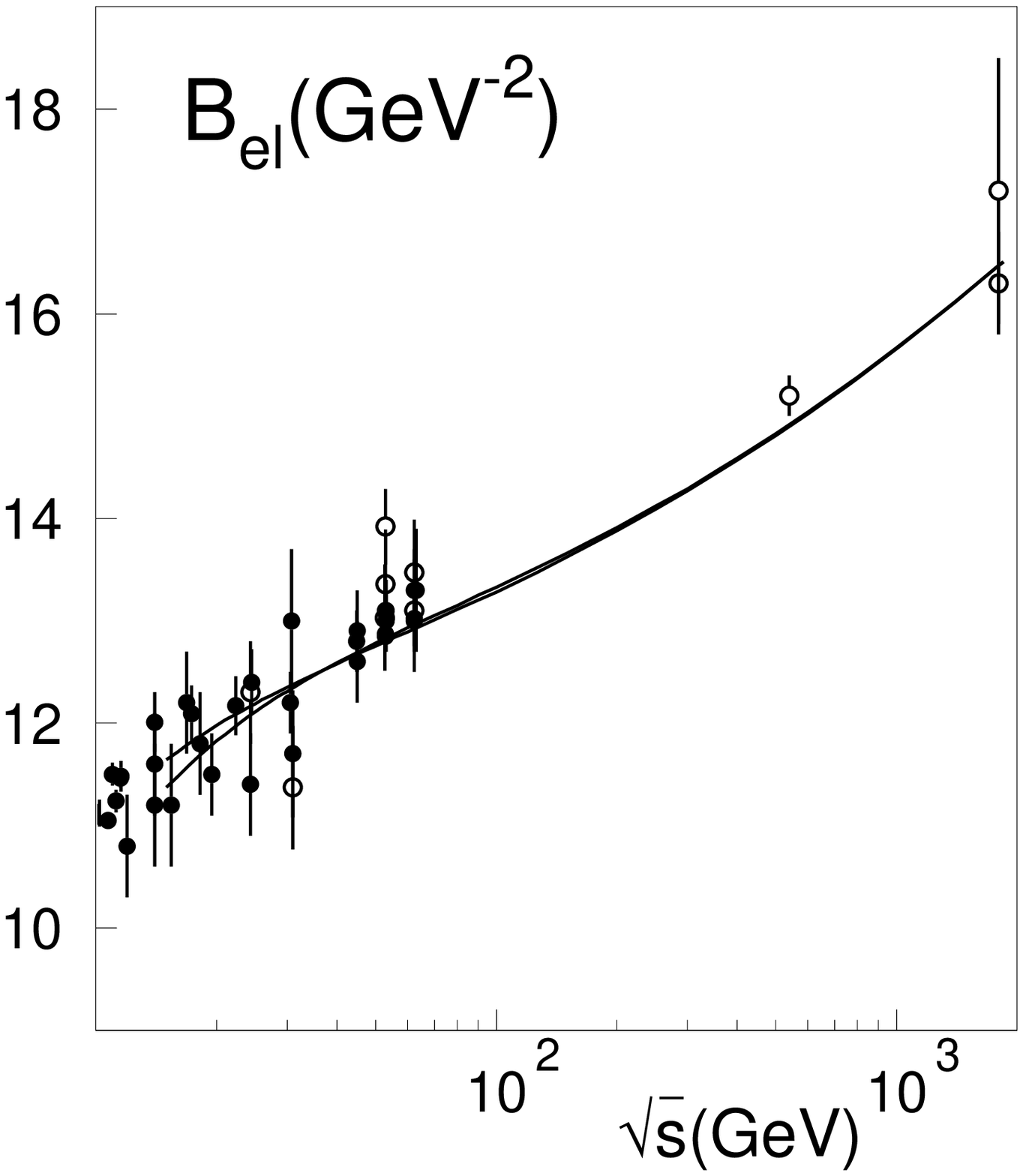}
 \caption{Dependence of the total cross-section (left) and elastic slope 
(right) on center of mass energy. Open and closed points correspond to $\bar 
pp$ and $pp$ collisions, respectively. The curves show calculations based on 
the Pomeron model of \cite{k3p}. Data are from Ref.~\cite{ref-k3p}.}
 \label{data}
 \end{figure}

The Pomeron parameters Eq.~(\ref{1000}) were extracted from data on elastic
scattering. The Pomeron intercept $\alpha_\Pom(0)$ comes from data on total
hadronic (mostly $pp$ and $\bar pp$) cross-section fitted with the energy
dependence (\ref{850}), while the parameter $\alpha_\Pom^\prime$ is related
to the elastic slope Eq.~(\ref{1200-c}). Corresponding data are shown in
Fig.~\ref{data}.

\subsection{Triple Regge phenomenology}

\begin{figure}[t]
\centering\includegraphics[width=.95\linewidth]{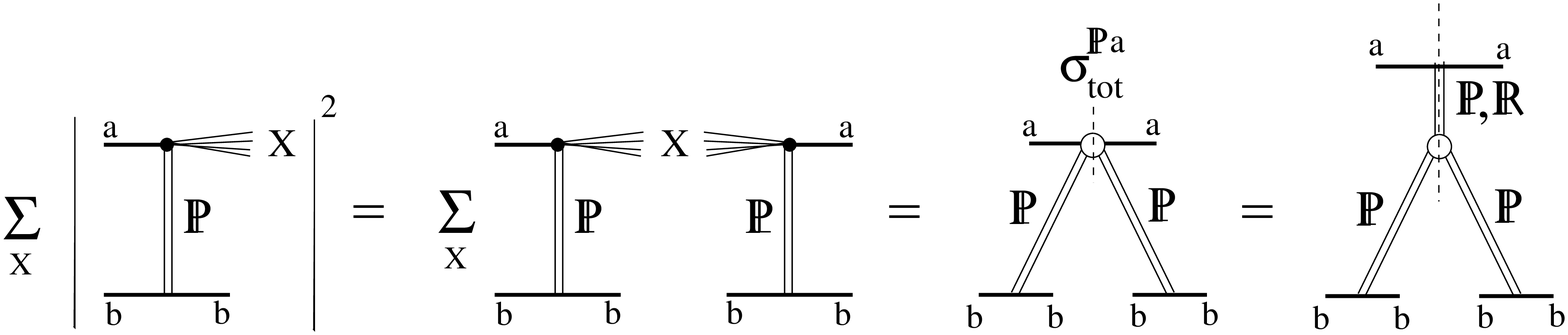}
 \caption{The cross-section of single diffraction, $a+b\to X+b$ summed 
over all excitation channels at fixed effective mass $M_X$.}
\label{3r}
 \end{figure}

The cross-section for the inclusive process, $a+b \to X +c$ can be
also expressed in terms of the Regge approach. Here we focus on the
most interesting case of diffractive excitation, $c=b$, via Pomeron
exchange.  To sum up all final-state excitations $X$, one can apply
the unitarity relation to the pomeron-hadron ($\Pom-a$) amplitude as
shown in Fig.~\ref{3r}. Provided that the effective mass of the
excitation is large (but not too much), $s_0\ll M_X^2\ll s$, one can
describe the Pomeron-hadron elastic amplitude via Pomeron or secondary
Reggeon exchanges in the $t$-channel. Then one arrives at the
triple-Regge graph, Fig.~\ref{3r}, which corresponds to the cross-
section, \beq
\frac{d\sigma_{sd}^{ab\to Xb}}{dx_F\,dt} =
\sum\limits_{r=\Pom,\Reg} G_{\Pom\Pom r}(t)
(1-x_F)^{\alpha_{r}(0)-2\alpha_\Pom(t)}
\left(\frac{s}{s_0}\right)^{\alpha_{r}(0)-1},
\label{1300-c}
 \eeq
 where $x_F$ is the Feynman variable for the recoil particle $b$ defined in the
center of mass, $x_F=2p^{||}_b/\sqrt{s} \approx 1-M_X^2/s$.

\begin{figure}[htbp]
\centering\includegraphics[width=.5\linewidth]{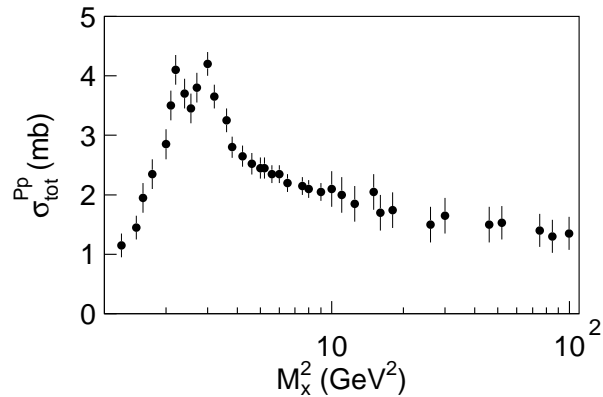}
 \caption{The Pomeron-proton total cross-section extracted from 
single-diffraction data, $pp\to pX$, as function of the invariant mass $M_X$ which 
is the center-of-mass energy in $\Pom p$ collision. Experimental data are from \cite{kaidalov}.}
\label{pom-p}
 \end{figure}

Equation (\ref{1300-c}) contains new phenomenological functions, effective 
triple-Regge vertices, $G_{\Pom\Pom\Pom}(t)$ and $G_{\Pom\Pom\Reg}(t)$.
The diffractive cross-section can also be expressed in terms of 
the Pomeron-hadron total cross-section $\sigma^{\Pom a}_{tot}(s'=M_X^2)$.
Most interesting is the asymptotic ($s'=M_X^2\gg s$) of this cross-section
related to the triple-Pomeron coupling,
 \beq
G_{3\Pom}(t)=\sigma^{\Pom a}_{tot}\,
N_{\Pom bb}(t)^2\ .
\label{1400-c}
 \eeq
 Here $N_{\Pom bb}(t)$ is the Pomeron-hadron vertex known from $bb$ elastic
scattering. Thus one can extract from data on single diffraction the Pomeron-hadron
total cross-section, $\sigma^{\Pom a}_{tot}$ \cite{kaidalov}, which carries unique
information about the properties of the Pomeron. The results shown in
Fig.~\ref{pom-p} demonstrate an amazingly small cross-section, less than $2\mb$.

 This is at least an order of magnitude less than one could expect. Indeed, the 
Pomeron as a gluonic object should interact more strongly than a meson, i.e., the 
Pomeron-proton cross-section could be about twice as big as the pion-proton one.
Such a weak interaction of the Pomeron is probably the strongest evidence for the location 
of the glue in hadrons within small spots \cite{spots}.

\subsection{Building the Pomeron}

It has been a natural and simple assumption made in the early years of 
the Regge theory that the Pomeron is a Regge pole with a linear 
trajectory and the intercept $\alpha_{\Pom}(t) = 1$. Nowadays, however, we
have a multi-choice answer, and it is still debated whether the Pomeron 
is:
\begin{itemize}
\item
a Regge pole  (probably not, since
$\alpha_{\Pom}(0)$
varies with $Q^2$ in DIS);
\item the DGLAP Pomeron \cite{dgl,book}, which corresponds to a specific
ordering for radiated gluons in the ladder graph in Fig.~\ref{dis33}, $p^2_{i+1}<p^2_i\leq Q^2$ (see Section 7.2);
\item the BFKL Pomeron \cite{bfkl} which does not have ordering in
transverse momenta of radiated gluons, but has no evolution with $Q^2$
either \cite{sardinia} (see Section 8);

\item
something else?

\end{itemize}

\begin{figure}[b]
\centering\includegraphics[width=0.8\linewidth]{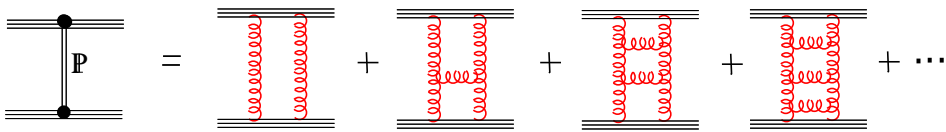}
 \caption{Perturbative Pomeron represented by the Born term, two-gluon exchange, 
and higher order terms having a form of gluonic latter graphs.}
 \label{ladder}
 \end{figure}

Gluons seem to be the most suitable building material: already the 
Born graph provides $\alpha_\Pom(0)=1$. The higher order corrections are 
expected to pull the intercept above one. These corrections are dominated by 
ladder type graphs shown in Fig.~\ref{ladder}.
 A ladder is a shadow of gluon bremsstrahlung according to the unitarity 
relation, Fig.~\ref{unitarity}. 

The leading-log approximation (LLA) corresponds to keeping those terms only,
where each coupling $\alpha_s$ has a big factor $\ln(s)$.
For fixed coupling, the BFKL result is not a Regge pole, but a cut with an
intercept
 \beq
\alpha_\Pom(0)-1=\frac{12\alpha_s}{\pi}\,\ln2\,.
\label{1500}
 \eeq 
 This result will be derived in Section 8, see 
 Eq.~(\ref{pom-s}). Unfortunately, the next-to-leading-log corrections
 (extra powers of $\alpha_s$) to the intercept are of the same order \cite{nnlb}, 
\beq
\alpha_\Pom(0)-1=\frac{12\alpha_s}{\pi(1-6.5\alpha_s)}\,\ln2\,,
\label{1500-cc}
 \eeq 
and it may be even negative for most reasonable values of $\alpha_s$. 

\begin{figure}[htbp]
\centering\includegraphics[width=0.4\linewidth]{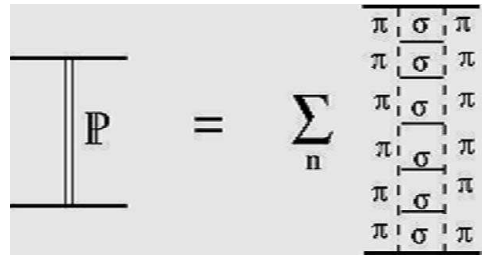}
 \caption{Ladder graphs built of pions and $\sigma$-mesons as a model for the 
soft Pomeron.}
 \label{pion-pom}
 \end{figure}

However, it does not look reasonable to describe a soft Pomeron, controlling soft
hadronic interaction at high energies perturbatively. Similar latter graphs,
but built of light hadrons, e.g. of pions and $\sigma$ mesons as depicted in
Fig.~\ref{pion-pom}, well describe many features of soft hadronic collisions
\cite{boreskov}.  
One can adjust the poorly known $\sigma$-pion coupling to
reproduce the Pomeron intercept. However, its closeness to one, which is very
natural in QCD, looks as an accidental coincidence in this model.

\subsection{Duality}

\begin{figure}[htbp]
\centering\includegraphics[width=0.5\linewidth]{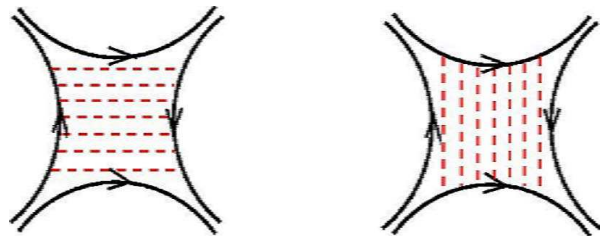}
 \caption{The amplitude of meson-meson interaction 
via quark exchanges. Dashed lines show intermediate interactions within a 
quark-antiquark pair in $t$-channel (left), or in $s$-channel (right).} 
 \label{dual}
 \end{figure}

Reggeons correspond to the exchange of valence quarks.
The descriptions of meson-meson scattering amplitude in terms of interacting
$\bar qq$ pairs in the $t$ channel (Reggeons), or in the $s$ channel (resonances) are 
dual \cite{collins}, as is illustrated in Fig.~\ref{dual}.
 
No $s$ channel resonances is possible in $pp$ and $K^+p$ elastic amplitudes.  
However, $t$ channel Reggeons are present. To comply with duality the Reggeons
must cancel each other in the imaginary part of the amplitude. For this reason,
pairs of leading Reggeons must be exchange-degenerate, $f$ with $\omega$, and
$\rho$ with $a_2$, i.e., their Regge trajectories and residue functions must be
identical, differing only in the signature factors (phases) \cite{collins}.
Data depicted on the Chew-Frautschi plot in Fig.~\ref{trajectories} indeed 
confirm this expectation.

The sums, $f+\omega$ and $a_2+\rho$ must be real for $pp$ and $K^+p$, but
imaginary for $\bar pp$ and $K^-p$. Data at low energies dominated by Reggeons
nicely confirm this. For the same reason spin effects are much stronger in $pp$
and $K^+p$, than in $\bar pp$ and $K^-p$.

%%%%%%%%%%%%%%%%%%%%%%%%%%%%%%%%%%%%%%%%%%%%%%%%%%%%%%%%%%%%%%%%%%%%%%%
%%%%%%%%%%%%%%%%%%%%%%%%%%%%%%%%%%%%%%%%%%%%%%%%%%%%%%%%%%%%%%%%%%%%%%%%%%%%%
\begin{figure}[!t]
       \centerline{\includegraphics[width=6.0 cm] {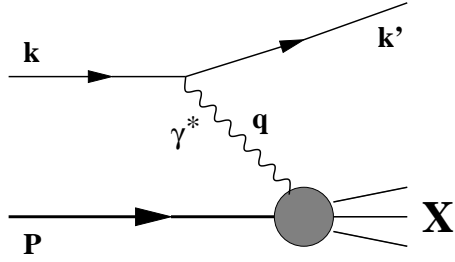}}
        \caption{ Electron-proton deep inelastic scattering\label{dis0}}
\end{figure}

\section{Deep inelastic scattering}
The basic idea of electron-proton Deep-inelastic scattering (DIS) is to use a lepton probe to study a hadron. A
lepton of momentum $k$ acquires momentum $k\prime$ by
exchanging a virtual photon of momentum $q$ with the proton (with a spin $1/2$) of momentum
$P$ and mass $m_{_N}$. After the collision, the rest of the energy is transferred to
the unobserved final state $X$ with mass $M_{_X}$. We ignore the lepton mass compared to the scale $Q^2=-q^2$. 
The kinematics of DIS is characterized by
a few Lorentz-invariant variables, see Fig.~\ref{dis0}.
\begin{eqnarray}
   \nu&\equiv& \frac{P\cdot q}{m_{N}},\nonumber\\
  W^2&\equiv& (P+q)^2,\nonumber\\
  s&\equiv& (P+k)^2.\
\end{eqnarray}

We define two other commonly used variables, namely the Bjorken variable 
\begin{equation} 
x=\frac{Q^2}{2P\cdot q}=\frac{Q^2}{2m_N\nu}\approx\frac{Q^2}{Q^2+W^2},
\end{equation}
where $Q^2=-q^2$, and the relative energy loss of the lepton
\begin{equation} 
y=\frac{P\cdot q}{P\cdot k}\approx\frac{Q^2+W^2}{s}.
\end{equation} 
The kinematic limits are $Q^2<W^2$ and $x>Q^2/W^2$ which leads to 
$0\le x\le 1$. The value of $x=1$ is reached when the proton is scattered elastically. 
The differential cross-section for inclusive scattering has the following form
\begin{equation} 
d\sigma(eP\to e^\prime X)=\frac{1}{2s}\frac{d^3k^\prime}{(2\pi)^3 2E^\prime}
\sum_X (2\pi)^4\delta^4(P+k-p_X-k^\prime)|{\cal{A}}|^2,
\end{equation} 
where one sums over all final hadronic states. The matrix element squared after summing over polarization of the virtual photon becomes
\begin{equation} 
{\cal{|A|}}^2=\frac{2\pi\alpha_{em}}{Q^2}\langle P|J^{\mu\dagger}(0)|X\rangle \langle X|J^{\nu}(0)|P\rangle L_{\mu\nu},
\end{equation} 
where the leptonic tensor is
\begin{eqnarray}
L^{\mu\nu}&\equiv& \langle \bar{u}(\vec{k\prime})\gamma^{\mu} u(\vec{k})\bar{u}(\vec{k})\gamma^{\nu}\bar{u}(\vec{k\prime})\rangle,
\nonumber\\
&=& 2(k^\prime_\mu k_\nu+k^\prime_\nu k_\mu-g_{\mu\nu}k^\prime\cdot k)
\; .
\end{eqnarray}
In the above expression, we ignored the electron mass. We define the
hadronic tensor as
\begin{eqnarray} 
W^{\mu\nu} & = & \sum_X 
\langle P|J^{\mu}(0)|X\rangle\langle X|J^{\nu}(0)|P\rangle
(2\pi)^4\delta^4(P+q-p_X),\\
\label{second}
& = & \int d^4x{\rm e}^{{\rm i}qx}\langle P|J^{\mu}(x)J^{\nu}(0)|P\rangle,
\end{eqnarray} 
where the second equation is obtained by using completeness 
for states of $X$. The hadronic tensor is directly related to the
imaginary part of the forward Compton scattering amplitude via the
optical theorem. We should stress that the hadronic tensor cannot be
computed by perturbative QCD. However, one can write down the most general tensor from
available momentum vectors $P^\mu, q^\mu$ and from $g^{\mu\nu}$ by using the transversality of
the electromagnetic current $q_\mu W^{\mu\nu}=q_\nu W^{\mu\nu}=0$ , and parity
and time-reversal symmetry $W^{\mu\nu}=W^{\nu\mu}$: 
\begin{equation} 
W^{\mu\nu}=\left(-g^{\mu\nu}+\frac{q^\mu q^\nu}{q^2}\right)F_1(x,Q^2)+
\left(P^\mu+\frac{q^\mu}{2x}\right)\left(P^\nu+\frac{q^\nu}{2x}\right)
\frac{F_2(x,Q^2)}{\nu}, 
\end{equation}
where $F_{1,2}(x,Q^2)$ are the so-called structure functions. Making use of the above expression, the DIS cross-section reads

\begin{equation} 
\frac{d^2\sigma}{dx dQ^2}=\frac{4\pi\alpha_{em}^2}{Q^4}\left\{
\left(1-y-\frac{x^2 y^2 m_N^2}{Q^2}\right)\frac{F_2(x,Q^2)}{x}
+y^2F_1(x,Q^2)\right\}.
\end{equation}
In the next section, we consider whether $F_{1}$ and $F_{2}$ can
be independent. The DIS can be viewed as $\gamma^\star p$
scattering. The cross-section for a virtual photon with helicity
$\lambda$ can be defined as
\begin{equation}
\sigma_\lambda=\frac{4\pi^2\alpha_{em}}{2s}
\epsilon_\mu(\lambda)\epsilon_\nu^*(\lambda)\,{\rm Im}T^{\mu\nu}, \label{cco}
\end{equation}
where the forward virtual Compton scattering
amplitude $T^{\mu\nu}$ is related to the hadronic tensor, 
\begin{equation}
W^{\mu\nu}=\frac{1}{2\pi}{\rm Im}T^{\mu\nu}
=\lim_{\epsilon\to 0}
\frac{1}{4\pi{\rm i}}(T^{\mu\nu}(q_0+{\rm i}\epsilon)-T^{\mu\nu}(q_0-{\rm i}\epsilon)),
\end{equation}
with
\begin{equation}\label{com}
T^{\mu\nu}={\rm i}
\int d^4x{\rm e}^{{\rm i}qx}\langle P|{\cal T}( J^{\mu}(x)J^{\nu}(0))|P\rangle.
\end{equation}
Here, ${\cal T}$ is the time ordering operator. Note that the relation between the
hadronic tensor and the Compton scattering amplitude is a manifestation of the optical
theorem. Using Eqs.~(\ref{cco},\ref{com}), one obtains the corresponding cross-section
for transverse and longitudinal photons,  
\begin{eqnarray} 
\sigma_T&=&\frac{4\pi^2\alpha_{em}}{Q^2(1-x)}2xF_1(x,Q^2),\nonumber\\
\sigma_L&=&\frac{4\pi^2\alpha_{em}}{Q^2(1-x)}\left[
\left(1+\frac{Q^2}{\nu^2}\right)F_2(x,Q^2)
-2xF_1(x,Q^2)\right]. \label{nnn}
\end{eqnarray} 
It is sometimes common to define linear combinations of the structure functions
\begin{eqnarray}
F_L(x,Q^2)&=&F_2(x,Q^2)-2xF_1(x,Q^2),\\
F_T(x,Q^2)&=& 2xF_1(x,Q^2).
\end{eqnarray}
The usefulness of the above definition is that $\gamma^\star p$ scattering
for transverse and longitudinal photons can be defined in terms of $F_{L,T}$. 
\begin{figure}[!t]
       \centerline{\includegraphics[width=9.0 cm] {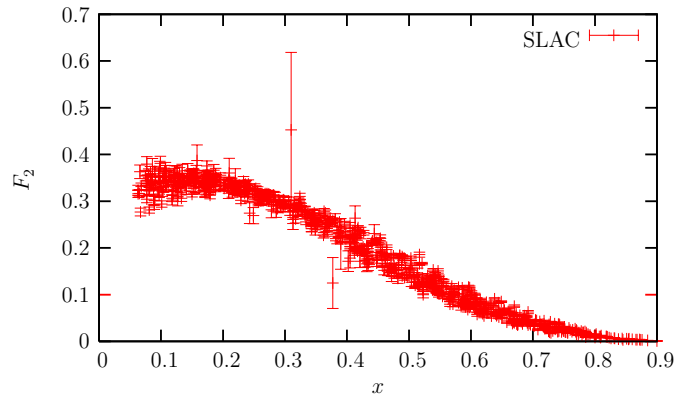}}
        \caption{SLAC data on the structure function $F_2$ in DIS \cite{sla}. \label{slac}}
\end{figure}
%%%%%%%%%%%%%%%%%%%%%%%%%%%%%%%%%%%%%%%%%%%%%%%%%%%%%%%%%%%%%%%%%%%%%%%%%%%%%%%%%%%%%
%%%%%%%%%%%%%%%%%%%%%%%%%%%%%%%%%%%%%%%%%%%%%%%%%%%%%%%%%%%%%%%%%%%%%%%%%%%%%%%%%%%%%%%
\subsection{Bjorken scaling and parton model}
In the late 1960's, experimental result from SLAC \cite{sla} surprisingly showed that
the structure function $F_{2}(x, Q^2)$ is nearly
independent of $Q^2$ at a fixed $x$. In Fig.~\ref{slac}, we show the
measured values of $F_{2}(x, Q^2)$ for various $Q^2$ as a function of
$x$. It is obvious that all data points seems to lie on a single curve
which show that within error bars $F_2$ is independent of $Q^2$. This
phenomenon is called Bjorken scaling \cite{xb,xb1}. 

\begin{figure}[t]
       \centerline{\includegraphics[width=6.0 cm] {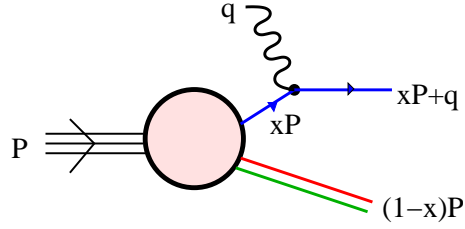}}
        \caption{ Parton picture of DIS and Bjorken x. \label{dis1}}
\end{figure}

An intuitive explanation of this phenomenon was given by Feynman
\cite{fey}, within the parton model. In the parton model, one assumes
that the proton is made of pointlike charged constituents, called partons
which interact incoherently. Then the total $\gamma^\star p$ cross-section can be written as an incoherent sum of photon-parton
cross-sections. We work in the Breit frame in which proton and virtual
photon are moving collinearly and the virtual photon does not carry
the energy, but only momentum. Assume that the scattering is elastic
and the parton of type $q$ carries a fraction $\eta$ of the proton's momentum, see Fig.~\ref{dis1}. For massless partons, we have
\begin{equation} 
(q+\eta P)^2=2\eta P.q-Q^2=0,  
\end{equation} 
which leads to 
\begin{equation} 
\eta=x.
\end{equation} 
This implies that in the Breit frame, the Bjorken $x$ is the momentum
fraction of the proton carried away by the struck quark.

In order to calculate the $\gamma^\star p$ scattering cross-section in
the parton model, one should first calculate $eq\to eq$ cross-section, 
which can be obtained from those for $e^{+}e^{-}\to q\bar{q}$ by
crossing symmetry. Equivalently, one may first calculate the cross-section for transverse and longitudinal photons scattering off spin-$1/2$ parton, see Fig.~\ref{dis1}, 
\begin{eqnarray}
\sigma_T^{\gamma^*q}&=&\frac{4\pi^2\alpha_{em}Z_f^2}{Q^2(1-x)}
\delta\left(1-\frac{x}{\eta}\right),\nonumber\\
\sigma_{L}^{\gamma^*q}&=&0.\label{pa-c}
\end{eqnarray}

For massless quarks, the longitudinal cross-section has to be zero
because of helicity conservation. Now, by comparing Eqs. (\ref{pa-c},\ref{nnn}),
one can define the structure function in parton language, 
namely by introducing the density $q_{f}(x)$ of quarks of flavor $f$ inside proton, 
\begin{eqnarray}\label{master}
F_2(x)&=&x \sum\limits_{f=u,d,..}^{}Z_f^2\left( q_f(x)+\bar q_f(x)\right),\\
F_L&=&0.
\end{eqnarray}
At this order, the structure functions depend only on $x$ and
not on $Q^2$. When the longitudinal structure function vanishes, one obtains the
Callan-Gross relation \cite{cal},
\begin{equation}
F_2-2x F_1=0.
\end{equation}
This equation is approximately confirmed by experiment and proved that partons are fermion with spin one half. 
It is common to define valence quarks $u_{v}, d_v, ..$, as
\begin{eqnarray}
u_v&=&q_u-\bar{q}_u,\nonumber\\
d_v&=&q_d-\bar{q}_d,\
\end{eqnarray}
where $\bar{q}_u, \bar{q}_d$ are called sea anti-quarks. It is also
possible to measure DIS on the neutron and extract the neutron
structure function. Assuming strong isospin symmetry, we have the
following relations between the parton distribution functions in proton
and neutron:
\begin{eqnarray}
\fl q_u/n=q_d/p, \hspace{1cm}\bar{q}_u/n=\bar{q}_d/p,\hspace{1cm}
\bar{q}_d/n=\bar{q}_u/p,\hspace{1cm}
q_s/n=q_s/p,\hspace{1cm}
q_c/n=q_c/p,\nonumber\\
\end{eqnarray}
and so on. The convention is always to refer to the proton PDFs and drop the subscript $/p$ or $/n$.   

At this stage, one may wonder why the proton form factor 
$F(Q)$ falls steep with $Q$, while the structure function does
not. The answer is that the form factor is the probability for the
proton to survive intact a kick of strength $Q$. The stronger the kick,
the less survival probability. However, in the case of inclusive
DIS, all final states are allowed, so the total probability saturates and
is independent of $Q$. A similar situation is known to happen in hadronic collisions, where the
$t$-slope of single diffraction is half that for elastic $pp$,
because of the disappearance of one of the proton form factors.

%\begin{figure}[!t]
%       \centerline{\includegraphics[width=8.0 cm] {pic6.eps}}
%        \caption{ partons within protons\label{dis1}}
%\end{figure}
%%%%%%%%%%%%%%%%%%%%%%%%%%%%%%%%%%%%%%%%%%%%%%%%%%%%%%%%%%%%%%%%%%%%%%%%%%%%%%%%%%%
%%%%%%%%%%%%%%%%%%%%%%%%%%%%%%%%%%%%%%%%%%%%%%%%%%%%%%%%%%%%%%%%%%%%%%%%%%%%%%%%%%%%%
\subsection{Scaling violation and DGLAP evolution equation}
\begin{figure}[!t]
       \centerline{\includegraphics[width=8.0 cm] {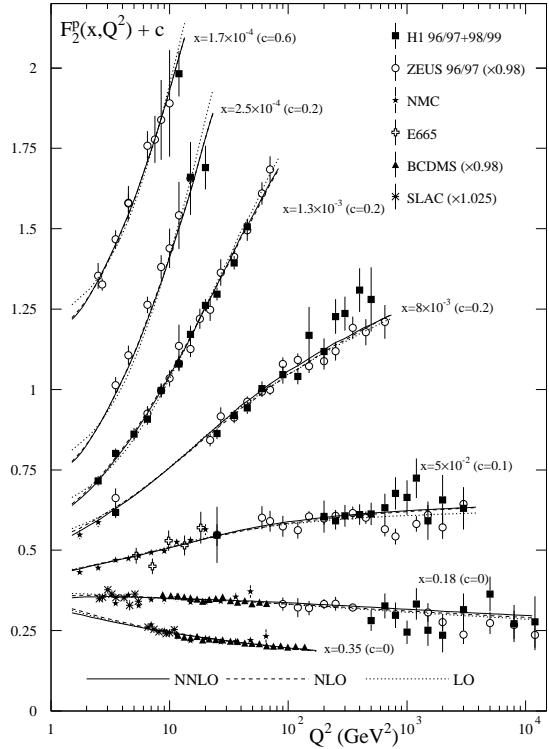}}
       \caption{ Comparison of the measured proton structure function $F_2$
       with QCD fits \cite{pdfnlo}\label{div}.}
\end{figure}

In the previous section we showed that at leading order the partonic
sub-process of DIS $e q\to eq$ is $Q^2$ independent which leads
to Bjorken scaling. This could be correct if the number of partons
were constant. However, they are not classical particles but quantum
fluctuations. A photon of virtuality $Q$ can resolve partons with
transverse momentum $k_{T}<Q$ but is blind to harder
fluctuations. Increasing $Q$, one can see more partons in
the proton. Correspondingly, the parton distribution slowly changes with
$Q$, shifting to smaller $x$ due to momentum conversation,
i.e. it is expected to rise with $Q$ at small $x$, but to fall at large
$x$, see Fig.~\ref{div}. The $Q^2$ dependence of the structure
function can be described by the DGLAP
(Dokshitzer-Gribov-Lipatov-Altarelli-Parisi) equations \cite{dgl}.

One of the major successes of QCD has been the prediction of the pattern of
Bjorken scaling violation as shown in Fig.~\ref{div}. We will explain
that the presence of gluon radiation controls the behaviour of Bjorken
scaling violation. At higher order in $\alpha_s$ one should also include gluon
radiation $eq\to eqg$. As in the previous section, one obtains at next order
\begin{eqnarray}
\fl \frac{F_2(x,Q^2)}{x}&=&\sum\limits_fZ_f^2\Bigg[
q_f\left(x\right)
+\frac{\alpha_s}{2\pi}\int\limits_{x}^1 \frac{dx_1}{x_1} 
g(x_1)\left\{P_{fg}\left(\frac{x}{x_1}\right)
\ln\left(\frac{Q^2}{\mu^2}\right)
+...\right\}
\Bigg], \label{dglp}
\end{eqnarray}
where $g(x_{1})$ denotes the gluon density of the proton. The origin
of the $\ln (Q^2/\mu^2)$ is easy to understand. The struck quark
acquires transverse momentum $p_{T}$ with probability $\alpha_s
\frac{d^2p_{T}}{p_{T}^{2}}$. On the order hand, partons with $p_{T}^2>Q^2$ are suppressed. Now, integrating over all phase space, $p_{T}$
produces the logarithmic term $\alpha_s \ln Q^2/\mu^2$. The parameter
$\mu$ was introduced as a cutoff regulator. The divergence when
$\mu\to 0$ corresponds to a case that the outgoing gluon becomes
exactly collinear with the incoming quark. This means that the
internal quark line becomes on-shell leading to the logarithmic
divergence. This is called {\it collinear divergence}. The function $P_{fg}$
is quark-quark splitting function \cite{dgl,split1}
\begin{equation}
\label{sp1}
P_{fg}(z)=\frac{4}{3}\left(\frac{1+z^2}{1-z}\right).
\end{equation}
The splitting function $P_{fg}$ shows the probability for a quark to turn
into a quark and a gluon. This function is independent
of the regularization and is universal. 
\begin{figure}[!t]
       \centerline{\includegraphics[width=14.0 cm] {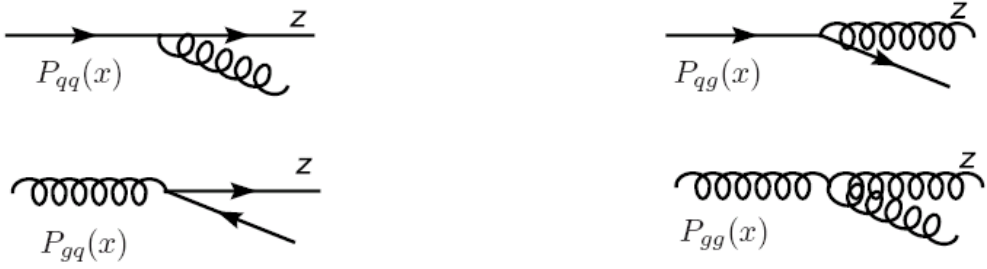}}
        \caption{Four type of diagrams corresponding to DGLAP splitting functions of QCD. \label{pp}}
\end{figure}

Note that the structure function equation (\ref{dglp}) is now obviously $Q^2$ dependent,
violating Bjorken scaling. We also introduced an ad hoc parameter $\mu$, called
the {\it factorization scale}, which separates the underlying
physics into two parts: all physics at scale below $\mu$ contained in
the parton distribution, and all calculable physics at scales above $\mu$ are 
part of the partonic scattering cross-section. It is important to
mention that although we have obtained Eq.~(\ref{dglp}) at the higher
order, the leading logarithmic behaviour is
universal and this factorization formula is valid at any order of
$\alpha_{s}$. Since $\mu$ is not a physical quantity, observables
should not depend on it. Therefore,
\begin{equation}
\frac{dF_2(x,Q^2,\mu)}{d\mu}=0,
\end{equation}
and 
\begin{equation}
Q^2\frac{dq_f(x,Q^2)}{dQ^2}=
\frac{\alpha_s}{2\pi}\int_{x}^1 \frac{dx_1}{x_1}
P_{fG}\left(\frac{x}{x_1}\right)g(x_1,Q^2). \label{ddgg}
\end{equation}
This is one of the DGLAP equations \cite{dgl} which describes the evolution of
the quark density. In the same fashion one can obtain the DGLAP
equations for gluon density $g(x_{1},Q^2)$.  Altogether one obtains
$N_f+1$ coupled equations (ignoring anti-quark for
simplicity) describing the $Q^2$ evolution of the singlet parton
densities $q_f(x_{1},Q^2)$ and $g(x_{1},Q^2)$,
\begin{equation}\label{DGLAP}
\fl Q^2\frac{d}{dQ^2}
\left(
\begin{array}{c}
q_f(x,Q^2)\\[1em]
g(x,Q^2)
\end{array}
\right)
=
\frac{\alpha_s}{2\pi}\int\limits_{x}^1 \frac{dx_1}{x_1}
\left(
\begin{array}{cc}
P_{ff}\left(\frac{x}{x_1}\right)
&P_{fg}\left(\frac{x}{x_1}\right)\\[1em]
P_{gf}\left(\frac{x}{x_1}\right)&P_{gg}\left(\frac{x}{x_1}\right)
\end{array}\right)
\left(
\begin{array}{c}
q_f(x_{1},Q^2)\\[1em]
g(x_{1},Q^2)
\end{array}
\right),
\end{equation}
where the splitting function $P_{ff}, P_{fg}, P_{gf}$ and $P_{gg}$ can
be computed from pQCD order by order. The analytic calculation of these splitting
functions to next-to-next-to-leading order has been carried out in Ref~\cite{split1}.
In Fig.~\ref{pp} we show the
lowest typical diagrams corresponding to the various splitting functions.

The steep rise of $F_2$ at small $x$ in Fig.~\ref{div} can be simply seen from  the double log DGLAP equation, 
\begin{equation}
\frac{\partial^2xg(x,Q^2)}{\partial\ln(1/x)\partial\ln Q^2}
=\frac{N_c\alpha_s}{\pi}xg(x,Q^2). \label{dla-dg}
\end{equation}
For a fixed coupling constant $\alpha_s$, the solution can be approximated by  
\begin{equation}\label{asys}
xg(x,Q^2)\propto\exp\left(2\sqrt{\frac{N_c\alpha_s}{\pi}
\ln(1/x)\ln(Q^2/Q^2_0)}\right).
\end{equation}
This equation clearly indicates that at small $x$ and high $Q^2$ the
gluon density rises.

The parton distribution function (PDF) cannot be calculated from the first
principles. However, their scale evolution can be perturbatively
computed via DGLAP equations. We therefore, calculate the $\mu^2$
dependence of the PDFs. In this way knowing the value of PDF at a
given scale by fitting data is sufficient to obtain
information about PDFs at all scales via DGLAP evolution
equations. The DGLAP equation is a special kind of a
renormalization equation. It is obvious from the $\ln Q^2/\mu^2$ term that
one should not chose $\mu$ too far from $Q^2$ since the log term will
become large enough to compensate the smallness of $\alpha_{s}$ and
perturbative computation will become questionable.
\begin{figure}[!t]
        \centerline{\includegraphics[width=15.0 cm] {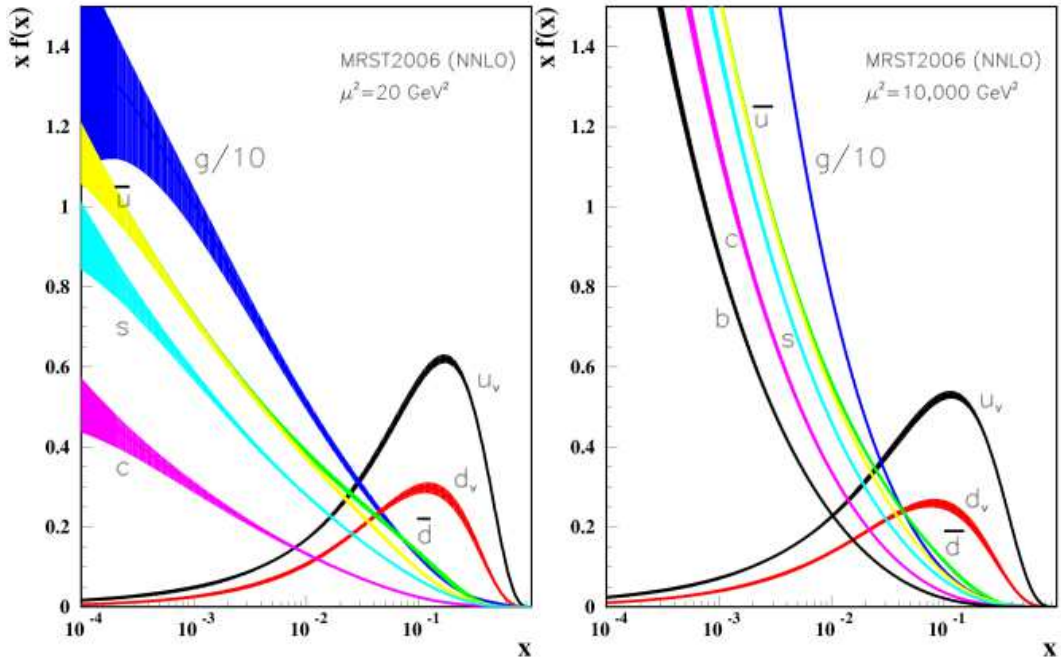}}
        \caption{PDFs $xf(x,\mu^2)$ at $\mu^2=20, 10^4 ~\hbox{GeV}^2$
        as a function of $x$. It is clearly seen that the gluons and the
        valence quarks are more important at small and large $x$,
        respectively. The curves are obtained from the NNLO global
        analysis \cite{mrst-c}. The figure is taken from
        Ref.~\cite{pdf-c}. \label{pf}}
\end{figure}

The typical strategy for extracting PDFs from DIS data is first to introduce ad hoc PDFs at some
scale and then to evolve them with DGLAP to other scales, and finally compare $F_2$ at higher values of
$Q^2$ with data and adjust the starting PDFs. Having good data with high statistics one can single out PDFs
for different parton species. Such parametrizations are provided by three collaborations: GRV \cite{grv}, MRST \cite{mrst} and CTEQ \cite{cteq}
in leading and next to leading orders. In Fig.~\ref{pf} we show
typical PDFs at $Q^2=10~\hbox{GeV}^2$ as a function of $x$.

Having results for the PDFs, one can check how much of proton's
momentum is carried by quarks and antiquarks. The data shows that, 
\begin{equation}
\int_{0}^{1}dx F_{2}\approx 0.5.
\end{equation}
This result is quite significant, since it shows that only half
of the total momentum is carried by all quarks and antiquarks in
the proton. Another half of the proton momentum is carried by partons
which do not interact with the photon, apparently gluons.
%%%%%%%%%%%%%%%%%%%%%%%%%%%%%%%%%%%%%%%%%%%%%%%%%%%%%%%%%%%%%%
\subsection{Factorization theorem}
A cross-section of any hadronic reaction with a hard scale generally gets contribution from short- and long-distance
interactions, and is hence not computable directly in perturbation theory
for QCD.  Factorization theorems \cite{collins1} allow one to derive predictions for
the hadronic cross-sections by writing the cross-section as 
a convolution product of factors, namely an infrared finite part for
the short distance which is calculable in perturbative theory, with a
nonperturbative function which is universal to many different
processes, but non-calculable at the perturbative level. The applications
and predictability of perturbative QCD rely on the factorization theorem.

There has been tremendous effort to examine factorization theorems
for various processes; for a review see Ref.~\cite{collins1}. For inclusive processes, it has been shown that
the factorization theorem holds if (1) all Lorentz invariants defining
the process are large and comparable, except for particle masses, and
(2) one counts all final states that include the specified outgoing
particles or jets, namely in processes as hadron $A + \hbox{hadron}~B \to \hbox{hadron}~C+X$, the $X$ denotes anything else, in addition
to the specified hadron $C$.  

For example, in DIS, the factorization theorem for the structure functions has the following form,
\begin{equation}
F_{i}(x,Q)=\sum_{a}\int_{x}^1 \frac{d\xi}{\xi} f_{a/H}(\xi,\mu)\mathcal{C}_{ia}(x/\xi,Q/\mu, \alpha_s(\mu))+... \label{fact}
\end{equation}
which is valid in the Bjorken limit in which $Q$ gets large with $x$
fixed. The sum is over all species of partons, namely gluon, quarks
and antiquarks of different flavours. The function $f_{a/H}$ denotes the PDF of
parton of type $a$ in hadron $H$. The hard process-dependent factor
$\mathcal{C}_{ia}$ is ultraviolet dominated, that is, it receives
important contributions only from momenta of order Q. This ensures
that one can perturbatively calculate $\mathcal{C}_{ia}$ in power of
$\alpha_s(Q)$ [see Eq.~(\ref{dglp})]. Notice that the factor $\mathcal{C}_{ia}$ depends only
on the parton type a, and not directly on our choice of hadron A.
The parameter $\mu$ in Eq.~(\ref{fact}) defines the limit between the short-distance
dynamics. The ability to calculate the $\mathcal{C}_{ia}$ leads to
great predictive power for factorization theorems. For instance, if we
measure $F_2(x,Q)$ for a particular hadron $A$, Eq.~(\ref{fact}) will
enable us to determine the PDFs $f_{a/A}$. Then we predict $F_1(x,Q)$
for the same hadron $A$, in terms of the same $f_{a/A}$ and calculable
$\mathcal{C}_{1a}$.

%%%%%%%%%%%%%%%%%%%%%%%%%%%%%%%%%%%%%%%%%%%%%%%%%%%%%%%%%%%%%%
\section{BFKL formalism}
\begin{figure}[!t]
       \centerline{\includegraphics[width=4.0 cm] {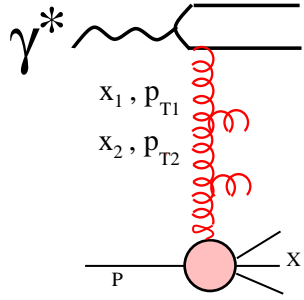}}
       \caption{The QCD improved parton model, $x_{i}$ denote the
       longitudinal momentum fraction of the partons with respect to
       the target. \label{dis33}}
\end{figure}

We recall that the DGLAP equations take into account all the contributions proportional
to 
\begin{equation}
[\alpha_{s}(Q^2)\ln\left(\frac{Q^2}{Q^{2}_{0}}\right)]^{n},
\end{equation}
 which arises from ladder type diagrams with strong ordering in the transverse momenta, see Fig.~\ref{dis33}, i.e.,
\begin{equation}
p^{2}_{T1}>>p^{2}_{T2}>>....
\end{equation}
%This corresponds to the leading logarithmic approximation valid in the limit
%\begin{equation}
%\alpha_{s}(Q^2)\ln\left(\frac{1}{x}\right)<<\alpha_{s}(Q^2)\ln\left(\frac{Q^2}{Q^{2}_{0}}\right)<1.
%\end{equation}
For processes where $x$ is small, but $Q^2$ is not sufficiently large to
make the double logarithmic approximation valid, the
Balitsky-Fadin-Kuraev-Kuraev-Lipatov (BFKL) equation \cite{bfkl} has been
proposed. In this scheme, the gluonic branching in the ladder diagrams has ordering in longitudinal momentum (see Fig.~\ref{dis33})
\begin{equation}
x_{1}>>x_{2}>>....
\end{equation}
This resums, in the so-called $\ln(1/x)$ approximation, the terms
\begin{equation}
[\alpha_{s}(Q^2)\ln\left(\frac{1}{x}\right)]^{n}.
\end{equation}
At the same time there is no ordering in transverse momentum, one may
have
\begin{equation}
p^{2}_{T1}\sim p^{2}_{T2}\sim ....\sim p^{2}_{Tn}. \label{t-or}
\end{equation}
%The BFKL formalism is therefore applicable in the limit which
%\begin{equation}
%\alpha_{s}(Q^2)\ln\left(\frac{Q^2}{Q^{2}_{0}}\right)<<\alpha_{s}(Q^2)\ln\left(\frac{1}{x}\right)<1.
%\end{equation}
In the high energy limit, the scattering processes are dominated by
partonic processes with gluon exchange in the $t$-channel. The BFKL
equation accounts for resummation of multiple gluon radiation when $s>>t$. 
%Here, for pedagogical reason, we closely follow Refs.~\cite{bfkl3,bfkl4}. 

The BFKL equation is more conveniently
written in terms of the unintegrated gluon density $\phi(x,k_{T}^{2})$ which 
relates to the gluon density $g(x, Q^2)$ introduced in the previous section by
\begin{equation}
xg(x, Q^2)=\int_{0}^{Q^2} dk_{T}^{2} \phi(x,k_{T}^{2}). \label{d-ugg}
\end{equation}
The unintegrated gluon distribution gives the probability of finding a gluon in
the hadron with longitudinal momentum fraction $x$ and transverse momentum $k_T$.
Note that there is no unique definition for the unintegrated gluon density in terms of gluon density \cite{unin-gg,has}. 
For comparison of various parametrizations for the unintegrated gluon distribution in different schemes, see Ref.~\cite{unin-gg}.

\begin{figure}[!t]
%       \centerline{\includegraphics[width=14.0 cm] {bbb.eps }}
        \includegraphics[height=.19\textheight]{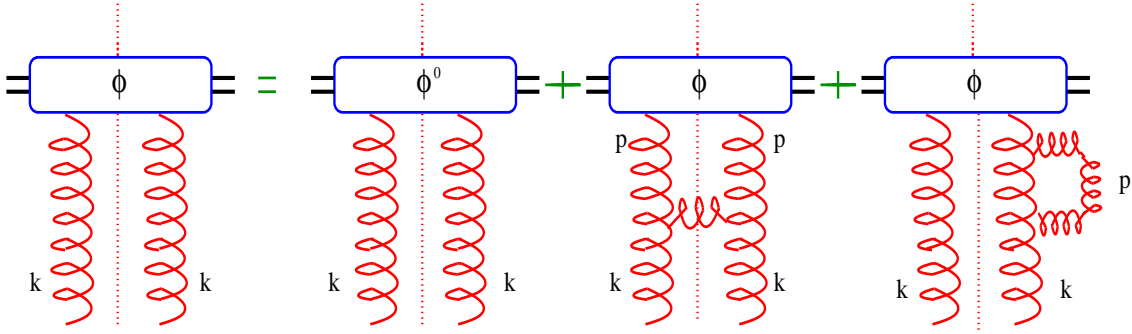}
       \caption{ Schematic representation of BFKL evolution for
       unintegrated gluon distribution. The dashed vertical denotes
       the cut.\label{bf1p}}
\end{figure}

At leading order in $\ln(1/x)$ the BFKL equation can then be written in the following simple form:
\begin{equation}
\frac{\partial \phi(x,k_{T}^{2})}{\partial \ln(1/x)}= \frac{N_c\alpha_{s}}{\pi^2}\int \frac{dp_{T}^2}{(k_{T}-p_{T})^2}
\left(\phi(x,p_{T}^2)-\frac{k_{T}^{2}\phi(x,k_{T}^2)}{p_{T}^2+(k_{T}-p_{T})^2} \right).  \label{bfkl-eq1}
\end{equation} 
 Equation (\ref{bfkl-eq1}) is illustrated in Fig.~\ref{bf1p}. The
 first term in Fig.~\ref{bf1p} corresponds to two gluon exchange, the
 initial condition for Eq.~(\ref{bfkl-eq1}). The first and second
 terms on the right hand side of Eq.~(\ref{bfkl-eq1}), correspond to the
 second (real) and third (virtual) terms in Fig.~\ref{bf1p},
 respectively. Iterating the BFKL kernel leads to the ladder diagrams shown in Fig.~\ref{ladder}. Note that the gluon propagators and vertices in Figs.~\ref{ladder},\ref{bf1p}
 are not the usual QCD vertices and propagator. The vertices are effective
 Lipatov vertices and propagator are the so-called reggeized gluon
 propagators generated by iterating the BFKL kernel \cite{bfkl}. We
 refer the interested readers to Refs.~\cite{bfkl,bfkl2,bfkl3} for
 derivation of the BFKL equation. For a recent review of the subject see
 Ref.~\cite{bfkl4}.

It is rather straightforward to see that for a fixed
coupling $\alpha_{s}$, the solution for the unintegrated gluon density up to a constant is
\begin{equation}
\phi(x,k_{T}^{2})\propto (1/x)^{\alpha_\Pom-1}, \hspace{2cm}  \alpha_\Pom-1=\frac{4\alpha_{s}N_{c}}{\pi}\ln2. \label{pom-s}
\end{equation}  
However, it is important to notice that, based on the BFKL equation,  
number of gluons rises with $1/x$ forever. This strongly indicates that
some physics must be missing here. That is because in QCD the gluon
fields cannot be stronger than $A_\mu\sim 1/g$ at very small coupling $g$. Therefore, when the gluon field reaches a density with
\begin{equation}
\frac{F_{\mu\nu}}{Q^2}\sim \frac{1}{g},
\end{equation}  
we expect some new physics to be at work in order to slow down the rise
of gluon density. We shall postpone to elaborate more on this problem in the next section.

The total cross-section of quarkonium-quarkonium scattering, in the
lowest order in the coupling $\alpha_s$, in the simplest model of
two-gluon exchange between two quarkonium is energy independent
\begin{equation}
\sigma\sim s^0,
\end{equation} 
where $s$ denote the center of energy of the system. However,
experimental data indicates that hadronic cross-sections 
increase as power of $s$, see Fig.~\ref{data}
\begin{equation}
\sigma\sim s^\Delta,
\end{equation} 
where $\Delta$ is called the {\it Pomeron intercept}. Finding an explanation for the
experimental value of $\Delta$ has been one of the remaining
challenges of QCD. The DGLAP equations has been 
successful in describing the DIS data, but it cannot reproduce the energy growth of hadronic cross-sections.  
One of the interesting features of the BFKL formalism is that it naturally leads to an energy-dependent
cross-section. One can show that cross-sections mediated by the BFKL exchange grows as a power of energy
\begin{equation}
\sigma\sim s^{\alpha_\Pom-1}, \label{fo1}
\end{equation}     
where $\alpha_\Pom$ is given in Eq.~(\ref{pom-s}). Unfortunately
the value of $\alpha_\Pom-1\approx 0.8$ is higher than experimental
value $0.2-0.3$ observed in DIS experiments. Before going to
higher-order corrections to the BFKL kernel, it is important to notice
that there is already a serious problem at lower order. That is due to the fact
that the power energy growth of the total cross-section (\ref{fo1})
violates the Froissart unitarity bound \cite{f-u} which put a limit on the growth rate of
total cross-sections with energy $s$ at asymptotically high energies 
\begin{equation}
\sigma\leq \hbox{const} \ln^2s.
\end{equation} 
This indicates that the BFKL kernel should be modified in order to
restore the unitarity at high energy.

\section{The GLR-MQ evolution equation and saturation}
In the previous section we pointed out that, based on the BFKL formalism,
the number of gluons rises sharply at small $x$ or high energy. At the
same time, the transverse sizes of the gluons $r_T\sim 1/p_T$, can be
similar, see Eq.~(\ref{t-or}). This means that at high energy, a hadron
produces many gluons with a similar size. As the energy increase, more
gluons are produced and eventually they start overlapping in
transverse plane. The crucial assumption behind both DGLAP and BFKL
evolution equations is that parton densities inside a hadron are small
enough, so that the only important partonic sub-process is
{\it splitting}. However, at very low value of $x$, the gluon density may
become so large that gluons start overlapping and the gluon recombinations
process becomes important. This phenomenon is generally known as
{\it parton saturation}, and it should limit the growth of the gluon density
generated by splitting.

Gribov, Levin and Ryskin (GLR) \cite{glr-ci} proposed that at high density of gluon
fields when nonlinear effects become important, there should be an
energy region where the gluon recombination becomes important. In the GLR scheme this recombination is described through
a modification of the linear BFKL equation with a quadratic correction
which gives rise to effective ladder merger vertices which are the triple Pomeron ones,
\begin{equation}
\fl \frac{\partial \phi(x,k_{T}^{2})}{\partial \ln(1/x)}= \frac{N_c\alpha_{s}}{\pi^2}\int \frac{dp_{T}^2}{(k_{T}-p_{T})^2}
\left(\phi(x,p_{T}^2)-\frac{k_{T}^{2}\phi(x,k_{T}^2)}{p_{T}^2+(k_{T}-p_{T})^2} \right)-\frac{\alpha_s^2\pi}{S_T}[\phi(x,k_{T}^{2})]^2,  
\label{GLR}
\end{equation} 
where $S_T=\pi R^2$ defined the geometrical cross-sectional area of
a hadron or a nucleus along the beam axis. In comparison with the BFKL equation
(\ref{bfkl-eq1}), only the last term is new. 
\begin{figure}[!t]
       \centerline{\includegraphics[width=10.0 cm] {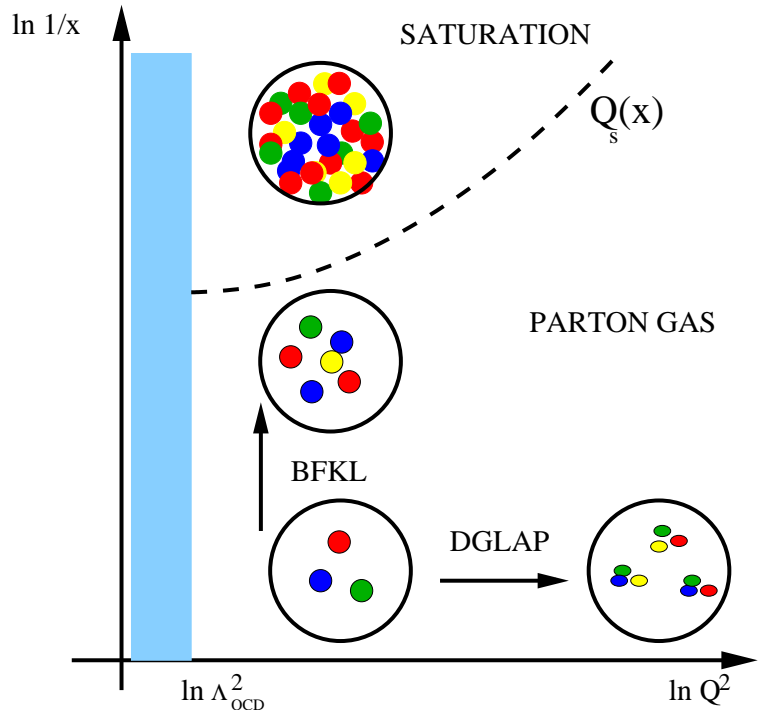}}
       \caption{ Saturation region in $x, Q^2$ plane. \label{sat}}
\end{figure}

Later, it was proved by Mueller and Qiu \cite{mq-ci} that the ansatz
Eq.~(\ref{GLR}) can be derived in the double leading logarithmic
approximation (DLLA) with a resummation of the type 
$\left(\alpha_s\ln(Q^2/\Lambda^2)\ln(1/x)\right)^n$. Muller and Qiu
\cite{mq-ci} showed that in the DLLA approximation, including diagrams
with two fusing DGLAP ladders, one arrives at the following nonlinear
equation for gluon density:
\begin{equation}
\frac{\partial^2xg(x,Q^2)}{\partial\ln(1/x)\partial\ln Q^2/\Lambda^2}
=\frac{N_c\alpha_s}{\pi}xg(x,Q^2)- \frac{\alpha_s^2\pi}{S_T}[xg(x,Q^{2})]^2. \label{glr-mq}
\end{equation} 
The above equation can be converted into Eq.~(\ref{GLR}) via the definition of an unintegrated gluon density Eq.~(\ref{d-ugg}). 
Eq.~(\ref{glr-mq}) is known as the GLR-MQ equation. 

Notice that in the DLLA approximation both the BFKL and DGLAP equations
are identical, since the resummations are the same.  This can be already
seen in Eq.~(\ref{dla-dg}), where the merging ladders were ignored. That equation is
identical to the first term of Eq.~(\ref{glr-mq}).
 
One of the remarkable properties of the GLR-MQ equation (\ref{glr-mq})
is that it introduces a scale $Q_s^2$ at which the non-linear effects become relevant. This may occur when
the linear and quadratic terms in Eq.~(\ref{glr-mq}) becomes equal:
\begin{equation}
Q^2_s \sim \frac{\alpha_{s} \pi^2 xg(x, Q^2_s)}{S_T N_c}.\label{sss}
\end{equation}
A quantitative condition for gluon
saturation can be obtained by comparing the gluon recombination
cross-section $\sigma \sim \alpha_{s}/Q^2$ with the surface density of
gluons $\rho\sim xg(x,Q^2)/\pi R^2$. Saturation takes place when
$\sigma \rho\sim 1$ which leads to Eq.~(\ref{sss}).

The saturation scale $Q^2_s$ separates the linear
(governed by DGLAP or BFKL equations) and non-linear evolution of QCD. 
The DGLAP, the BFKL and saturated regimes are sketched in
Fig.~\ref{sat}. At low energy, colour screening is due to confinement
with typical colour screening distance $\Lambda_{\hbox{QCD}}^{-1}$, and
thus non-perturbative. At high energy (or small $x$), partons are much
more densely packed, and colour neutralization occurs in fact over
distances of the order $Q_s<<\Lambda^{-1}$. This means that small
$x$ physics seems to be universal, and all hadrons and nuclei should behave in
the same way at very high energy.

The basic physics of saturation is to introduce higher twist terms \cite{glr-ci,mq-ci} in
the factorization formula like Eq.~(\ref{dglp}). This is difficult to
implement. During last decade there has been some progress along these lines
and some models has been proposed \cite{bfkl4}. For example, the  description of this
non-linear evolution has been given in the so-called {\it  Colour Glass
Condensate} \cite{cgc} scheme in terms of a classical field theory of dynamical
gluon fields coupled to static stochastic sources. The evolution of
multi-parton correlators with energy is described by the JIMWLK
renormalization group equations \cite{jamal}. At large $N_c$ and large nuclei, one
recovers the Balitsky-Kovchegov (BK) equation \cite{bk-ci} for forward colour dipole
cross-section. The recombination effect is taken into account by the
non-linear term of the BK equation \cite{bk-ci,bask}.

%%%%%%%%%%%%%%%%%%%%%%%%%%%%%%%%%%%%%%%%%%%%%%%%%%%%%%%%%%%%%%%%%%%%%%%
%%%%%%%%%%%%%%%%%%%%%%%%%%%%%%%%%%%%%%%%%%%%%%%%%%%%%%%%%%%%%%%%%%%%%%%
\section{The colour dipole approach and low-$x$ DIS}
\begin{figure}[!t]
\centerline{\includegraphics[width=5.0 cm] {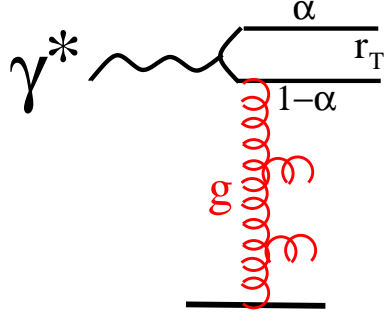}} 
\caption{ Photon virtual dissociation to a $\bar qq$ pair with transverse
separation $r_T$ and fractional the light-cone momenta $\alpha$ and
$1-\alpha$.\label{fr1}}
\end{figure}

The parton model description is not Lorentz invariant, only
observables have to be Lorentz invariant. One cannot even say where a
sea parton has originated, who is the owner, the beam or the
target, see Fig.~\ref{fr1}.  In the domain of small $x$, sea quarks and
gluons dominate, and the rest frame of the proton is more
convenient. In this frame, the photon can convert
into a quark-antiquark pair which then develops a parton cloud, see
Fig.~\ref{fr1}. Photons can also hit a quark inside the target without
being split to $q\bar{q}$-pair.  However, in target rest frame this
process is strongly suppressed. Therefore the former contribution is
the dominant one. The lifetime $t_{c}$ of such $q\bar{q}$-pair
fluctuation can be estimated via the uncertainty relation $t_{c}\approx
\frac{1}{2m_N x}$, where $m_N=1$ GeV is the mass of a nucleon. The
smaller the Bjorken $x$, the larger the coherence time. For the
lowest value of $x$ accessible at HERA the coherence time in the
proton rest frame is about $10^5$ fm. Therefore the coherence time or
lifetime of such a pair creation $t_{c}$ can be larger than nuclear
radius at low $x$ and pairs can experience multiple scattering within
the coherence length. This is very important point for understanding
the phenomenon of nuclear shadowing. The total $\gamma^\star-p$
cross-section, or the forward amplitude, is described as the interaction
of a $q\bar{q}$ fluctuation of the photon with the target, as is shown
in Fig.~\ref{fr1},

\begin{figure}[!t]
\centering\includegraphics[width=.8\linewidth]{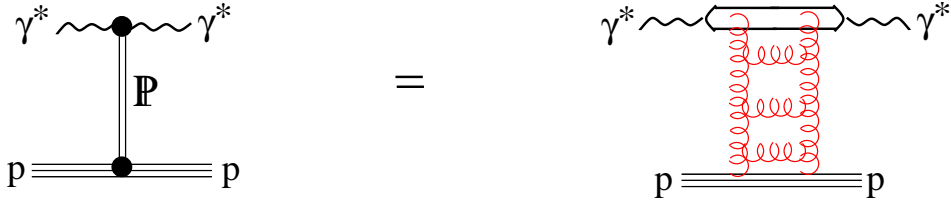}
\caption{The virtual 
photon interacts via its hadronic fluctuations
which are $\bar qq$ dipoles and more complicated Fock states.
The Pomeron exchange is illustrated as a perturbative ladder.}
\label{sigma-dipole}
 \end{figure}

The cross-section for the transverse and longitudinal photons is given by the factorized formula, \cite{zkl,nz},
\begin{equation}\label{ma}
\sigma_{T,L}^{\gamma^*p}=\int_0^1d\alpha\int d^2 r_T
\left|\Psi_{q\bar q}^{T,L}(\alpha,r_T)\right|^2\sigma_{q\bar q}(r_T),
\end{equation}
 where $r_T$ is the distance between the quark and antiquark in the
 transverse plane and $\alpha$ is the fraction of the photon energy
 carried by the quark, see Figs.~\ref{fr1} and \ref{sigma-dipole}. The
 cross-section for scattering a $q\bar q$-dipole off the proton is
 denoted by $\sigma_{q\bar q}(r_T)$. The light-cone (LC) distribution
 functions $\Psi_{q\bar q}^{T,L}(\alpha,r_T)$ for the transition
 $\gamma^*\to q\bar q$ can be calculated in perturbation theory and
 read to first order in the QED coupling constant $\alpha_{em}$ \cite{bks}:
\begin{eqnarray}\label{psit}
\fl\left|\Psi_{q\bar q}^{T}(\alpha,r_T)\right|^2&=&
\frac{2N_c\alpha_{em}}{(2\pi)^2}\sum\limits_{f=1}^{N_f}Z_f^2
\left\{\left[1-2\alpha(1-\alpha)\right]\epsilon^2{\rm K}^2_1(\epsilon r_T)
+m_f^2{\rm K}^2_0(\epsilon r_T)\right\},\\
\label{psil}
\fl\left|\Psi_{q\bar q}^{L}(\alpha,r_T)\right|^2&=&
\frac{8N_c\alpha_{em}}{(2\pi)^2}\sum\limits_{f=1}^{N_f}Z_f^2
Q^2\alpha^2(1-\alpha)^2{\rm K}^2_0(\epsilon r_T),
\end{eqnarray}
where ${\rm K}_{0,1}$ are the modified Bessel functions of the second
kind.  Note that the above distribution functions are not normalized
and can be even divergent. That is why we avoided to call them {\it wave
functions}. Although the transverse part of the distribution function
is divergent at $r_T\to 0$, the dipole cross-section vanishes in this
limit as $\sigma_{q\bar q}(r_T)\propto r_T^2$, so the result of
Eq.~(\ref{ma}) remains finite. We have also introduced a parameter
\begin{equation}\label{disextension}
\epsilon^2=\alpha(1-\alpha)Q^2+m_f^2,
\end{equation}
where the parameter $m_f$ is quark mass. The mean transverse
$q\bar{q}$ separation for a virtual photon is controlled by the Bessel
functions,
\begin{equation}
\langle r_{T}^2 \rangle \sim \frac{1}{\epsilon^2}= \frac{1}{\alpha(1-\alpha)Q^2+m_f^2}. \label{1800-cc}
\end{equation}
Thus the separation is about as small as $1/Q^2$ except at the end points
$\alpha\to 0, 1$. This implies that even a highly virtual photon can
create a large $q\bar{q}$ fluctuation although with a small probability. This is
an important point for the aligned jet model \cite{alig}. Notice that $m_{f}\sim
\Lambda_{\hbox{QCD}}$ plays here the role of an infra-red cutoff.

The incoming photon (or hadron) is not an eigenstate of the
interaction, since it can be diffractively excited. Therefore one
should switch to the eigenstate representation. The choice of the
eigenstate basis depends on the underlying theory. It was first
realized in Ref.~\cite{zkl} that the eigenstates of interaction in QCD
are colourless dipoles. Such dipoles cannot be excited during the
interaction time and can experience only elastic scattering. Indeed,
high-energy dipoles have no definite mass, but only separation $\vec
r_T$ which cannot be altered during soft interaction. The eigenvalues
of the total cross-section $\sigma_{q\bar q}(r_T)$ depend on $r_T$,
but may also depend on energy.
%One should also note that the same phenomena
%may be described from other approaches.  ~rst 

At the level of two-gluon exchange (Born approximation), the dipole cross-section is independent
of energy and related to the two-quark form factor of the proton via
 \cite{zkl}
\begin{equation}\label{bd}
\fl \sigma_{q\bar q}(r_T)=\frac{16\alpha_s^2}{3}\int d^2p_{T}
\frac{\left[1-\langle p|\exp({\rm i}\vec p_{T}\cdot(\vec r_1-\vec r_2))|p\rangle\right]
\left[1-\exp({\rm i}\vec p_{T}\cdot\vec r_T)\right]}
{p_{T}^4},
\end{equation}
Notice the colour screening factor $[1-\exp({\rm i}\vec
p_{T}\cdot\vec r_T)]$ in Eq.~(\ref{bd}), which makes the dipole cross-section vanishes as $r_{T}^2$ at $r_T\to 0$. This is an important
property of the dipole cross-section which is the cornerstone of the colour transparency phenomenon.

The energy dependence of the dipole cross-section is generated by higher order QCD corrections. 
For small distances $r_T\to 0$, one can relate $\sigma_{q\bar q}(r_T)$ to the phenomenological gluon density \cite{a10-ci}
\begin{equation} \label{gdd1}
\sigma_{q\bar q}(x,r_T)= \frac{\pi^2}{3}r_{T}^2\alpha_s(Q^2\sim 1/r_{T}^2)xg(x,Q^2 \sim 1/r_{T}^2).
\end{equation}
 When the dipole cross-section is proportional to the gluon density of the target, only quarks generated
from gluon splittings are taken into account in the cross-section Eq.~(\ref{ma}).  In other words, the valence
quark contribution (or the reggeons in the Regge phenomenology) are neglected and therefore
 Eq.~(\ref{ma}) is only applicable
when sea quarks dominate, i.e.\ at low $x$. Having said that, the
master Eq.~(\ref{ma}) is quite general and does not rely on the
applicability of the pQCD.

\begin{figure}[!t]
       \centerline{\includegraphics[width=6 cm] {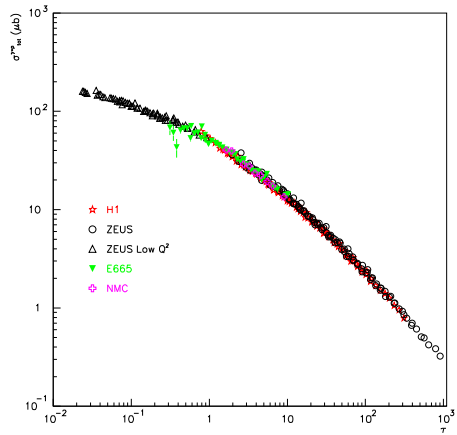}}
       \caption{Photon-proton total cross-section as a function of
       $\tau=Q^2/Q^2_s$ for $x<0.01$. The data are from the H1
       \cite{ref-h1}, ZEUS \cite{ref-z}, E665 \cite{ref-e} and NMC
       \cite{ref-nmc} collaborations.
\label{hera-s}}
\end{figure}

The dipole cross-section is theoretically difficult to predict, but several
parametrizations have been proposed in the literature. For our
purposes, here we consider two parametrizations, the saturation
model of Golec-Biernat and W\"usthoff (GBW) \cite{gbw} and the
modified GBW coupled to DGLAP evolution (GBW-DGLAP) \cite{gbw-d}.

\subsection{GBW model}
In the GBW model \cite{gbw} the dipole cross-section is parametrized as,
\begin{equation}
\sigma_{q\bar{q}}(x,r)=\sigma_{0}\left(1-e^{-\frac{1}{4}r^{2}Q_{s}^{2}(x)}\right), \label{gbw}
\end{equation}
where the parameters, fitted to DIS HERA data at small $x$, are given
by $\sigma_{0}=23.03$ mb, $Q_{s}(x)=1 \hbox{GeV }\times
(x/x_{0})^{-\lambda/2}$, where $x_{0}=3.04\times 10^{-4}$ and
$\lambda=0.288$. This parametrization gives a quite good description
of DIS data at $x<0.01$. One of the interesting feature of the HERA data is
a geometrical scaling \cite{gem-ci}; namely all available data for the
inclusive virtual photon-proton cross-section for $\leq 0.01$ and
various $Q^2$ seems to scale as a function of $\tau=Q^2/Q^2_s$, see Fig.~\ref{hera-s}.  This
might indicate that the semi-hard scale $Q^2_s$ which is also present
in the saturation region (See Section 9), plays a role already at 
the kinematics of HERA. However, one should be aware that the DGLAP evolution describes the same data as well.
So far it is not clear how much saturation is relevant to available DIS data.

%The saturation radius $R_{0}$ is analogous
%to the gluon distribution and determines the growth of cross section
%with decreasing $x$.

A salient feature of the model is that, for decreasing $x$, the dipole
cross-section saturates for smaller dipole sizes, and that at small
$r$, as perturbative QCD implies, the colour transparency
phenomenon $\sigma\sim r^{2}$, is at work.

\begin{figure}[!t]
       \centerline{\includegraphics[width=14 cm] {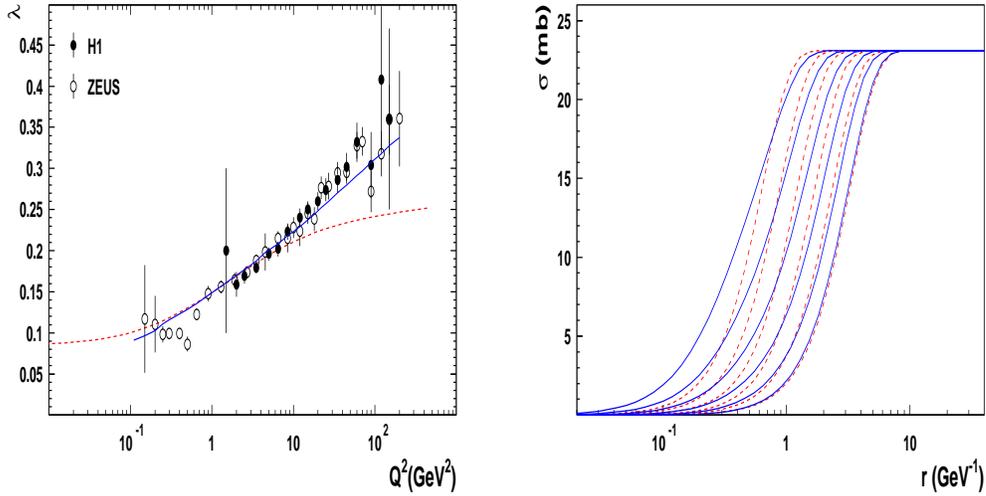}}
       \caption{Right: The dipole cross-section for $x=10^{-2},
       10^{-3},... 10^{-7}$ from left to right. The solid and dotted
       lines show results from the model with the DGLAP evolution
       Eq.~(\ref{gbw1}) and the saturation model Eq.~(\ref{gbw}),
       respectively. Left: The effective slope $\lambda(Q^2)$ from the
       parametrization $F_2\sim x^{-\lambda(Q^2)}$ as a function of
       $Q^2$.  The lines are the same as the right panel. The figure
       is taken from \cite{gbw-d}.  \label{fig-1}}
\end{figure}

One of the shortcomings of the GBW model is that it does not
match QCD evolution (DGLAP) at large values of $Q^{2}$. This
failure can be seen in the energy dependence of
$\sigma^{\gamma^{\star} p}_{tot}$ for $Q^{2}> 20~\hbox{GeV}^{2}$,
where the model predictions are below the data
\cite{gbw,gbw-d}. Apparently, the simple $r_T^2$ behaviour at small $r_T$ should be corrected.

\subsection{GBW coupled to DGLAP equation and dipole evolution}
A modification of the GWB dipole parametrization model,
Eq.~(\ref{gbw}), was proposed in Ref.~\cite{gbw-d}:
\begin{equation}
\sigma_{q\bar{q}}(x,\vec{r})=\sigma_{0}\left(1-exp\left(-\frac{\pi^{2}r^{2}
\alpha_{s}(\mu^{2})xg(x,\mu^{2})}{3\sigma_{0}}\right)\right), \label{gbw1}
\end{equation}
where the scale $\mu^{2}$ is related to the dipole size by
\begin{equation}
\mu^{2}=\frac{C}{r^{2}}+\mu_{0}^{2}. \label{scale}
\end{equation}
Here the gluon density $g(x,\mu^{2})$ is evolved to the scale
$\mu^{2}$ with the leading order (LO) DGLAP equation (\ref{DGLAP}).
Moreover, the quark contribution to the gluon density is neglected
in the small-$x$ limit, and therefore
\begin{equation}
\frac{\partial xg(x,\mu^{2})}{\partial \ln\mu^{2}}=\frac{\alpha_{s}(\mu^{2})}
{2\pi^{2}}\int_{x}^{1} dz P_{gg}(z)\frac{x}{z}
g(\frac{x}{z},\mu^{2}).
\end{equation}
The initial gluon density is taken at the scale $Q_{0}^{2}=1
\hbox{GeV}^{2}$ in the form
\begin{equation} xg(x,\mu^{2})=A_{g}x^{-\lambda_{g}}(1-x)^{5.6},
\end{equation}
where the parameters $C=0.26$, $ \mu_{0}^{2}=0.52 \hbox{GeV}^{2}$,
$A_{g}=1.20$ and $\lambda_{g}=0.28$ are fixed from a fit to DIS
data for $x<0.01$ and in a range of $Q^2$ between $0.1$ and $500$
$\hbox{GeV}^2$ \cite{gbw-d}.  We use the LO formula for the running
coupling $\alpha_{s}$, with three flavors and for
$\Lambda_{\hbox{QCD}}=0.2~\hbox{GeV}$. The dipole size determines
the evolution scale $\mu^{2}$ through Eq.~(\ref{scale}). The
evolution of the gluon density is performed numerically for every
dipole size $r$. Therefore
the DGLAP equation is now coupled to the master equation (\ref{ma}). It is important to stress that the GBW-DGLAP
model preserves the successes of the GBW model at low $Q^{2}$ and
its saturation property for large dipole sizes, while incorporating
the evolution of the gluon density by modifying the small-$r$
behaviour of the dipole size, Fig.~\ref{fig-1}.

To highlight the failure of GBW parametrization, in Fig.~\ref{fig-1} we
show the effective slope $\lambda(Q^2)$ from the parametrization
$F_2\sim x^{-\lambda(Q^2)}$ as a function of $Q^2$. It is seen that
the GBW-DGLAP parametrization is essential in order to describe the data.

%%%%%%%%%%%%%%%%%%%%%%%%%%%%%%%%%%%%%%%%%%%%%%%%%%%%%%%%%%%%%%%%%%%%%%%%%%%%%%%%
\section{The Drell-Yan process and direct photons }
\subsection{The partonic description}
\begin{figure}[t]
  \centerline{\scalebox{0.5}{\includegraphics{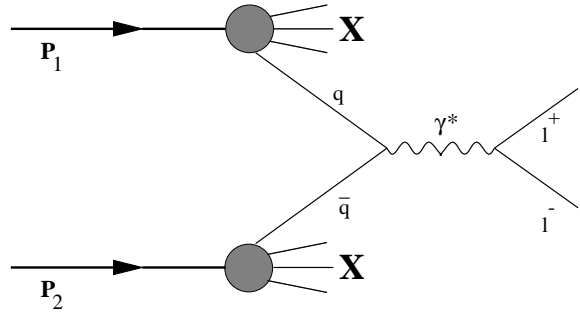}}}
    \caption{
      \label{dyfig}
      Partonic picture of the DY process in the leading order. Two hadrons collide and a quark from one
      hadron annihilates with an antiquark from the other hadron into a timelike
      photon, which decays into a lepton pair.
    }  
\end{figure}

 As we already mentioned, the PDFs are universal. Therefore one can
 use the DIS data to extract PDFs and then make prediction for other
 hard processes. The most prominent example of hadron~hadron
 collisions is the so-called Drell~Yan (DY) process \cite{dy-ci},
 where lepton pairs are produced:
\begin{equation}
h_1+h_2\to \mu^{+}+\mu^{-}+X,
\end{equation}
where $X$ can be any undetected particles. In the parton model, this
process looks like a quark and an anti-quark from two hadrons annihilating
into into a lepton pair, see Fig.~\ref{dyfig}

The kinematic of the DY process can be conveniently defined via
light-cone momentum fractions of the projectile (target) parton,
$x_1$ ($x_2$),
\begin{equation}
x_1=\frac{2P_2\cdot q}{s}\quad,\quad x_2=\frac{2P_1\cdot q}{s}.  \label{kin-c}
\end{equation}
where $P_1$ and $P_2$ are the four momenta of hadron $1$ and hadron
$2$, respectively and $s$ denotes the square of the center-of-mass
energy of the colliding hadrons is $s=(P_1+P_2)^2$. The Feynman
variable $x_F$ is related to other kinematics variables as,
\begin{equation}
x_F=\frac{2p_L^{cm}}{\sqrt{s}}\approx x_1-x_2,
\end{equation}
where $p_L^{cm}$ is the longitudinal momentum of the dilepton in the 
hadron-hadron center-of-mass frame. Another relation is
\begin{equation}\label{scaling}
\tau=x_1x_2=\frac{M^2}{s},
\end{equation}
Where $M^2=q^2>0$ denotes the mass of the spacelike photon and the transverse momentum 
of the virtual photon has been neglected.
The partonic annihilation cross-section for Fig.~\ref{dyfig} reads
\begin{equation}\label{dycross}
\frac{d\widehat\sigma}{dM^2}=\frac{4\pi\alpha^2_{em}Z_f^2}{3N_cM^2}
\delta(x_1x_2s-M^2).
\end{equation}
The hadronic cross-section can be then written as the convolution of
PDFs with the partonic cross-section, like in DIS,
\begin{eqnarray}
\frac{d\sigma}{dM^2}&=&\int_0^1dx_1dx_2\sum_f
\left\{q_f(x_1)\bar q_{f}(x_2)+(1\leftrightarrow 2)\right\}
\frac{d\widehat\sigma}{dM^2},\label{dylo}\
\end{eqnarray}
where $q_f(x_1)$ is the probability to find a quark of flavor $f$ with 
light-cone momentum
fraction $x_1$ in hadron $a$, and $\bar q_f$ is the analog for antiquarks.
In the second line, we have plugged the partonic cross-section Eq.~(\ref{dycross}) and
performed one of the integrals. It is interesting to note that the right-hand side of 
Eq.~(\ref{dylo}) depends only on $\tau$ and not separately on $M^2$ and $s$.
This scaling property is confirmed experimentally \cite{ex-ci}.

\begin{figure}[t]
  \scalebox{0.7}{\includegraphics*{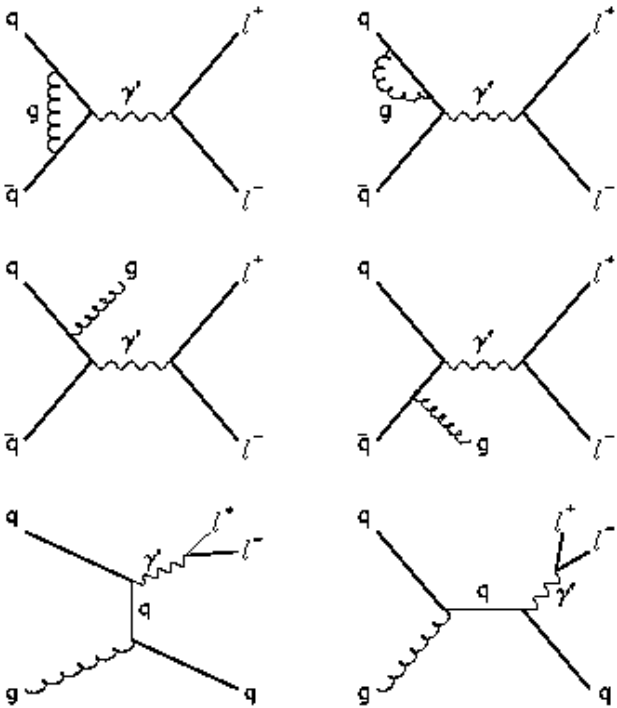}}\hfill
  \raise0.cm\hbox{\parbox[b]{3.0in}{ \caption{ \label{dyho} Higher
  order QCD corrections to the DY process. The diagrams for
  virtual corrections, the annihilation process, and the Compton
  process are depicted in the upper, middle, and last row,
  respectively.  These higher order corrections account for most of
  the K factor and explain data at large transverse momenta. The
  figure is taken from Ref.~\cite{figdy-ci}.  } } }
\end{figure}
     
  Some features of dilepton production cannot be understood in the
  lowest order picture. The cross-section given by Eq.~(\ref{dylo})
  is $2-3$ times smaller than the measured value. This
  discrepancy is usually treated by introducing an ad hoc
  normalization factor, the so-called $K$ factor. The $K$ factor is
  approximately independent of $M^2$. Another obvious problem is that the transverse momentum spectrum in the
naive parton model cannot describe data. Phenomenologically, one can introduce a primordial momentum distribution of the quarks, but what is observed in experiments about $1-2$ GeV is much larger than
what one would expect from Fermi motion.

These problems can be partially resolved by taking into account the
next-order QCD corrections, shown in Fig.~\ref{dyho}. Owing to the
radiation of the gluon (the second row diagrams), the quark acquires a
transverse momentum. In this way, the pQCD correction provides the
missing mechanism for the production of lepton pairs with large
transverse momentum $p_T$. However, the transverse momentum spectrum
is not described well in this order, and obviously somethings is still
missing. In particular, at low $p_T$ the pQCD result diverges. There have
been attempts to overcome this problem by a resummation of soft gluons
radiated from the quark and antiquark \cite{res-ci}. The last row in Fig.~\ref{dyho} displays the diagrams
for the QCD Compton process, where a quark in one hadron picks up a
gluon from the other hadron and radiates a photon. This mechanism is
dominant at large $p_T$ \cite{app-ci}.

\subsection{The colour dipole description} 
\begin{figure}[t]
  \centerline{\scalebox{0.3}{\includegraphics{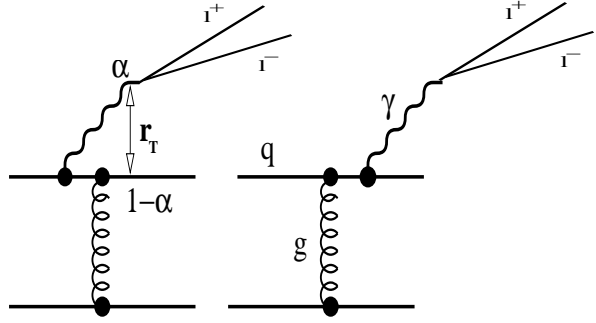}}}
    \caption{ \label{bo} In the target rest frame, the DY process looks like
  bremsstrahlung. A projectile quark (or antiquark) scatters off the target and radiates a massive photon which subsequently decays
  into the lepton pair. The photon can also be radiated before the quark hits the target. Both diagrams are important.
    }
\end{figure}

Similar to DIS, the DY process can be viewed in the target rest frame
where it looks like bremsstrahlung rather than parton annihilation,
see Fig.~\ref{bo}. A quark or an antiquark from a projectile hadron
radiates a virtual photon while hitting the target. This radiation can
occur before and after the quark scatters off the target. The impact
parameter representation of the cross-section for such a process can
be written in the factorized form similar to DIS \cite{k,bhq,p6}, 
\begin{equation}\label{dylctotal}
\frac{d\sigma(qp\to q\gamma^*p)}{d\ln\alpha}
=\int d^2 r_{T}\, |\Psi^{T,L}_{\gamma^* q}(\alpha,r_{T})|^2
    \sigma_{q\bar q}(x,\alpha r_{T}),
\end{equation}
where $\alpha$ is the light-cone momentum fraction of the quark,
carried away by the photon and $r_{T}$ the transverse separation
between $\gamma^\star$ and $q$. The dipole cross-section $\sigma_{q\bar
q}(x,\alpha r_{T})$ with transverse separation $\alpha r_{T}$ is
a universal quantity like PDFs and has already been introduced in the DIS section.

Where does the dipole cross-section come from if there is no dipole in
diagrams given in Fig.~\ref{bo}, and why is the transverse dipole size
$\alpha r_{T}$? The dipole cross-section appears because the quark is
displaced in the impact parameter plane after radiation of the
photon. The antiquark enters after taking the complex conjugate of the
amplitude. Therefore the dipole in equation (\ref{dylctotal}) is not a
real $q\bar{q}$-dipole. As in the real dipole in DIS where colour
screening is provided by interactions with either the quark or the
antiquark, in the case of radiation the two amplitudes for radiation
prior or after the interaction screen each other, leading to
cancellation of the infra-red divergences. Now, back to the second
question, if $r_{T}$ is the transverse separation between the quark
and the photon, and $\alpha$ is the fractional momentum of radiated
photon. Then, the transverse separation between the photon and center
of gravity is $(1-\alpha)r_{T}$ and the distance between the quark and
center of gravity will be $\alpha r_{T}$. Before radiation, the center
of gravity of the quark coincide with the incident quark, after
radiation the relative distance between quark and the center of
gravity is shifted to $\alpha r_{T}$. Taking the complex conjugate of
the amplitude, it looks as if the transverse size between $q$ and
$\bar{q}$ is $\alpha r_{T}$ which is the argument of the dipole
cross-section.

The wave function of the $\gamma^*q$ fluctuation 
in Eq.~(\ref{dylctotal}) for transversely and longitudinally
polarized photons reads,
\beq
\Psi^{T,L}_{\gamma^*q}(\alpha,\vec r_T)=
\frac{\sqrt{\alpha_{em}}}{2\,\pi}\,
\chi_f\,\widehat O^{T,L}\,\chi_i\,K_0(\eta r_T).
\label{dylcl}
\eeq
Here $\chi_{i,f}$ are the spinors of the initial and final quarks and 
$K_0(x)$ is the modified Bessel function.
The operators $\widehat O^{T,L}$ have the form,
\beq
\widehat O^{T} = i\,m_f\alpha^2\,
\vec {e^*}\cdot (\vec n\times\vec\sigma)\,
 + \alpha\,\vec {e^*}\cdot (\vec\sigma\times\vec\nabla)
-i(2-\alpha)\,\vec {e^*}\cdot \vec\nabla\ ,
\eeq
\beq
\widehat O^{L}= 2M(1-\alpha)\ ,
\eeq
where $\vec e$ is the polarization vector of the photon,
$\vec n$ is a unit vector along the projectile momentum,
and $\vec\nabla$ acts on $\vec r_T$. For radiation of prompt photons $M=0$. A parameter 
\begin{equation}\label{dyextension}
\eta^2 = m_f^2 \alpha^2 + M^2 \left(1-\alpha\right), 
\end{equation}
is the analog of the parameter $\epsilon$ Eq.~(\ref{disextension}) in DIS. 

In order to obtain the hadronic cross-section from the elementary
partonic one, Eq.~(\ref{dylctotal}), one should sum up the
contributions from quarks and antiquarks weighted with the
corresponding parton distribution functions (PDFs) in the projectile hadron. The hadronic
cross-section then reads \cite{k,p6}
\begin{eqnarray}\nonumber
\fl \frac{d\sigma}{dM^2dx_F}=\frac{\alpha_{em}}{3\pi M^2}
\frac{x_1}{x_1+x_2}\int_{x_1}^1\frac{d\alpha}{\alpha^2}
\sum_fZ_f^2\left\{q_f\left(\frac{x_1}{\alpha},Q^2\right)+
q_{\bar f}\left(\frac{x_1}{\alpha},Q^2\right)\right\}
\frac{d\sigma(qp\to q\gamma^*p)}{d\ln\alpha}\\
\lo =\frac{\alpha_{em}}{3\pi M^2}
\frac{1}{x_1+x_2}\int_{x_1}^1\frac{d\alpha}{\alpha}
F_2^p\left(\frac{x_1}{\alpha},Q^2\right)
\frac{d\sigma(qp\to q\gamma^*p)}{d\ln\alpha}.\label{ma22}
\end{eqnarray}
The PDFs of the projectile enter in a combination which can be
written in terms of the proton structure function $F_{2}^{p}$. Notice
that with our definitions the fractional quark charge $Z_{f}$ is not
included in the LC wave function of Eq.~(\ref{dylcl}), and that the
factor $\frac{\alpha_{em}}{3\pi M^{2}}$ in Eq.~(\ref{ma22}) accounts
for the decay of the photon into the lepton pair. We use the
standard notation for the kinematical variables $x_1$ and $x_2$ defined in Eq.~(\ref{kin-c}).

The transverse momentum $p_{T}$ distribution of photon
bremsstrahlung in quark-nucleon interactions, integrated over the
final quark transverse momentum, was derived in Ref.~\cite{p6}: 
 \begin{eqnarray}
\fl \frac{d \sigma^{qN}(q\to q\gamma)}{d(ln \alpha)d^{2}\vec{p}_{T}}=\frac{1}{(2\pi)^{2}}
\sum_{in,f}\sum_{L,T}
\int d^{2}\vec{r}_{1}d^{2}\vec{r}_{2}e^{i \vec{p}_{T}.(\vec{r}_{1}-\vec{r}_{2})}\Psi^{T,L}_{\gamma^*q}(\alpha, \vec{r}_{1})
\Psi^{T,L}_{\gamma^*q}(\alpha, \vec{r}_{2})
\Sigma_{\gamma}(x,\vec{r}_{1},\vec{r}_{2},\alpha),  \label{m1}\nonumber\\
 \end{eqnarray}
 where
 \begin{equation}
\Sigma_{\gamma}(x,\vec{r}_{1},\vec{r}_{2},\alpha)=\frac{1}{2}\{
\sigma_{q\bar{q}}(x,\alpha r_{1})+\sigma_{q\bar{q}}(x,\alpha r_{2})-\sigma_{q\bar{q}}(x,\alpha(\vec{r}_{1}-\vec{r}_{2}))\}.
\label{di}
\end{equation}
%The appearance of the dipole cross section $\sigma_{q\bar{q}}(x,
%\alpha r)$ in the above equation is due to the quark displacement in
%the impact parameter plane after radiation of the photon.
 and $\vec{r}_{1}$ and $\vec{r}_{2}$ are the quark-photon transverse
 separations in the two radiation amplitudes contributing to the cross-section, Eq.~(\ref{m1}), which correspondingly contains
 double-Fourier transformations. The parameter $\alpha$ is the
 relative fraction of the quark momentum carried by the photon, and is
 the same in both amplitudes, since the interaction does not change
 the sharing of longitudinal momentum. The transverse displacement
 between the initial and final quarks is $\alpha r_{1}$ and $\alpha
 r_{2}$, respectively.  After integrating the above equation (\ref{di}) over $p_{T}$, one recovers Eq.~(\ref{dylctotal}), as one should.

The hadronic cross-section can then be obtained in the same fashion as
given in Eq.~(\ref{ma22}) by convolution with the proton structure
function. Next we calculate the inclusive direct photon spectra within
the same framework. For direct photon we have $M=0$; the transverse
momentum distribution of direct photons production from hadron-hadron
collision reads
\begin{equation}
\frac{d \sigma^{\gamma}(pp\to \gamma X)}{dx_{F}d^{2}\vec{p}_{T}}=
\frac{1}{x_{1}+x_{2}}\int_{x_{1}}^{1}\frac{d\alpha}{\alpha} F_{2}^{p}(\frac{x_{1}}{\alpha},Q)\frac{d \sigma^{qN}(q\to q\gamma)}{d(ln \alpha)d^{2}\vec{p}_{T}}. \
\label{con}
\end{equation}

%\begin{figure}[!t]
%       \centerline{\includegraphics[width=13 cm] {pl3.eps}}
%       \caption{Inclusive direct photon spectra obtained from the GBW
%       and the GBW-DGLAP dipole models for midrapidity $\eta=0$ at
%       RHIC energy $\sqrt{s}=200$ GeV and CDF energy $\sqrt{s}=1.8$
%       TeV. Experimental data are from Ref.~\cite{p3,p4}.  The figure is taken from \cite{direct1} \label{me2}}
%\end{figure}
We also need to identify the scale $Q$ entering in the proton
structure function in Eqs.~(\ref{ma22},\ref{con}), and relate the variable
$x$ of the dipole cross-section entered in Eqs.~(\ref{dylctotal},\ref{di}) to
measurable variables. From our previous definition, and following
previous works \cite{m2,mq}, we have that $x=x_{2}$. At zero
transverse momentum, the dominant term in the LC wavefunction
Eq.~(\ref{dylcl}) is the one that contains the modified Bessel
function $K_{1}(\eta r)$. This function decays exponentially at
large values of the argument, so that the mean distances which
numerically contribute are of order $1/\eta$. On the other hand,
the minimal value of $\alpha$ is $x_{1}$, and therefore the
virtuality $Q^2$ which enters into the problem at zero transverse
momentum is $\sim (1-x_{1})M^{2}$. Thus the hard scale at which the
projectile parton distribution is probed turns out to be
$Q^{2}=p^{2}_{T}+ (1-x_{1})M^{2}$. Notice that in the previous
studies, $M^{2}$ \cite{mq} and $(1-x_{1})M^{2}$ \cite{m2} were used
for the scale $Q^{2}$. Nevertheless, these different choices for
$Q^{2}$ account for less than a $20\%$ effect at small $x_{2}$
values.  
\begin{figure}[!t]
       \centerline{\includegraphics[width=14 cm] {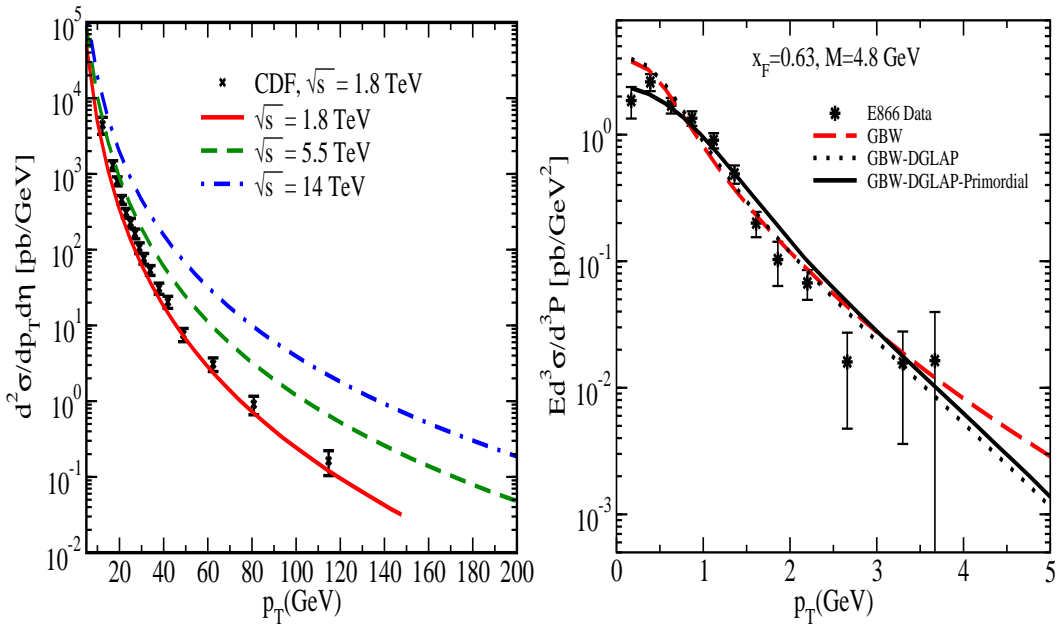}}
       \caption{Left: Inclusive direct photon spectra obtained
       from the GBW-DGLAP dipole models at CDF and CERN
       energies. Right: The dilepton spectrum with $800$-GeV beam energy in $pp$
       collisions from the E866 \cite{p2} fixed target experiment. We show the result of the
       GBW dipole model (dashed line) and the GBW-DGLAP model (dotted
       line). We also show the result when a constant primordial
       momentum $\langle k_{0}^{2}\rangle=0.4 \hbox{GeV}^2$ is
       incorporated within the GBW-DGLAP dipole model (solid line).
       Experimental data are from Refs.~\cite{p2,p4}.  The figures are
       taken from \cite{direct1,alhc}. \label{me1}}
\end{figure}

As example, in Fig.~\ref{me1} we show the dilepton and inclusive
direct photon spectra for different experiments. For the dipole
cross-section, we use two parametrizations introduced in Section
10. It is remarkable that both direct photon production and DY
dilepton pair production processes can be described within the same
colour dipole approach without any free parameters. From this study,
it is seen that the colour dipole formulation coupled to the DGLAP
evolution provides a better description of data at large transverse
momentum compared to the GBW dipole model.

The colour dipole predictions for the direct photons at the LHC is given
in Ref.~\cite{alhc}. In the same framework the azimuthal asymmetry of
the prompt photons was computed in Ref.~\cite{ph-i}, for the predictions of other approaches
at the LHC see Ref. \cite{aalhc}. 

%%%%%%%%%%%%%%%%%%%%%%%%%%%%%%%%%%%%%%%%%%%%%%%%%%%%%%%%%%%%%%%%%%%%%%%%%%%%%%%%%%%%%%%%%%%%%%%%%%%%%%%%%%%
\section{Diffraction}
Diffraction is associated with the optical analogy, which is elastic scattering of light caused by absorption.
A new feature of diffraction in quantum mechanics is the possibility of inelastic diffraction, which is
nearly elastic scattering with the excitation of one or both colliding hadrons to effective masses which
are much smaller that the c.m. energy of the collision. The main bulk of diffractive events originate
from soft interactions. Therefore it is still a challenge to describe these processes starting from the first
principles of QCD. Unavoidably, one faces the problem of confinement which is still a challenge for the
theory. Nevertheless, the ideas of QCD help to develop quite an effective phenomenology for diffractive
reactions, i.e., to establish relations between different observables.

\subsection{Diffraction in nonabelian theories}

Elastic and inelastic diffraction are large rapidity gap (LRG) processes.
Since they emerge as a shadow of inelastic interactions, their amplitudes
are nearly imaginary. This observation is a direct evidence for 
the underlying theory to be {\it nonabelian}.

\begin{figure}[t]
\centering\includegraphics[width=.4\linewidth]{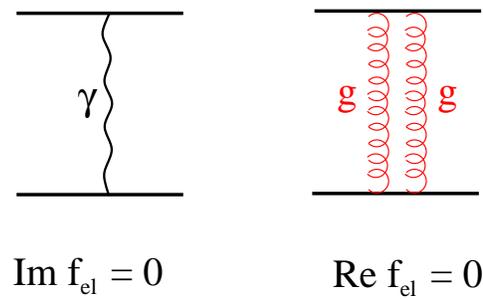}
 \caption{Born 
approximation for elastic scattering in abelian (left) and
nonabelian (right) theories.}
\label{abel}
 \end{figure}

 Indeed, the elastic amplitude can be mediated only by a neutral exchange
in $t$ channel, therefore the Born graphs in the abelian and nonabelian
cases look like as shown in Fig.~\ref{abel}.

The striking difference between these two amplitudes is in their phases.
In the abelian case (e.g. in QED) the Born amplitude is real, while in the
nonabelian theory (QCD) the amplitude is imaginary.

Data for elastic hadron scattering show that the real part of the   
elastic amplitude is small, and this is a direct evidence for the
nonabelian underlying dynamics. This is a remarkable
observation, since we have so far very few manifestations of
nonabelian features in the data.

\begin{figure}[htbp]
\centering\includegraphics[width=.6\linewidth]{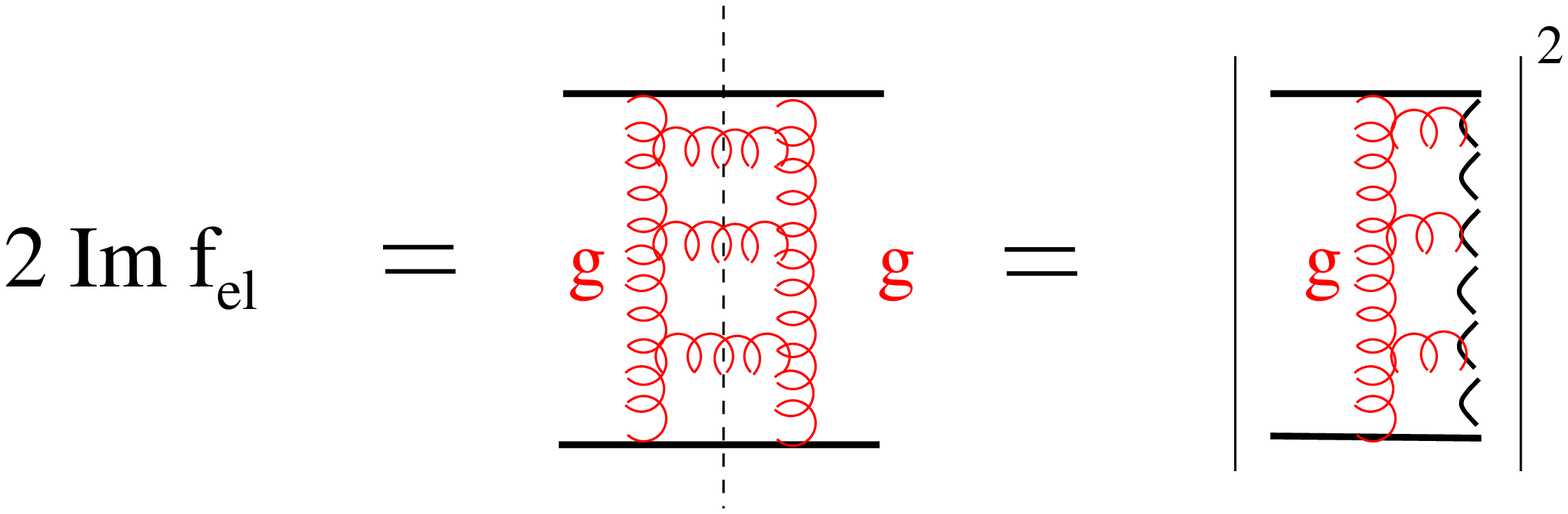}
 \caption{The unitarity relation for the Pomeron amplitude in terms of   
perturbative QCD}
\label{unitarity}
 \end{figure}

The Born amplitude depicted in Fig.~\ref{abel} is independent of energy.
Gluon radiation gives rise to the energy dependence of the total cross-section through the unitarity relation illustrated in Fig.~\ref{unitarity}.

Elastic scattering reaches maximal strength at the unitarity limit of
black disc, ${\rm Im}\, f_{el}(b)=1$, 
 \beq
\sigma_{el}=\sigma_{in}=\pi\,R^2,
\label{100-c}  
 \eeq
 where $R$ is the radius of interaction. The unitarity relation tells us that the
imaginary part of the partial amplitude ${\rm Im}\, f_{el}(b)$ cannot rise for ever.
After the unitarity bound is reached, the total cross-section can rise only due to
an energy dependence of the interaction radius $R(s)$.  Froissart theorem imposes a
restriction on this, the interaction radius cannot rise with energy faster than
$R\propto \ln(s)$. Then, the total and elastic cross-section rise with energy as
$\propto \ln^2(s)$ in the Froissart regime of unitarity saturation.

\subsection{Quantum mechanics of diffraction}
 
Diffractive excitation is a nontrivial consequence of the presence of quantum fluctuations in hadrons. In
classical mechanics only elastic scattering is possible. An example is diffractive scattering of electromagnetic waves.

One can understand the origin of diffractive excitation in terms of
elastic diffraction \cite{fp,gw}. Since a hadron has a composite structure, different hadronic constituents interact differently causing a
modification of the projectile coherent superposition of states. Such a modified wave packet is no longer
orthogonal to other hadrons different from the incoming one. This makes possible the production of new
hadrons, i.e., diffractive excitations.

To simplify the picture, one can switch to the basis of eigenstates of
interaction. Since a hadron can be excited, it cannot be an eigenstate
of interaction, and can be expanded over the complete set of eigen
states $|\alpha\ra$ \cite{kl,mp,kst2}:
 \beq
|h\ra = \sum\limits_{\alpha=1}C^h_{\alpha}\,|\alpha\ra\ ,
\label{700-c}
 \eeq
 which satisfy the condition, $\hat f_{el}|\alpha\ra =
f_\alpha\,|\alpha\ra$, where $\hat f_{el}$ is the elastic amplitude
operator.

Owing to completeness and orthogonality of each set of states, the
coefficient $C^h_{\alpha}$ in Eq.~(\ref{700-c}) satisfy the relations, 
 \beqn
\la h'|h\ra  &=&
\sum\limits_{\alpha=1}(C^{h'}_{\alpha})^*C^h_{\alpha} =
\delta_{hh'},
\nonumber\\
\la \beta|\alpha\ra  &=&
\sum\limits_{h'}(C^{h'}_{\beta})^*C^{h'}_{\alpha} =
\delta_{\alpha\beta}.
\label{800-c}
 \eeqn
The elastic and single diffraction amplitudes can be thus expressed via
the eigen amplitudes as,
 \beqn
f_{el}^{h\to h} &=& \sum\limits_{\alpha=1}|C^h_{\alpha}|^2\,f_\alpha,  
\nonumber\\
f_{sd}^{h\to h'} &=&
\sum\limits_{\alpha=1}(C^{h'}_{\alpha})^*C^h_{\alpha}\,f_\alpha.
\label{900}
 \eeqn
 Using these expressions and the completeness relations, Eqs.~(\ref{800-c}), 
one can calculate the forward single diffraction cross-section without
knowledge of the properties of $|h'\ra$,
 \beqn
\left.
\sum\limits_{h'\neq h}\frac{d\sigma^{h\to h'}_{sd}}
{dt}\right|_{t=0} &=&
\frac{1}{4\pi}\left[\sum\limits_{h'}|f_{sd}^{hh'}|^2
-|f_{el}^{hh}|^2\right],\nonumber\\
&=&
\frac{1}{4\pi}\left[\sum\limits_{\alpha}|C^h_{\alpha}|^2\,
|f_\alpha|^2 -\left(\sum\limits_{\alpha}
|C^h_{\alpha}|f_\alpha\right)^2\right],\nonumber\\
&=& \frac{\la f_\alpha^2\ra - \la f_\alpha\ra^2}
{4\pi}.
\label{1011}
 \eeqn
Thus the forward diffractive cross-section is given by the dispersion
of the eigenvalues distribution. For some specific distributions
the dispersion may be zero. For instance if all the eigenamplitudes are
equal, or one of them is much larger than others.

According to Eqs.~(\ref{900})-(\ref{1011}) one can calculate the total and
diffractive cross-sections on the same footing, provided that the
eigenstates $|\alpha\ra$, their weights $|C^h_{\alpha}|^2$ and the 
eigenvalues $f_\alpha$ are known. Notice that the eigenamplitudes
$f_\alpha$ are the same for different hadronic species $|h\ra$. This
remarkable property of eigen amplitudes is employed later on.   

 In the Froissart regime all the partial eigen amplitudes reach the
unitarity limit, ${\rm Im}\,f_\alpha=1$. Then, according
to the completeness conditions,
 \beqn
f_{el}^{hh}
&\Rightarrow&
\sum\limits_{\alpha=1}|C^h_{\alpha}|^2=1,
\nonumber\\
f_{sd}^{hh'} &\Rightarrow&
\sum\limits_{\alpha=1}(C^{h'}_{\alpha})^*C^h_{\alpha}
=0.
\label{1100-c}
 \eeqn
Diffraction is impossible within a black disc, but only on its periphery,
$b\sim R$. Since in the Froissart regime $R\propto \ln(s)$,
 \beqn
\sigma_{tot}&\propto& \sigma_{el}
\propto \ln^2(s),
\nonumber\\ 
\sigma_{sd}&\propto& \ln(s)\ ,
\label{1200}
 \eeqn
  i.e. $\sigma_{sd}/\sigma_{tot}\propto 1/\ln(s)$.
The total and single diffractive cross-sections in terms of the colour-dipole cross-section read,
 \beqn
\fl \sigma_{tot}^{hp} =
\sum\limits_{\alpha=1}|C^h_{\alpha}|^2\,\sigma_\alpha=\int d^2r_T\left|\Psi_h(r_T)\right|^2\sigma(r_T) =\la\sigma(r_T)\ra\ ,
\label{1300}
 \eeqn

\beqn
\fl \left.
\sum\limits_{h'}\frac{d\sigma^{h\to h'}_{sd}}
{dt}\right|_{t=0} = \sum\limits_{\alpha=1}|C^h_{\alpha}|^2\, \frac{\sigma_\alpha^2}{16\pi} = \int d^2r_T\left|\Psi_h(r_T)
\right|^2\frac{\sigma^2(r_T)}{16\pi} =
\frac{\la\sigma^2(r_T)\ra}{16\pi},
\label{1400}
 \eeqn
where the eigenvalue of the cross-section for a simplest $\bar qq$ dipole
$\sigma_{\bar qq}(r_T)$ was already introduced in Section 10.

\subsection{Diffractive DIS}

The contribution of diffractive quasielastic production of vector mesons
is a tiny fraction, vanishing as $1/Q^2$, of the total
inclusive DIS cross-section. However, the fraction of all diffractive
events associated with large rapidity gaps in DIS is large, about $10\%$,
and is nearly independent of $Q^2$. This turns out to be the result of a
contribution of rare soft fluctuations in the hard photon. According to 
Eq.~(\ref{1800-cc}) a longitudinally asymmetric $\bar qq$ pair with $\alpha$ or
$1-\alpha\sim 1/Q^2$ has a large hadronic size and experience soft
diffractive interactions like hadrons. Although the admixture of such soft
fluctuations in the virtual photon is tiny, that may be compensated by a
large interaction cross-section. This interplay between the fluctuation
probability and the cross-section is illustrated for inclusive and
diffractive DIS in Table.~\ref{dis-diff} \cite{kp-soft}.
 \begin{table}[htbp]
\caption{Interplay between the probabilities of hard and soft fluctuations
in a highly virtual photon and the cross-section of interaction of these
fluctuations.}\vspace*{0.3cm}
\centering\begin{tabular}{|c|c|c|c|c|}
 \hline
 \vphantom{\bigg\vert}
    &
$|C_\alpha|^2$
  & $\sigma_\alpha$
  &
$\sigma_{tot}\!=\!\!\!\!\!\sum\limits_{\alpha=soft}^{hard}|C_\alpha|^2
\sigma_\alpha$
  &
$\sigma_{sd}\!=\!\!\!\!\!\sum\limits_{\alpha=soft}^{hard}|C_\alpha|^2
\sigma^2_\alpha$
   \\
[3mm]
\hline &&&&\\
%[-2mm]
Hard& $\sim 1$ & $\sim\frac{1}{Q^2}$ &
$\sim \frac{1}{Q^2}$
& $\sim \frac{1}{Q^4}$  \\
[3mm]
\hline &&&&\\
%[-6mm]
Soft & $\sim \frac{m_q^2}{Q^2}$ &
$\sim\frac{1}{m_q^2}$ &
$\sim\frac{1}{Q^2}$ &
$\sim\frac{1}{m_q^2Q^2}$
  \\
[3mm]  
\hline
\end{tabular}
\label{dis-diff}
 \end{table}

 Hard fluctuations of the photon have large weight, but vanishing as
$1/Q^2$  in the cross-section, while soft fluctuations have a small, $m_q^2/Q^2$,
weight, but interact strongly, $\sigma\sim 1/m_q^2$. The latter factor
compensates the smallness of the probability in the case of DIS, and
over-compensates it for diffraction.

 Thus we conclude that inclusive DIS is semi-hard and semi-soft, and the
soft component is present at any high $Q^2$. On the other hand,
diffractive DIS (called sometimes "{\it hard diffraction}") is predominantly a
soft process. This is why its fraction in the total DIS cross-section is  
nearly $Q^2$ independent. One can test this picture studying the $Q^2$  
dependence of the diffractive DIS \cite{beatriz1}.

Since diffraction is a source of nuclear shadowing \cite{gribov},
that also should scale in $x$. Indeed, most of experiment have not
found any variation with $Q^2$ of shadowing in DIS on nuclei. Only the NMC
experiment \cite{ref-nmc1,ref-nmc2} managed to find a weak scaling violation which agrees with   
theoretical expectations \cite{krt}.

Notice that in spite of independence of $Q^2$, both diffraction and
shadowing are higher twist effects. This is easy to check considering
photoproduction of heavy flavors. In this case the hard scale is imposed
by the heavy quark mass, and diffraction becomes a hard process with cross-section vanishing as $1/m_Q^4$. Nuclear shadowing also vanishes as
$1/m_Q^2$.

The true leading twist diffraction and shadowing are associated with gluon
radiation considered below.

\subsection{Diffractive Drell-Yan reaction}
     
The dipole description of the Drell-Yan reaction in many respects
is similar to DIS, see Sections 10 and 11.2. This is not a surprise, since the two processes are
related by QCD factorization. 

There is an important difference between DIS and DY reaction. In the
inclusive DIS cross-section one integrates over $0<\alpha<1$, this is why 
this cross-section is always a mixture of soft and hard contributions (see
Table~1). In the case of DY reaction there is a new variable, $x_1$, which
is the fraction of the proton momentum carried by the dilepton. Since $\alpha
> x_1$, one can enhance the soft part of the DY cross-section selecting
events with $x_1\to 1$. This soft part of the DY process is subject to
unitarity corrections \cite{beatriz3} which are more important than in  
DIS \cite{beatriz4}.

Another distinction between DIS and DY is suppression of the DY
diffractive cross-section. Namely, the forward cross-section of
diffractive radiation $qp\to \bar llqp$ is zero \cite{kst1}. Indeed,
according to Eq.~(\ref{1011}) the forward diffractive cross-section is given
by the dispersion of the eigenamplitude distribution. However, in both
eigen states $|q\ra$ and $|q\gamma^*\ra$ only quark interacts. So the two
eigenamplitudes are equal, and the dispersion is zero.

Nevertheless, in the case of hadronic collisions the diffractive DY 
cross-section does not vanish in the forward direction. In this case the two
eigen states are $|\bar qq\ra$ and $|\bar qq\gamma^*\ra$ (for the sake of 
simplicity we take a pion). The interacting component of these Fock states
is the $\bar qq$ dipole, however, it gets a different size after the $q$ or
$\bar q$ radiate the photon. Then the two Fock states interact
differently, and this leads to a nonvanishing forward diffraction. Notice 
that the diffractive cross-section is proportional to the dipole size
\cite{ks-t}.

\subsection{Diffractive Higgs production}

\begin{figure}[t]
\includegraphics[width=.45\linewidth]{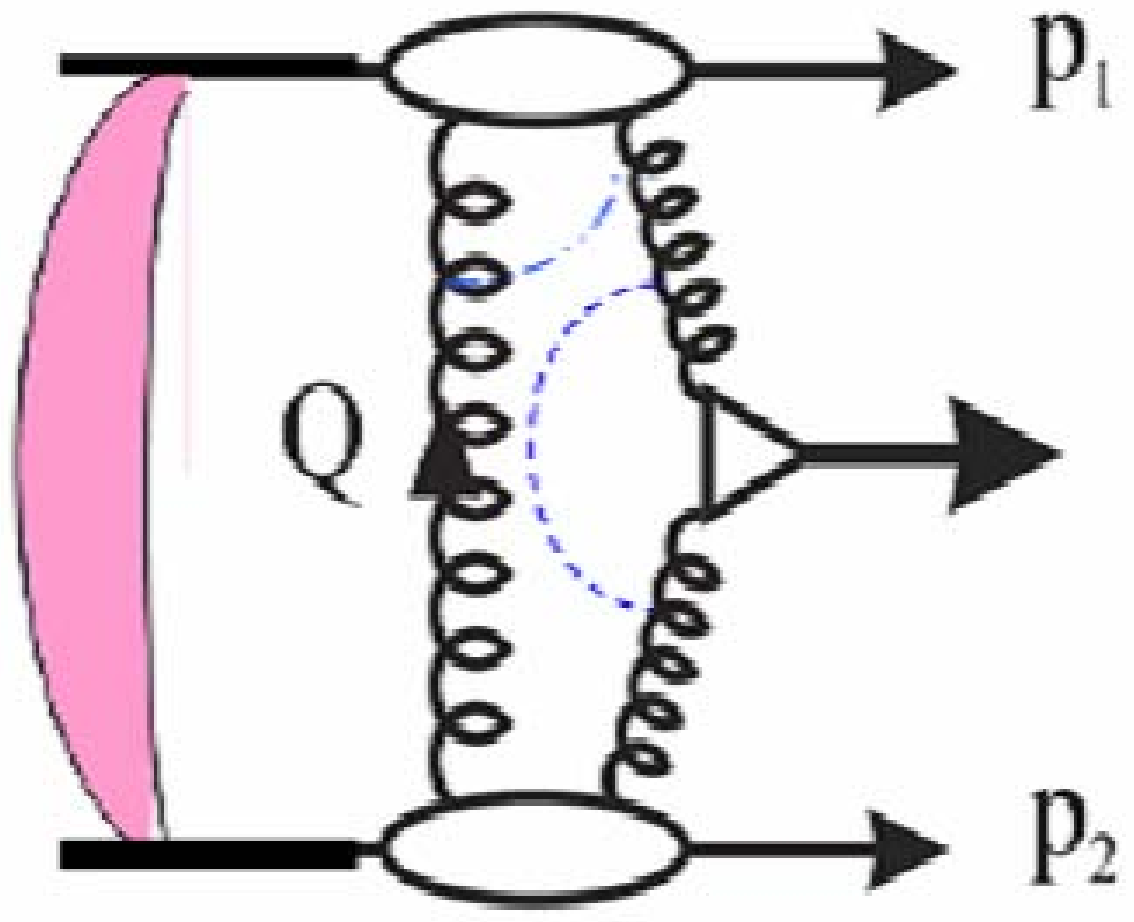}
\hspace{0.5cm}
\includegraphics[width=.45\linewidth]{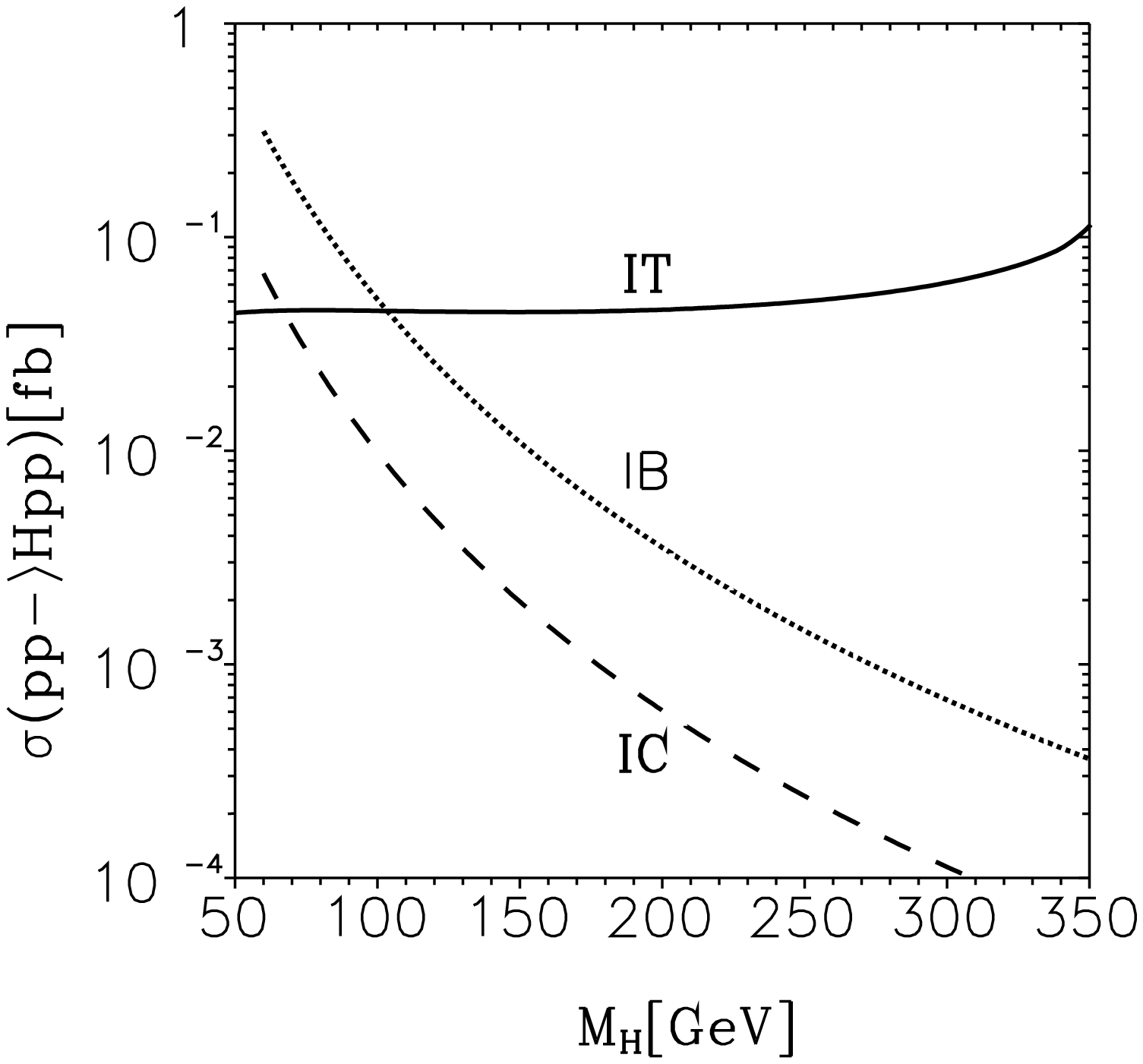}
 \caption{Left: Double diffractive Higgs production $pp\to Hpp$. Right: Cross-section of exclusive diffractive Higgs production, $pp\to
Hpp$, from intrinsic charm (IC), bottom (IB) and top (IT) \cite{bkss}.}
\label{higgs}
 \end{figure}

Detection of Higgs particle is the main challenge of the forthcoming experiments 
of the LHC at CERN. The most difficult problem here is to single out a weak signal from high 
backgrounds. One possible process to study is a double diffractive production of Higgs,
$p+p\to p+H+p$, with two large rapidity gaps, as illustrated in 
Fig.~\ref{higgs}(left). 

Like other diffractive processes, this reaction is strongly 
suppressed by the small survival probability of the gaps. Namely, initial- and 
final-state inelastic interactions of the colliding protons, can easily cause 
multiparticle production which will fill the gaps. The probability of 
no-interaction is usually called absorptive corrections, which are illustrated 
in Fig.~\ref{higgs}(left) by shaded strip. Recent calculations of the cross-section of 
this reaction \cite{kmr-c} led to a rather small cross-section, which, nevertheless, may be observed due to smallness of the 
background which is also suppressed by the absorptive corrections.

Another possible mechanism for Higgs production could be a direct
diffractive higgsstrahlung similar to diffractive DY. In 
both cases the radiated particle does not take part in the interaction
\cite{ks-t}. However, the Higgs coupling to a quark is proportional to the
quark mass, therefore the cross-section of higgsstrahlung by light
hadrons is vanishingly small.

A larger cross-section may emerge due to the admixture of heavy flavors in
light hadrons. A corresponding mechanism of exclusive Higgs production, $pp\to H
pp$, due to direct coalescence of heavy quarks, $\bar QQ\to H$ was
proposed in Ref.~\cite{bkss}. In this case the Higgs is produced not at the mid 
rapidities, but in the fragmentation region of the proton, at large Feynman 
$x_F$ where backgrounds are very small. The cross-section of Higgs production 
was evaluated assuming $1\%$ of intrinsic charm (IC) \cite{stan-c} and that
heavier flavors scale as $1/m_Q^2$ \cite{maxim}. The results are shown in
Fig.~\ref{higgs}(right) as a function of the Higgs mass for different intrinsic heavy 
flavors. The cross-section is small, but can be detected by dedicated measurements.

%%%%%%%%%%%%%%%%%%%%%%%%%%%%%%%%%%%%%%%%%%%%%%%%%%%%%%%%%%%%%%%%%%%%%%%%%%%%%%%%%%%%%%%%%%%%%%%%%%%%%%%%%%
\section{Quark and Gluon shadowing}
One may naively expect that the cross-section for scattering a lepton
off a nucleus with mass number $A$ must be $A$ times bigger than the
cross-section for the lepton-proton collision.  However, several
experiments show that the nuclear DIS cross-section at small $x<<1$ is smaller, 
\begin{equation}
\sigma^{\gamma^\star A}_{tot}<A\sigma^{\gamma^\star N}_{tot}.
\end{equation}
This phenomenon is called {\it shadowing}. Shadowing has been investigated by various experiments
in different kinematics ranges. For a review of the experimental and theoretical results, see Ref.~\cite{sh1}.

A particle thrown on a nuclear target has many possibilities of
interaction with different bound nucleons. However, the total
probability of interactions should not exceed 1. Therefore, a
probability of each interaction must be reduced which can be viewed as
a result of shadows produced by the preceding collisions. 

\begin{figure}[t]
       \centerline{\includegraphics[width=5.0 cm] {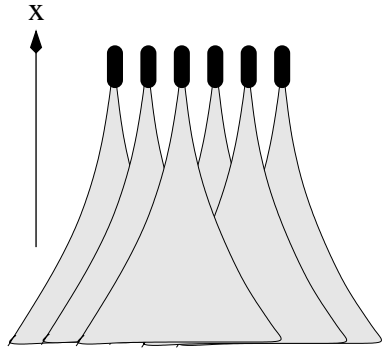}}
       \caption{Even when nucleons are well separated in the
       longitudinal direction in the infinite momentum frame, gluons
       fluctuation at small $x$ overlap. \label{c11}}
\end{figure}

Both the Colour Glass Condensate \cite{cgc} and shadowing
have the same origin: longitudinal overlap of gluon clouds
originating from different bound nucleons. This is illustrated in
Fig.~\ref{c11}. Bound nucleons in the nucleus do not overlap much,
either in the rest frame, or in the infinite momentum frame,
since both the nucleon size and internucleon spacing are subject
to Lorentz contraction. However, gluons carrying a small
fraction x of the proton momentum have a smaller gammafactor
and are less compressed in the longitudinal direction.
Then, the longitudinal propagation of small-$x$ partons is
large. They overlap and do talk to each other, i.e. they fuse and
reduce parton density at small $x$. The cross-section decreases and
this is shadowing. Fig.~\ref{c11} shows how gluonic clouds overlap at small x.

\begin{figure}[t]
       \centerline{\includegraphics[width=9.0 cm] {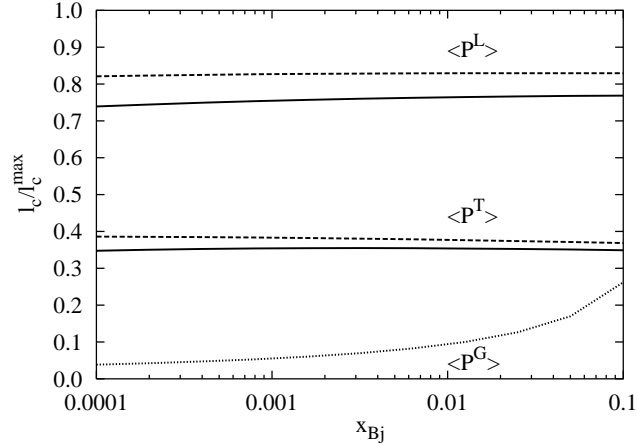}}
    \caption{ \label{l-x}
       Bjorken $x$ dependence of  $P$  
        defined in (\ref{1.1}), corresponding to the coherence length 
        for shadowing of transverse and longitudinal
        photons and gluon shadowing, respectively \cite{mqq}. 
           Solid and dashed curves
        correspond to $Q^2=4$ and $40$ GeV$^2$. The bottom curve represent $P$ for gluons.\label{c22} }  
\end{figure}

At small $x$, nuclear scattering is governed by coherence effects
which are better understood in the target rest frame. As described above, a virtual photon with virtuality $Q^2$ and energy $\nu$ splits into a $q\bar{q}$ pair with a
coherence length
\begin{equation}
l_c=\frac{2\nu}{Q^2+M^2_{q\bar{q}}}=\frac{P}{xM_N}, \label{1.1} 
\end{equation}
where $M^2_{q\bar{q}}$ is the effective mass of the fluctuation, and
the factor $P^{-1}=(1+M^2_{q\bar{q}}/Q^2)$.  The usual prescription is
that $M^2_{q\bar{q}}\sim Q^2$ since $Q^2$ is the only scale available,
which leads to $P=1/2$. Then, the coherence length can be bigger than
nuclear radius at low $x$. This means that $q\bar{q}$-pair can
experience multiple scatterings off different nucleons within
coherence length. In the infinite momentum frame this corresponds to
the overlap of parton clouds of different nucleons which leads to
diffusion of gluons and consequently a reduction of the gluonic
density in nuclei. A more careful analysis, however, shows that $P$,
even for quarks, depends on the polarization of the photon
\cite{mqq}. The factor $P$ for gluons is about one order of
magnitude smaller, see Fig.~\ref{c22}. Therefore, gluons need much
smaller $x$ in order to overlap in the longitudinal direction. This
simple observation leads to a remarkable prediction that the onset of
gluon shadowing occurs at smaller $x$ compared to quark shadowing.

The quark and gluon shadowing can be estimated within a simple model
at high energy. At high energies, dipoles $q\bar{q}$ are frozen by
Lorentz time dilation during propagation through the
nucleus. Therefore at very small $x$, it is possible to write entire
multiple rescattering which occurs during propagator of the $q\bar{q}$ with
fixed transverse size $r$ in a eikonal form \cite{tra-c}, 
\begin{equation}
\frac{q_{A}(x)}{Aq_{N}(x)}=\frac{2}{\langle  \sigma_{q\bar{q}}(r)\rangle}\int d^2b\left(1-\langle e^{-\frac{1}{2}\sigma_{q\bar{q}}(r)T_A(b)}\rangle \right), 
\end{equation}
where the nuclear thickness function is defined as an integral of the
nuclear density along the projectile trajectory,
$T_A(b)=\int_{-\infty}^{\infty}dz \rho(z)$.  
%A simple estimation of the above expression for lead gives $\frac{q_{A}}{Aq_{N}}\sim 0.35$. 
Similar calculations can be carried out for gluons \cite{mul-c},
\begin{equation}
\frac{g_{A}(x)}{Ag_{N}(x)}=\frac{2}{\langle  \sigma_{gg}(r)\rangle}\int d^2b\left(1-\langle e^{-\frac{1}{2}\sigma_{gg}(r)T_A(b)}\rangle\right).
\end{equation} 
where the gluon-gluon dipole is related to the quark-antiquark dipole
cross-section by the Casimir factor
$\sigma_{gg}(r)=\frac{9}{4}\sigma_{q\bar{q}}(r)$. Assuming the
gluon-gluon fluctuation of the projectile have the same distribution
function as for $q\bar{q}$, one may conclude that the effective
absorption cross-section providing shadowing is $9/4$ times larger
than for a $q\bar{q}$ fluctuation of a photon. Such a simple result
cannot be true because of the strong gluon-gluon interaction which
makes their distribution function quite different . Moreover, the spin
structure of the gluon-gluon distribution function is also
different. It turned out that in fact gluon shadowing is weaker
\cite{kst2}. That is because gluons in the proton are located within
small spots \cite{spots}, so they have a little chance to overlap in
the transverse plane, even in heavy nuclei.  If the mean value of
quark-gluon separation $r_{0}$, the mean number of other dipoles
overlapping with this one is
\begin{figure}[!t]
       \centerline{\includegraphics[width=7.0 cm] {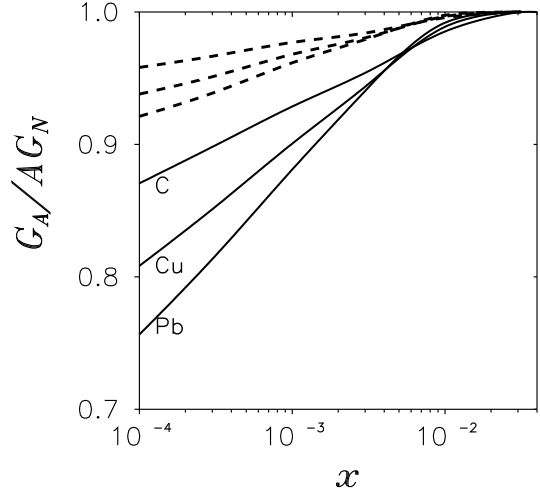}}
       \caption{ Gluon shadowing, for carbon, cooper and lead at
       $Q^2=4~\hbox{GeV}^2$ (solid) and $Q^2=40~\hbox{GeV}^2$
       (dashed) \cite{kst2}.  \label{glu}}
\end{figure}
\begin{equation}
\langle n_{g}\rangle=\frac{3}{4}\pi r^2_{0}\langle T_A\rangle \sim 0.3.
\end{equation} 
This indicates that, even at very small $x$, gluon shadowing must be
quite small, see Fig.~\ref{glu}. From experimental data it is very
difficult to extract gluon shadowing. For the only existing experimental data NMC \cite{ref-nmc1,ref-nmc2}, a leading
order analysis failed to extract the gluon distribution, and the NLO
fit turned out to be quite sensitive to gluons \cite{nmc-c}. Nevertheless, the results
indicate a very weak gluon shadowing. 

\section{Cronin Effect and nuclear broadening}
\begin{figure}[!t]
   \centerline{\includegraphics[width=8.0 cm] {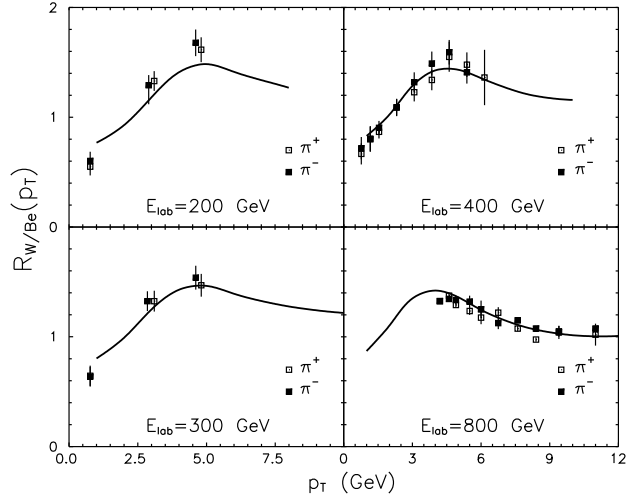}}
\caption{Ratio of the charged pion production cross-sections
for tungsten and beryllium as a function of the transverse momentum of the
produced pions \cite{cor-2}. Experimental data are from Ref.~\cite{ref-cor}. }
\label{ft} 
 \end{figure}

Back in 1973, Cronin's group discovered that nuclei may not only
suppress reactions, but also enhance them \cite{cor-1}. A considerable enhancement was
found for production of hadrons with large transverse momentum. This
effect is measured by the ration $R$, of the inclusive differential
cross-sections for proton scattering on two different targets normalized to the respective atomic numbers $A$ and $B$,
\begin{equation}
R(p_T)=\frac{Bd\sigma_{pA}/d^2p_T}{Ad\sigma_{pB}/d^2p_T}.
\end{equation} 
If there were no any nuclear effect, then we had $R(p_T)=1$; however,
for $A>B$ a suppression is observed experimentally at small $p_T$ and
an enhancement at intermediate $p_T$, and eventually at very
high $p_T$ the ratio seems to approach $R(p_T)=1$, see Fig.~\ref{ft}.

 Experimental data from RHIC \cite{cor-3,cor-33} for high-$p_T$ hadrons in gold-gold collisions
 raised again the long standing problem of quantitative
 understanding the Cronin effect. In nucleus-nucleus collisions this
 effect has to be reliably calculated as a baseline for a signal of new
 physics in heavy ion collisions. The only possibility to test
 models is to make comparisons with available data for $pA$ collisions since in
 $pA$ collisions no hot and dense medium is created.

\begin{figure}[t]
   \centerline{\includegraphics[width=8.0 cm] {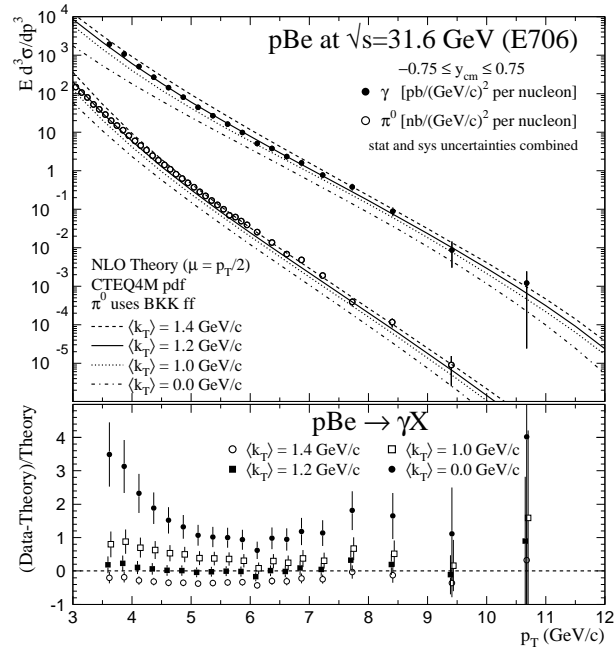}}
\caption{The photon and $\pi^0$ production cross-sections from the E706
experiment at $\sqrt{s}=31.6$ GeV, compared to $k_T$-corrected NLO
calculations \cite{kt-c}. Bottom: the ratio
(Data-Theory)/Theory for direct photon production. Theory is the NLO
calculations with primordial parton momentum $\langle k_T\rangle$. \label{kt-ph} }
\end{figure}

Soon after the discovery of the Cronin effect, it was understood that the
nuclear enhancement is a result of multiple interactions in the nucleus
\cite{cor-1,cor-4}. However, in the parton model based on QCD
factorization, this should be interpreted as a modification of
PDFs in the nucleus. In the parton model, inclusive particle
production for $pA$ collisions can be presented  in a factorized
form,

\begin{equation}
    \frac{d\sigma^h_{pA}}{d^2p_T} = K \sum\limits_{i,j,k,l}
        F_{i/p} \otimes F_{j/A} 
        \otimes \frac{d\hat\sigma}{d\hat t}(ij\to kl) \, \otimes D^h_k
        \ ,
 \label{pQCD}
\end{equation}
where $d\hat\sigma/d\hat t(ij\rightarrow kl)$ is the pQCD
parton-parton cross-section and $D_k^h(z,{Q'}^2)$ are the
fragmentation functions of a parton $k$ into a hadron $h$ with a
fraction $z$ of the parton momentum. The $K$-factor simulates the NLO
contributions. The proton and nucleus parton distribution
functions were parametrized as 
\begin{equation} 
\fl    F_{i/p} = f_{i/p}(x_i,Q^2) \, 
        \frac{e^{-k_{iT}^2/\langle{k_T^2}\rangle_{p}(b)}}
          {\pi\langle{k_T^2}\rangle_{pA}}  
        \hspace*{.4cm}{\rm{and}}\hspace*{.4cm}
    F_{j/A} = T_A(b) \, f_{j/p}(x_j,Q^2) \, 
        \frac{e^{-k_{jT}^2/\langle{k_T^2}\rangle_{Ap}(b)}}
             {\pi\langle{k_T^2}\rangle_{A}}  \ .
 \label{Fsoft}
\end{equation}
where $f_{i/p(A)}(x,Q^2)$ are the parton distribution functions of the
proton (nucleus). Isospin imbalance was taken into account and nuclear
shadowing is included by the HIJING parametrization
\cite{hij-c}. The results of the calculations \cite{kt-c} are depicted in Fig.~\ref{kt-ph}

Partons were assumed to have an intrinsic transverse momentum with
an average squared value $\langle{k_T^2}\rangle_{pA(Ap)}$ and a Gaussian
distribution. At a soft scale one does not resolve the gluonic
structure of a hadron, but only the valence quarks. The mean
transverse Fermi momentum of these quarks is small $\langle k_0\rangle
\sim \Lambda_{\hbox{QCD}}$.  At higher scale, relevant to hard
reactions, one can resolve the structure of the valence quarks,
i.e. the presence of gluons and sea quarks. Since those are located at
small separations, $r_0$ \cite{spots}, from the valence quark, both have more
intensive intrinsic Fermi motion,

\begin{equation}
\langle k_0^2\rangle \sim 1/r_0^2. \label{ink}
\end{equation}
This is obviously bigger than the scale associated with nucleon size due to
confinement. In the parton model it has been shown that even within
the next-to-leading order (NLO) pQCD correction, experimental data of
heavy-quark pair production \cite{w1}, direct photon production
\cite{kt-c} and DY lepton pair production \cite{w3} can only be 
described if an average primordial momentum as large as $1$ GeV is
included (see also Ref.~\cite{wang}). For example in Fig.~\ref{kt-ph}, the
NLO calculations and data for both direct photons and neutral pion
production are shown \cite{kt-c}. A primordial momentum $\langle k_T\rangle \sim
1.2$ GeV seems to provide the best description of data. 

A projectile parton propagating through a nucleus
experiences multiple interactions increasing its transverse
momentum. Then the parton participating in a hard collision inside the
nucleus has an increased transverse momentum compared to
Eq.~(\ref{ink}), which corresponds to the interaction with a free
proton, 
\begin{equation}
\langle k_{T}^2\rangle_{pA}(b, \sqrt{s})  = \langle k_0^2\rangle + \Delta k^2(b, \sqrt{s}), \label{ink2}
\end{equation}
where $\Delta k^2(b, \sqrt{s})$ is the nuclear broadening.  The nuclear
broadening is crucial for understanding the Cronin effect. Apparently, the
strength of the effect depends on the relative values of the two terms
in Eq.~(\ref{ink2}). In the limit of a weak primordial motion the
effect should be strongest, while in the case of $\langle
k_0^2\rangle>> \Delta k^2$ the effect will disappear. One may expect
$\Delta k^2(b, \sqrt{s})$ to be a function of the number of $pp$
collisions, i.e. $\Delta k^2(b, \sqrt{s})\propto \sigma_{pp}(\sqrt{s})
T_A(b)$, where $\sigma_{pp}$ denotes the nucleon-nucleon inelastic
cross-section. Different parametrizations exist for $\Delta k^2(b,
\sqrt{s})$, thought all seem to be rather ad hoc.  Here we present a
prescription in the framework of colour dipole approach which is free from any
arbitrary parameters.

As we already mentioned in the previous section, the coherence length $l_c$
is an important quantity to understand the effect of multiple parton
rescattering \cite{amirlast}. Therefore the underlying mechanisms of the Cronin
enhancement should also depend on the coherence length. In the case of
incoherent hard interaction, the incoming projectile and outgoing
partons experience multiple soft rescattering leading to a high-$p_T$
enhancement. At very small $x$, we are in the large coherence length
$l_c>>R_A$ regime. Such a coherent regime is relevant for hadron production
at medium large $p_T$ at RHIC, and it dominates a large range of $p_T$
at LHC energies. In addition, in the latter case the Cronin effect is
substantially reduced by shadowing.

In the short-coherence length regime $l_c<<R_A$, one can rely on the
factorized expression Eq.~(\ref{pQCD}) corrected for broadening Eq.~(\ref{ink2}). The letter can be computed within the dipole approach as propagation
of a $q\bar q$ pair through the target nucleus. The final parton
transverse-momentum distribution $dN_i/d^2k_{iT}$ is written as
\cite{tra-c,amirlast}:
\begin{equation}
\fl \frac{dN_{j=q}}{d^2k_{iT}} = \!
        \int \! d^2r_1d^2 r_2\,e^{i\,\vec k_T\,(\vec r_1 - \vec r_2)} 
        \left[ \frac{\langle{k_0^2}\rangle}{\pi}\,
        e^{-\frac12(r_1^2+r_2^2) \langle{k_0^2}\rangle}
        \right] 
        \left[ e^{-\frac12\,\sigma_{\bar qq}(\vec r_1-\vec r_2,x)
        \,T_A(b)} \right].
 \label{dNdkt}
\end{equation}
The first bracket in the above equation represents the contribution of
the proton intrinsic momentum, while the second bracket takes into
account the soft parton rescatterings on target nucleons. We use the
dipole cross-section $\sigma_{\bar qq}$ introduced in Section 10, 
fitted to DIS data. For a gluon when $j=g$ in Eqs.~(\ref{pQCD},\ref{dNdkt}), we
have $\sigma_{\bar qq}\to \sigma_{ gg}=\frac{9}{4}\sigma_{\bar qq}$.

\begin{figure}[!t]
   \centerline{\includegraphics[width=8.0 cm] {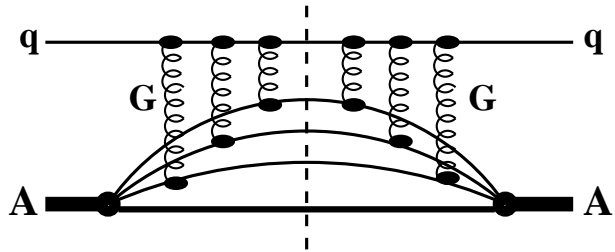}}
\caption{The probability of multiple interactions via one gluon exchange
for the quark in the nucleus. The dashed line shows the unitarity cut. }
\label{ba} 
 \end{figure} 

Notice that
the simple exponential in Eq.~(\ref{dNdkt}) should not be confused
with the Glauber eikonal multiple scattering introduced in the previous
section. Thus if one needs to establish a relation between the
expansion of the exponential in the second bracket of Eq.~(\ref{dNdkt})
and the multiple quark interaction, it would be incorrect to think
that the $n$-th order term of this expansion corresponds to the
probability to have $n$-fold quark multiple scattering (we recall the
probability cannot be negative!).  The appearance of the dipole cross-section in Eq.~(\ref{dNdkt}), is the result of a product of the amplitude and the time-conjugated one, which describe the quarks with different impact parameters. Clearly, the object participating in the scattering is
not a $q\bar{q}$ dipole but rather a single coloured quark, see
Fig.~\ref{ba}. The above prescription describes the fixed
target experiments rather well, see Fig.~\ref{ft}.

In the case of a coherence length $l_c>>R_A$, a hard
fluctuation in the incident proton containing a high-$p_T$ parton
propagates through the whole nucleus and may be freed by the interaction.
Since multiple interactions in the nucleus supply a larger momentum
transfer than a nucleon target, they are able to resolve harder
fluctuations, i.e., the average transverse momentum of produced hadrons
increases. In this case broadening looks like colour filtering rather than
Brownian motion. We employ the light-cone dipole formulation in the target rest frame
which leads to \cite{amirlast}, 
\beq
\sigma^{l_c\gg R_A}_{pA}(p_T) = f_{g/p}\otimes
\sigma(gA\to g_1g_2X)\otimes D_{h/g_1}\ . 
\label{70-co}
 \eeq 
 We assume that high-$p_T$ hadrons originate mainly from radiated
 gluons at such small $x$. The cross-section of gluon radiation reads
\cite{kst2,kst1,kov1}
 \beqn
\fl \frac{d\sigma(gA\to g_1g_2X)}
{d^2p_T\,dy_{1}} =
\int d^2b\int d^2r_1d^2r_2\,
e^{i\vec p_T(\vec r_1-\vec r_2)}\ 
\overline{\Psi_{gg}^*(\vec r_1,\alpha)
\Psi_{gg}(\vec r_2,\alpha)}\nonumber\\
\lo \times\left[1 - e^{-{1\over2}R_g\sigma^N_{3g}(r_1,x)T_A(b)}- e^{-{1\over2}R_g\sigma^N_{3g}(r_2,x)T_A(b)} +
e^{-{1\over2}R_g\sigma^N_{3g}(\vec r_1-\vec r_2,x)T_A(b)} 
\right]\ ,
\label{80-co}
 \eeqn 
where $\alpha = p_+(g_1)/p_+(g)$ is the momentum fraction of
 the radiated gluon. The function $R_g$ incorporates the shadowing effect
 which originates from the higher Fock components $|3g\ra,\ |4g\ra$, etc., missing in the naive
 eikonalization \cite{kst2,cor-2}. $\sigma^N_{3g}(r,\alpha)$ is the dipole cross-section for a
three-gluon colourless system, where $\vec r$ is the transverse separation
of the final gluons $g_1$ and $g_2$. It can be expressed in terms of the
usual $\bar qq$ dipole cross-sections,
 \beq
\sigma^N_{3g}(r) = {9\over 8}\Bigl\{
\sigma_{\bar qq}(r) + \sigma_{\bar qq}(\alpha r) 
+ \sigma_{\bar qq}[(1-\alpha)r]\Bigr\}\ .
\label{90-co}
 \eeq 
 The variable $x$ in $\sigma^N_{3g}(r,\alpha)$ and
 $R_g$ is implicit. The light-cone wave function of the $g_1-g_2$ Fock
 component of the incoming gluon including the nonperturbative
 interaction of the gluons reads \cite{kst2}, 
\beqn  \Psi_{gg}(\vec
 r,\alpha) &=& \frac{\sqrt{8\alpha_s}}{\pi\,r^2}\,
\exp\left[-\frac{r^2}{2\,r_0^2}\right]\,
\Bigl[\alpha(\vec e_1^{\,*}\cdot\vec e)(\vec e_2^{\,*}\cdot\vec r) 
+(1-\alpha)(\vec e_2^{\,*}\cdot\vec e)(\vec e_1^{\,*}\cdot\vec r)\nonumber\\
&-&
\alpha(1-\alpha)(\vec e_1^{\,*}\cdot\vec e_2^{\,*})(\vec e\cdot\vec r)
\Bigr]\
\label{100-co}
 \eeqn
 where $r_0=0.3\fm$ is the parameter characterizing the strength of the
nonperturbative interaction which was fitted to data on diffractive $pp$
scattering. The product of the wave functions is averaged in (\ref{80-co})
over the initial gluon polarization, $\vec e$, and summed over the final
ones, $\vec e_{1,2}$. 
\begin{figure}[!t]
   \centerline{\includegraphics[width=8.0 cm] {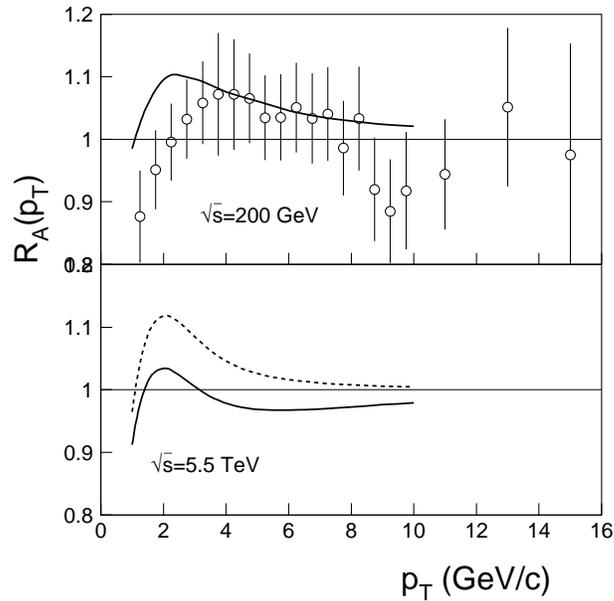}}
\caption{Up panel: Ratio of $p-Au$ to $pp$ cross-sections as function 
of transverse momentum of produced pions at the energy of RHIC $\sqrt{s}=200$. Down panel: Predictions for LHC $\sqrt{s}=5.5$ TeV calculated using Eq.~(\ref{70-co}).
The dashed and solid curves correspond to calculations 
without and with gluon shadowing respectively. The theoretical curves are taken from \cite{cor-2}.
Data are for $d+Au$ collisions from the PHENIX collaboration \cite{ref-cord}.
 }
\label{RL1} 
 \end{figure}

In the upper panel Fig.~\ref{RL1} we show the results for RHIC energy
$\sqrt{s}=200\GeV$. In the lower 
panel we show the prediction for the ratio of pion production rates
in $pA$ and $pp$ collisions obtained using
Eqs.~(\ref{70-co})-(\ref{80-co}) for mid-rapidity at the energy of LHC
$\sqrt{s}=5.5$ TeV \cite{cor-2}. It is seen that the inclusion of the
shadowing effect (solid line) leads to a reduction of the Cronin
effect. Note that this curve approaches to unity from below at high
$p_T$.  We stress that all phenomenological parameters in the above
prescription are fixed in reactions different from $p-A$
collisions. Therefore, these results may be considered as a
free-parameter predictions.

\section{Summary}
During the last half a century, QCD survived through many experimental
tests leading to a consensus that this is a correct theory of strong
interactions. In the asymptotically free region, perturbative QCD has
been quite successful and many QCD perturbative computational tools and
techniques have been developed. This is particularly useful in order to have a detailed understanding of backgrounds
for the search for signals of new physics at the LHC.

Unfortunately, we still have a rather poor understanding of soft
nonperturbative physics which is never avoidable.  Nevertheless, 
QCD-based phenomenology is well developed. Nowadays we are able to
calculate many reaction without having to fit to the data that we want to explain.
On the other hand, the current phenomenology of strong interaction
looks far more complicated and messy than the first principles (QCD
Lagrangian) we started with.

In these lectures we introduced two different approaches to 
high-energy QCD phenomenology: the parton model and the colour dipole
formalism. We discussed the relevance of both methods as an
efficient way to include the non-perturbative features of QCD via fitting
to some experimental data and predicting others. In the case of parton 
model one fits the universal parton distributions, which then allow 
one to predict other reactions by combining these PDFs with perturbative 
calculations. Next-to-leading-order corrections and higher-twist effects make 
this program more difficult. In the case of the dipole approach, the universal 
phenomenological function is the dipole-proton cross-section, which is 
mainly fitted to DIS data from HERA. This description by default includes 
the higher-order and higher-twist corrections. However, this is expected to 
work only at very small Bjorken $x$ and is not useful at large $x$ where 
valence quarks dominate the PDFs.

The LHC is expected to become a laboratory for gluo-dynamics, which should
settle many of the controversies in our understanding of small Bjorken $x$ physics.
LHC data should bring forth important information on the gluonic
structures in the proton.
The currently observed steep rise of the gluon density is expected to be slowed down by
 saturation. This is still debatable, since even in $pp$ at the 
Tevatron saturation is reached only for central collisions.

The forthcoming LHC data with nuclear beams will reveal the gluonic
structure of nuclei. They should resolve the controversy about the
magnitude of gluon shadowing. The saturation scale in nuclei is expected
to reach values of a few GeV, leading to strong observable effects.

%%%%%%%%%%%%%%%%%%%%%%%%%%%%%%%%%%%%%%%%%%%%%%%%%%%%%%%%%%%%%%%%%%%%%%%%%%%%%%%%%%%%%%%%%%%%%%%%%%%%%%%%%
%%%%%%%%%%%%%%%%%%%%%%%%%%%%%%%%%%%%%%%%%%%%%%%%%%%%%%%%%%%%%%%%%%%%%%%%%%%%%%%%%%%%%%%%%%%%%%%%%%%%%%%%%%

%%%%%%%%%%%%%%%%%%%%%%%%%%%%%%%%%%%%%%%%%%%%%%%%%%%%%%%%%%%%%%%%%%%%%%%%%%%%%%%%%%%%%%%%%%%%%%%%%%%%%%%%%%
%%%%%%%%%%%%%%%%%%%%%%%%%%%%%%%%%%%%%%%%%%%%%%%%%%%%%%%%%%%%%%%%%%%%%%%%%%%%%%%%%%%%%%%%%%%%%%%%%%%%%%%%%%

\ack
This work was supported in part by Fondecyt (Chile) grants 1070517 and
1050589 and by DFG (Germany) grant PI182/3-1.

%%\endrefs

\end{document}